\documentclass{aa}
\usepackage{graphicx, amsmath, color, mathrsfs}
\graphicspath{{/Users/p_salati/Dropbox/super_code_positrons/article_CRAC_1_as_is/figures/},{figures/},{./}}
\usepackage{txfonts}
\usepackage{natbib,twoopt}
\usepackage[breaklinks=true]{hyperref} 
\bibpunct{(}{)}{;}{a}{}{,}             
\makeatletter

\newcommandtwoopt{\citeads}[3][][]{\href{http://adsabs.harvard.edu/abs/#3}
	{\def\hyper@linkstart##1##2{}
	\let\hyper@linkend\@empty\citealp[#1][#2]{#3}}}

\newcommandtwoopt{\citepads}[3][][]{\href{http://adsabs.harvard.edu/abs/#3}
	{\def\hyper@linkstart##1##2{}
	\let\hyper@linkend\@empty\citep[#1][#2]{#3}}}

\newcommandtwoopt{\citetads}[3][][]{\href{http://adsabs.harvard.edu/abs/#3}
	{\def\hyper@linkstart##1##2{}
	\let\hyper@linkend\@empty\citet[#1][#2]{#3}}}

\newcommandtwoopt{\citeyearads}[3][][]
	{\href{http://adsabs.harvard.edu/abs/#3}
	{\def\hyper@linkstart##1##2{}
	\let\hyper@linkend\@empty\citeyear[#1][#2]{#3}}}

\makeatother
\usepackage{hyperref}
\usepackage{hypcap}
\definecolor{cyan}{cmyk}{1.,0.,0.,0.2}
\definecolor{vert}{cmyk}{0.5,0.,0.5,0.5}
\definecolor{magenta}{cmyk}{0.,1.,0.,0.1}
\definecolor{verdatre}{cmyk}{0.5,0.,0.5,0.5}
\definecolor{vert_clair}{cmyk}{0.5,0.,0.5,0.2}
\definecolor{yellow}{cmyk}{0.,0.,1.,0.0}
\definecolor{yellow_1}{cmyk}{0.,0.,0.5,0.0}
\definecolor{rouge}{cmyk}{0.,0.4,0.6,0.0}
\definecolor{orange}{cmyk}{0.,0.5,0.5,0.05}
\definecolor{violet}{rgb}{0.5,0.,0.5}
\definecolor{darwin_box}{rgb}{0.988,0.878,0.77}
\definecolor{darwin_text}{rgb}{0.1,0.07,0.02}
\newcommand{\beq}{\begin{equation}}
\newcommand{\eeq}{\end{equation}}
\newcommand{\bea}{\begin{eqnarray}}
\newcommand{\ena}{\end{eqnarray}}

\newcommand{\ie}{i.e.}
\begin{document}
\title{A new look at the cosmic ray positron fraction}

\author{
M.~Boudaud \inst{1}
\and S.~Aupetit \inst{1}
\and S.~Caroff \inst{2}
\and A.~Putze \inst{1,2}
\and G.~Belanger \inst{1}
\and Y.~Genolini \inst{1}
\and C.~Goy \inst{2}
\and V.~Poireau \inst{2}
\and V.~Poulin \inst{1}
\and S.~Rosier \inst{2}
\and P.~Salati \inst{1}
\and L.~Tao \inst{2}
\and M.~Vecchi \inst{2,3}
\fnmsep\thanks{The authors are members of the Cosmic Ray Alpine Collaboration.}
\fnmsep\thanks{Contact authors are mathieu.boudaud@lapth.cnrs.fr, antje.putze@lapth.cnrs.fr, sami.caroff@lapp.in2p3.fr and yoann.genolini@lapth.cnrs.fr.}
}

\institute{
LAPTh, Universit\'e de Savoie \& CNRS, 9 Chemin de Bellevue, B.P.110 Annecy-le-Vieux, F-74941, France\\
%
\and
LAPP, Universit\'e de Savoie \& CNRS, 9 Chemin de Bellevue, B.P.110 Annecy-le-Vieux, F-74941, France\\
%
\and
Instituto de Fisica de Sa\~{o} Carlos - Av. Trabalhador sa\~{o}-carlense, 400
CEP: 13566-590 - Sa\~{o} Carlos (SP), Brazil\\
}
%

\date{Received; accepted\\
Preprint numbers : LAPTH-224/14}
%

%
%
\abstract
%
{The positron fraction in cosmic rays has recently been measured with improved accuracy up to 500 GeV, and
it was found to be a steadily increasing function of energy above $\sim$ 10 GeV. This behaviour contrasts with standard
astrophysical mechanisms, in which positrons are secondary particles, produced in the interactions of primary cosmic rays during
their propagation in the interstellar medium. The observed anomaly in the positron fraction triggered a lot of excitement, as
it could be interpreted as an indirect signature of the presence of dark
matter species in the Galaxy, the so-called weakly interacting massive particles or WIMPs.
Alternatively, it could be produced by nearby sources, such as pulsars.}
{
These hypotheses are probed in light of the latest AMS-02 positron fraction measurements. As
regards dark matter candidates, regions in the annihilation cross section to mass plane, which best fit the
most recent data, are delineated and compared to previous measurements. The explanation of the
anomaly in terms of a single nearby pulsar is also explored.}
%
%
{
The cosmic ray positron transport in the Galaxy is described using a semi-analytic two-zone model.
Propagation is described with Green functions as well as with Bessel expansions. For consistency, the
secondary and primary components of the positron flux are calculated together with the same propagation
model. The above mentioned
explanations of the positron anomaly are tested using $\chi^{2}$ fits.
The numerical package MicrOMEGAs is used to model the positron flux generated by dark matter species.
The description of the positron fraction from conventional astrophysical sources is based on the pulsar observations included in the
Australia Telescope National Facility (ATNF) catalogue.
}
%
%
{
The masses of the favoured dark matter candidates
are always larger than 500~GeV, even though the results are very sensitive to the lepton flux.
The Fermi measurements point systematically to much heavier candidates than the recently released
AMS-02 observations. Since the latter are more precise, they are much more constraining.
A scan through the various individual annihilation channels disfavours leptons as the final state.
On the contrary, the agreement is excellent for quark, gauge boson, or Higgs boson pairs, with best-fit
masses in the 10 to 40 TeV range.
The combination of annihilation channels that best matches the positron fraction is then determined
at fixed WIMP mass. A mixture of electron and tau lepton pairs is only acceptable around 500~GeV.
Adding $b$-quark pairs significantly improves the fit up to a mass of 40~TeV.
Alternatively, a combination of the four-lepton channels provides a good fit between 0.5 and 1~TeV,
with no muons in the final state.
Concerning the pulsar hypothesis, the region of the distance-to-age plane that best fits the positron fraction
for a single source is determined.
}
%
%
{
The only dark matter species that fulfils the stringent gamma ray and cosmic microwave background bounds is a particle annihilating
into four leptons through a light scalar or vector mediator, with a mixture of tau (75\%) and electron
(25\%) channels, and a mass between 0.5 and 1~TeV.
The positron anomaly can also be explained by a single pulsar, and a list of five pulsars
from the ATNF catalogue is given. We investigate how this list could evolve when more statistics are
accumulated.
Those results are obtained with the cosmic ray transport parameters that best fit the B/C ratio.
Uncertainties in the propagation parameters turn out to be very significant. In the WIMP annihilation
cross section to mass plane for instance, they overshadow the error contours derived from the positron
data.
}

\keywords{cosmic ray positron anomaly -- dark matter particles -- WIMP indirect signatures -- pulsars}
\maketitle
%
%
\section{Introduction}

The cosmic ray positron flux at the Earth exhibits above 10~GeV an excess with respect to the astrophysical background
produced by the interactions of high-energy protons and helium nuclei with the interstellar medium (ISM). Observations
by the High-Energy Antimatter Telescope (HEAT)
collaboration\citepads{1997ApJ...482L.191B,2001ApJ...559..296D,2004PhRvL..93x1102B} already hinted
at a slight deviation of the flux with respect to a pure secondary component. The anomaly was clearly established
by\citetads{2009Natur.458..607A} with measurements of the positron fraction by the
Payload for Antimatter Matter Exploration and Light-nuclei Astrophysics (PAMELA) satellite, up to 100~GeV.
Drawing any definite conclusion about the nature of the positron excess requires nevertheless precise measurements. The
release by \citet{Aguilar:2013qda} of data with unprecedented accuracy can be seen as a major step forward, which opens
the route for precision physics. Recently, the Alpha Magnetic Spectrometer (AMS-02) collaboration has published \citep{Accardo:2014lma} an update on the
positron fraction based on high statistics with measurements extending up to 500~GeV.

\vskip 0.1cm
The observed excess of positrons was readily interpreted as a hint of the presence of dark matter particles in the Milky Way halo.
A  number of dark matter candidates have been proposed so far. The most favoured option is a weakly interacting massive particle
(WIMP), whose existence is predicted by several theoretical extensions of the standard model of particle physics.
The typical cross section for WIMP annihilation in the early Universe has to be close to $3 \times 10^{-26}$ cm$^{3}$ s$^{-1}$ to match the cosmological value of
$\Omega_{\rm DM} \approx 0.27$.
Although marginal, WIMP annihilations are still going on today, especially in the haloes of galaxies where dark matter (DM)
has collapsed, and where they produce various cosmic ray species. The hypothesis that the positron anomaly could be
produced by the annihilation of DM particles is supported by the fact that the energy of the observed
excess lies in the GeV to TeV range, where the WIMP mass is expected.

\vskip 0.1cm
The initial enthusiasm for interpreting the positron anomaly as an indirect signature for DM particles has
nevertheless been dampened by several observations.
First, since positrons rapidly lose energy above 10~GeV as they propagate in the Galactic magnetic
fields, the positron excess is produced near the Earth, where the DM density $\rho_{\chi}(\odot)$ is known to be of order
$0.3$ GeV cm$^{-3}$ as shown by\citetads{2012ApJ...756...89B}. Bearing in mind the benchmark values for the
WIMP annihilation cross section and density, one finds that the signal is too small to account for the observed excess.
For a WIMP mass $m_{\chi}$ of 1~TeV, the positron production rate needs to be enhanced by a factor of a thousand
to match the measurements.
The second difficulty lies in the absence, up to 100~GeV, of a similar excess in the antiprotons, since the PAMELA
measurements of the antiproton-to-proton ratio \citep{Adriani:2008zq} and of the absolute flux
\citep{Adriani:2010rc,Adriani:2012paa} are consistent with the expected astrophysical background of secondary species.
DM particles cannot couple with quarks under the penalty of overproducing antiprotons as shown by\citetads{2009NuPhB.813....1C}
and confirmed by\citetads{2009PhRvL.102g1301D}. Therefore, besides an abnormally large annihilation rate today,
WIMPs should preferentially annihilate into charged leptons, a feature which is unusual in supersymmetry.
An additional obstacle against the WIMP interpretation of the positron anomaly arises from the
lack of DM signatures in the electromagnetic radiation measurements.
Regardless of their origin, electrons and positrons undergo inverse Compton scattering on the cosmic microwave background (CMB) and
on stellar light. The emission of gamma rays, observable by atmospheric Cherenkov detectors such as
the High Energy Stereoscopic System (H.E.S.S.),
the Major Atmospheric Gamma-ray Imaging Cherenkov (MAGIC) telescope, and
the Very Energetic Radiation Imaging Telescope Array System (VERITAS),
as well as satellite-borne devices like
Fermi-LAT, is expected as a result of these processes.
Photons can also be directly produced by WIMP annihilation or radiated
by final state charged leptons. Finally, electrons and positrons also spiral in Galactic magnetic fields. The resulting
synchrotron emission should not outshine what is already collected by radio telescopes.
%

%
Several analyses have been carried out on these messengers. They partially and sometimes completely exclude the regions
of the WIMP annihilation cross section to mass plane that are compatible with the positron anomaly, even though their conclusions
may be found to be dependent on the astrophysical assumptions on which they are based.
For example, the limits derived by\citetads{2012JCAP...01..041A} from the H.E.S.S. measurements of the Galactic
centre\citepads{2011PhRvL.106p1301A} vanish if the DM profile is taken to be isothermal. This is in agreement with the
conclusions drawn by\citetads{2010NuPhB.840..284C}.
Constraints from\citetads{2013arXiv1310.0828T} based on observations of dwarf spheroidal satellites of the Milky
Way\citepads[see also]{2011PhRvL.107x1303G,2011PhRvL.107x1302A} depend on the DM profiles assumed for these objects.
These limits have been improved by VERITAS\citepads{2012PhRvD..85f2001A} and MAGIC\citepads{2014JCAP...02..008A}
with dedicated searches for DM in Segue 1, and by the H.E.S.S. collaboration\citepads{2014arXiv1410.2589H}. The $J$ factors of
the satellites are somewhat uncertain but the bounds are very stringent.
Following\citetads{2010NuPhB.840..284C},\citetads{2012ApJ...761...91A} restrained WIMP properties with
Fermi-LAT observations of the gamma ray diffuse emission from regions located at moderate Galactic latitude.
This procedure alleviates the sensitivity of the constraints to the DM profile. The limits are quite robust if only
the inverse Compton and final state radiation photons from DM annihilation are considered. They become more stringent once the
Galactic gamma ray diffuse emission is taken into account, but then are sensitive to how it is modelled.
The analysis by\citetads{2010JCAP...04..014A} of the extra-Galactic gamma ray background (EGB) could jeopardise
the DM interpretation of the positron excess, but depends on how the Galactic diffuse emission has been subtracted and
is extremely sensitive to the DM structure scenario used to derive the WIMP contribution to the EGB. Unresolved blazars
and millisecond pulsars also need to be withdrawn from the gamma ray measurements to yield the EGB. This is also
the case for star-forming galaxies and secondary electromagnetic showers induced by ultra-high energy cosmic rays,
both components being highly model dependent as mentioned by\citetads{2012PhRvD..85b3004C}.
The IceCube collaboration has set limits on the DM annihilation cross section by searching for high energy neutrinos in nearby galaxies and galaxy
clusters\citepads{2013PhRvD..88l2001A}. The properties of WIMP are constrained between 300~GeV and 100~TeV
for a variety of annihilation channels including neutrino pairs. The DM explanation of the positron excess is
partially challenged by the IceCube limits derived for the Virgo cluster, in the case of the
$\tau^{+} \tau^{-}$ and $\mu^{+} \mu^{-}$ channels, even though a very large boost of nearly a thousand from DM
subhaloes needs to be assumed for this conclusion to hold. In addition, independent studies foresee severe limits on
the properties of decaying WIMPs using the IceCube experiment (see for example \citeads{2010PhRvD..81d3508M}).
Finally, the CMB provides stringent constraints on WIMP annihilation and mass~{\citepads{2009PhRvD..80b3505G, 2009PhRvD..80d3526S, 2009JCAP...10..009C}.

\vskip 0.1cm
Many solutions have been proposed to circumvent part of these difficulties.
As regards the first problem, one possibility is to artificially increase the annihilation cross section well above
its thermal value.
To commence, it is conceivable that the universe was not radiation dominated at WIMP freeze-out,
contrary to what is commonly assumed, but
underwent a period of very fast expansion. This would have been the case, for instance, if a fast rolling scalar field
had taken over radiation at that time, during a stage of so-called kinetic quintessence\citepads{2003PhLB..571..121S}.
The larger the Hubble rate at WIMP decoupling, the larger the annihilation cross section at fixed relic abundance.
Another option is to disentangle the annihilation cross section from the relic abundance. Dark matter species could have
been produced by the decay of a heavier partner after freeze-out, as suggested by\citetads{2008arXiv0807.1508G},
but the most popular solution relies on the Sommerfeld effect. In this scenario, independently proposed
by\citetads{2009PhLB..671..391P} and\citetads{2009PhRvD..79a5014A}, DM species couple to a light
scalar or vector boson, which makes them attract each other when they annihilate, hence an enhanced cross
section especially at low relative velocities. 
A quarkophilic DM candidate heavier than about 10~TeV also permits us to overcome the antiproton
problem as mentioned by\citetads{2009NuPhB.813....1C}, although future measurements could
jeopardise that possibility as discussed by\citetads{2013JCAP...04..015C}.
Another solution to intensify the production of positrons is based on the existence of DM clumps in the vicinity
of the Earth. In this case, the annihilation is enhanced. However, the astrophysical boost on the positron flux cannot exceed a factor of $\sim 20$~\citepads{2008A&A...479..427L}, unless a DM clump is located near the Earth, although this is very unlikely~\citepads{2009PhRvD..80c5023B}.
Finally, although the above-mentioned gamma ray constraints are fairly stringent, there is still room for a WIMP
explanation of the positron excess. Most of the limits so far derived concentrate on species lighter than 10~TeV and
rely on specific assumptions. The case of secluded DM\citepads{2009PhLB..671..391P,2009PhRvD..79a5014A}
is also rarely addressed.

\vskip 0.1cm
A completely different approach relies on the existence of pulsars in the Earth vicinity. These conventional astrophysical sources
are known to exist and a few of them have been detected nearby. Highly-magnetised neutron stars can emit
electromagnetic radiation as they spin, even if the magnetic field dipole is aligned on the rotation
axis\citepads{1983bhwd.book.....S}, and very strong electric fields are generated. At the surface, they extract and accelerate
charged particles, which subsequently interact with the magnetic field or the thermal emission of the pulsar to trigger an electromagnetic
cascade\citepads{1974MNRAS.167....1R}. This yields an electron-positron plasma which, for a pulsar wind nebula, drifts away
from the star to form a shock on the surrounding medium. Acceleration takes place there until the reverse shock from the supernova
explosion releases in the ISM the positrons and electrons confined so far.
Shortly after\citetads{2009Natur.458..607A} confirmed the positron anomaly,\citetads{2009JCAP...01..025H} showed
that the observations could easily be explained in this framework. Their conclusion was confirmed
by\citetads{2012CEJPh..10....1P}.
Recently,\citetads{2013ApJ...772...18L} concluded that either Geminga or Monogem, two well-known nearby pulsars,
could produce enough positrons to account for the AMS-02 precision measurements \citep{Aguilar:2013qda}.

\vskip 0.1cm
It is timely to reanalyse the cosmic ray positron excess in the light of the latest AMS-02 release \citep{Accardo:2014lma}
and to thoroughly explore whether or not DM particles or a local conventional astrophysical source can account for this anomaly.
We are motivated to conduct this analysis for three reasons.
%
First, the AMS-02 data are of unprecedented accuracy and extend up to 500~GeV, a region so far unexplored.
As their quality improves, measurements become more and more constraining and can rule out scenarios that previously
matched the observations, as already noticed by\citetads{2013PhRvD..88b3013C}.
Then, while most of the analyses rely on a theoretical prediction of the electron flux at the Earth, we have used available
measurements of the total electron and positron flux $\Phi_{e^{+}} + \Phi_{e^{-}}$ (hereafter lepton flux) when deriving
the positron fraction. Our procedure alleviates the uncertainty arising from primary electrons, a component that is
generated by supernova shocks and cannot be derived from first principles. The lepton flux introduced at the denominator
of the positron fraction ${\Phi_{e^{+}}}/({\Phi_{e^{+}} + \Phi_{e^{-}}})$ has been actually measured  and it is not just
provided by theory. This could lead to differences in the preferred WIMP cross section and mass regions with respect
to previous studies.
Finally, uncertainties in modelling the cosmic ray transport are also expected to affect the results.
Radio data, for instance, preclude thin Galactic magnetic haloes as shown by\citetads{2012JCAP...01..049B}.
This is confirmed by\citetads{2013JCAP...03..036D} who analyse the positron flux below a few GeV and set a limit of
2~kpc on the half-height $L$. This limit has been improved recently by\citetads{2014PhRvD..90h1301L}, who conclude
that the PAMELA positron flux\citepads{2013PhRvL.111h1102A} disfavours values of $L$ smaller than 3~kpc,
and excludes the so-called MIN
propagation model defined in\citetads{2004PhRvD..69f3501D}
\citepads[see also]{2001ApJ...555..585M}. We have therefore investigated the sensitivity of the WIMP and pulsar parameter space
on cosmic ray propagation.

\vskip 0.1cm
The paper is organised as follows.
In Sec.~\ref{sec:CR_transport}, we recall the salient features of the cosmic ray transport model, which we
use to derive the primary (signal) and secondary (background) components of the positron flux.
Our analysis is based on a semi-analytic approach that makes clear the various processes at stake
and whose rapidity allows us to efficiently explore the parameter space. More comprehensive codes such as
GALPROP\citepads{1998ApJ...509..212S,1998ApJ...493..694M} and DRAGON\citepads{2013PhRvL.111b1102G}
offer a more realistic description of diffusion and energy losses, but are heavier to manipulate and appear less suited for
performing chi-square fits and fast scans over extended regions of parameters.
Different measurements of the lepton flux are also presented. The fitting procedure is detailed, with a description of
the errors, which come into play in the calculation of the $\chi^{2}$.
Section~\ref{sec:DM_analysis} is devoted to the DM interpretation of the positron excess. We scan a variety of
annihilation channels and delineate for each of them the preferred regions in the WIMP cross section to mass plane.
Our results strongly depend on the lepton flux data that are used to compute the expected positron fraction.
The transition from the Fermi-LAT
data\citepads{2010PhRvD..82i2004A} to the latest measurements by AMS-02 presented at the ICHEP 2014
conference \citep{leptons_ICHEP_2014} implies a significantly lower WIMP mass.
We then look for the best combination of branching ratios at fixed WIMP mass and explore several possibilities.
In Sec.~\ref{sec:single_pulsar}, we challenge the statement by\citetads{2013ApJ...772...18L} that Geminga
or Monogem could alone account for the positron anomaly. We explore whether or not a single source is still a
viable explanation given the precision reached by the latest AMS-02 observations.
The study of the uncertainty arising from cosmic ray transport is discussed in Sec.~\ref{sec:CR_uncertainty}.
A best-fit model is found for each annihilation channel and is presented with the associated WIMP parameters.
We then discuss in Sec.~\ref{sec:discussion_conclusion} our results in the light of the above-mentioned
astrophysical constraints and look for DM scenarios that still evade them and finally we conclude.

%
\section{Cosmic ray transport and the positron fraction}
\label{sec:CR_transport}

Charged cosmic rays propagate through the magnetic fields of the Milky Way and are deflected by
its irregularities: the Alfv\'en waves. Although the magnetic turbulence is strong, cosmic ray transport
can still be modelled  as a diffusion process\citepads[see for instance]{2002PhRvD..65b3002C} and
Fick's law applies. The diffusion coefficient can be taken of the form
\beq
K(E) = K_{0} \; \beta \,
\left( {\mathcal R}/{\rm 1 \; GV} \right)^{\delta} \, ,
\label{space_diffusion_coefficient}
\eeq
where $\beta$ denotes the positron velocity ${v}/{c}$, expressed in units of the speed of light $c$,
and $K_{0}$ is a normalisation constant. The diffusion coefficient $K$ increases as a power law
with the rigidity ${\mathcal R} = {p}/{q}$ of the particle.
Positrons also lose energy as they diffuse. They spiral in the Galactic magnetic fields, emitting synchrotron
radiation, and they undergo Compton scattering on the CMB and stellar light. Energy losses occur at a rate
$b(E)$ which, in the energy range considered in this analysis, can be approximated by
\beq
b(E) = \frac{E_{0}}{\tau_{E}} \, \epsilon^{2} \, ,
\label{eq:b_loss}
\eeq
where $E_{0} =$ 1 GeV and $\epsilon = {E}/{E_{0}}$. For simplicity, the typical energy loss timescale
has been set equal to the benchmark value of $10^{16}$~s.
More sophisticated modellings of the energy loss rate $b(E)$ are available in public
codes such as GALPROP\citepads{1998ApJ...493..694M} and DRAGON\citepads{2013PhRvL.111b1102G}.
As thoroughly discussed by\citetads{2010A&A...524A..51D}, the loss rate $b(E)$ does not vary quadratically
with energy. It also depends on the exact value assumed for the local magnetic field and depends on the position
inside the magnetic halo. However,
between 1 and 6~$\mu$G, relation~(\ref{eq:b_loss}) yields
the correct answer within a factor of 2 and is a very good approximation for 3~$\mu$G.
We will investigate in a forthcoming publication
how the constraints on DM species and pulsars are affected by the modelling of the loss rate $b(E)$.
%
Taking diffusion and energy losses into account, the time dependent transport equation can be written as
\beq
\frac{\partial \psi}{\partial t} \, - \, \pmb{\nabla} \cdot \left\{ K(E) \, \pmb{\nabla} \psi \right\}
\, - \, \frac{\partial}{\partial E} \left\{ b(E) \, \psi \right\} = q(\vec{x} , t , E) \, ,
\label{eq:CR_prop_1}
\eeq
where $q$ denotes the production rate of positrons and
$\psi = {d^{4}N}/{{d^{3} \vec{x}} \, {dE}} \equiv {dn}/{dE}$ is the cosmic ray positron density per unit
of volume and energy. Diffusion has been assumed to be homogeneous
everywhere inside the Galactic magnetic halo (MH) so that the coefficient $K$ depends only on the
energy $E$ and not on the position $\vec{x}$.

\vskip 0.1cm
The stationary version of the transport equation~(\ref{eq:CR_prop_1}) can be reformulated as
a heat diffusion problem by translating the energy $E$ into the pseudo-time $\tilde{t}$ through the identity
\beq
K_{0} \, \tilde{t}(E) = {\displaystyle \int_{E}^{+ \infty}} \frac{K(E')}{b(E')} \, dE' \, .
\eeq
This leads to
\begin{align}
\frac{\partial \tilde{\psi}}{\partial \tilde{t}} \, - \, K_{0} \, \Delta \tilde{\psi} &=
\tilde{q}(\vec{x} , \tilde{t}) \label{eq:CR_prop_2}  \\
\rm{where} \quad
\tilde{\psi} &= \frac{b(E)}{b(E_{0})} \, \psi \nonumber \\
\rm{and} \quad
\tilde{q} &= \frac{b(E)}{b(E_{0})} \, \frac{K(E_{0})}{K(E)} \, q \, . \nonumber
\end{align}
The solution of the heat diffusion equation~(\ref{eq:CR_prop_2}) is obtained by modelling the MH as a thick
disc that matches the circular structure of the Milk Way. The Galactic disc of stars and gas, where primary
cosmic rays are accelerated, lies in the middle. Primary species, such as protons, helium nuclei, and electrons,
are presumably accelerated by the shock waves driven by supernova explosions. These take place mostly in the
Galactic disc, which extends radially 20~kpc from its centre, and has a half-thickness $h$ of 100~pc. Confinement
layers, where cosmic rays are trapped by diffusion, lie above and beneath this thin disc of gas. The intergalactic
medium starts at the vertical boundaries $z = \pm L$, as well as beyond a radius of $r = R \equiv 20$ kpc.
Within the MH, the steady production of positrons with energy $E_{S}$ at position $\vec{x}_{S}$ with rate $q$
leads at position $\vec{x}$ to the density $\psi$ of positrons with energy $E$
\beq
\psi(\vec{x} , E) = \frac{1}{b(E)} \, {\displaystyle \int_{E}^{+ \infty}} dE_{S} \,
{\displaystyle \int_{\rm MH}} d^{3}{\vec{x}_{S}} \,
\tilde{G}(\vec{x} \leftarrow {\vec{x}_{S}} ; \lambda_{D}) \; q(\vec{x}_{S} , E_{S}) \, ,
\eeq
where $\tilde{G}$ denotes the Green function associated with the heat equation~(\ref{eq:CR_prop_2}). This
propagator depends on the energies $E$ and $E_{S}$ through the typical diffusion length $\lambda_{D}$
such that
\beq
\lambda_{D}^{2} = 4 \, K_{0} \left\{ \tilde{t}(E) - \tilde{t}(E_{S}) \right\} .
\label{eq:lambdaD}
\eeq

\vskip 0.1cm
The derivation of the Green function $\tilde{G}$ is detailed in\citetads{2009A&A...501..821D,2010A&A...524A..51D}
\citepads[see also]{1974Ap&SS..29..305B,1990acr..book.....B}
where the MH is pictured as an infinite slab without radial boundaries.
In the regime where the diffusion length $\lambda_{D}$ is small with respect to the MH half-thickness $L$,
the method of the so-called electrical images consists of
implementing \citepads{1999PhRvD..59b3511B} an infinite series over
the multiple reflections of the source as given by the vertical boundaries at $+L$ and $-L$.
In the opposite regime, a large number of images needs to be considered and the convergence of the series is a problem.
Fortunately, the diffusion equation along the vertical axis boils down to the Schr\"{o}dinger equation, written in imaginary
time, that accounts for the behaviour of a particle inside an infinitely deep 1D potential well that extends from $z = - L$ to
$z = + L$. The solution may be expanded as a series over the eigenstates of the corresponding Hamiltonian~\citepads{2007A&A...462..827L}.
None of those methods deal with the radial boundaries at $r = R$. The MH is not an infinite slab but is modelled
as a flat cylinder. The Bessel approach presented by\citetads{2008PhRvD..77f3527D} solves that problem by
expanding the positron density $\psi$ and production rate $q$ along the radial direction as a series of Bessel
functions of zeroth-order $J_{0}(\alpha_{i} r / R)$. Since $\alpha_{i}$ is the $i^{\mathrm{th}}$ zero of $J_{0}$, the density
vanishes at $r = R$. The method also makes use of a Fourier expansion along the vertical axis.

\vskip 0.1cm
The astrophysical background
consists of the secondary positrons produced by
the collisions of high-energy protons and helium nuclei on the atoms of the ISM. The corresponding flux at the Earth may
be expressed as the convolution over the initial positron energy
\beq
\Phi_{e^{+}}^{\rm sec}(\odot , E) = \frac{c}{4 \pi} \, \frac{1}{b(E)} \, {\displaystyle \int_{E}^{+ \infty}} dE_{S} \,
{\cal I}_{\rm disc}(\lambda_{D}) \; q_{e^{+}}^{\rm sec}(\odot , E_{S}) \, .
\eeq
The disc integral ${\cal I}_{\rm disc}$, which comes into play in that expression, is defined as
\beq
{\cal I}_{\rm disc}(\lambda_{D}) \equiv {\displaystyle \int_{\rm MH}} d^{3}{\vec{x}_{S}} \,
\tilde{G}(\odot \leftarrow {\vec{x}_{S}} ; \lambda_{D}) \, ,
\eeq
and depends on the energies $E$ and $E_{S}$ through $\lambda_{D}$, and also on the half-height $L$.
The production rate of secondary positrons $q_{e^{+}}^{\rm sec}$ can be safely calculated at the Earth
and has been derived as in\citetads{2009A&A...501..821D}. In particular, Green functions have been used
instead of the Bessel method, which would have required too many Fourier modes along the vertical axis.

\vskip 0.1cm
The DM signal consists in the primary positrons produced by the WIMP annihilations taking place in the MH.
Assuming that the DM species are identical particles, like Majorana fermions, leads to the source term
\beq
q_{e^{+}}^{\rm DM}(\vec{x}_{S} , E_{S}) = \frac{1}{2} \, \langle \sigma v \rangle \,
\left\{ \frac{\rho_{\chi}(\vec{x}_{S})}{m_{\chi}} \right\}^{2}
\left\{ g(E_{S}) \equiv \sum_{i} B_{i} \left. \frac{dN_{e^{+}}}{dE_{S}} \right|_{i} \right\} ,
\eeq
where $m_{\chi}$ and $\rho_{\chi}$ denote the WIMP mass and density, respectively, while
$\langle \sigma v \rangle$ is the annihilation cross section averaged over the momenta of the
incoming DM species. The sum runs over the various possible annihilation channels $i$ with branching
ratio $B_{i}$ so that $g(E_{S})$ is the resulting positron spectrum at the source.
Once propagation is taken into account, the positron flux at the Earth is given by
\beq
\Phi_{e^{+}}^{\rm DM}(\odot , E) = \frac{c}{4 \pi} \, \frac{\Gamma(\odot)}{b(E)} \,
{\displaystyle \int_{E}^{\displaystyle m_{\chi}}} dE_{S} \, {\cal I}_{\rm halo}(\lambda_{D}) \; g(E_{S}) \, ,
\label{eq:flux_DM_1}
\eeq
where $\Gamma(\odot)$ stands for the DM annihilation rate per unit volume in the solar neighbourhood,
\beq
\Gamma(\odot) = \frac{1}{2} \, \langle \sigma v \rangle \,
\left\{ \frac{\rho_{\chi}(\odot)}{m_{\chi}} \right\}^{2} .
\eeq
The halo integral, which comes into play in the convolution~(\ref{eq:flux_DM_1}), is defined as
\beq
{\cal I}_{\rm halo}(\lambda_{D}) \equiv {\displaystyle \int_{\rm MH}} d^{3}{\vec{x}_{S}} \,
\tilde{G}(\odot \leftarrow {\vec{x}_{S}} ; \lambda_{D}) \,
\left\{ \frac{\rho_{\chi}(\vec{x}_{S})}{\rho_{\chi}(\odot)} \right\}^{2} .
\eeq
%
%
%
%
The Green method is used to compute $\tilde{G}$ for small values of the diffusion length $\lambda_{D}$
whereas the Bessel expansion is preferred in the opposite situation, when the positron sphere starts to probe
the radial boundaries of the MH. The transition between these two regimes occurs at 3~kpc. This
value is lowered at 0.3~kpc for a half-height $L$ of the MH smaller than 3~kpc. In that case, the convergence
of the Bessel series is achieved by taking 200 orders along the radial direction and 50 vertical harmonics.
The halo integral, which is a function of $\lambda_{D}$, also depends on $L$ and on the DM
distribution within the Galaxy. Throughout this analysis, we have assumed for the latter
the parameterisation described in~\citetads[][hereafter NFW]{1997ApJ...490..493N} profile, with
\beq
\rho_{\chi}(r) = \rho_{\chi}(\odot) \, \left( \frac{r_{\odot}}{r} \right)
\left\{ \frac{r_{s} + r_{\odot}}{r_{s} + r} \right\}^{2} \, ,
\eeq
where $r$ denotes the radius in spherical coordinates and $r_{s} = 20$~kpc is the typical NFW scale radius.
%
%
%
The galactocentric distance of the solar system has been set equal to $ r_{\odot} = 8.5$~kpc while a fiducial value of
$0.3$ GeV cm$^{-3}$ has been taken for the local DM density $\rho_{\chi}(\odot)$\citepads[following]{2012ApJ...756...89B}.
The NFW distribution exhibits a ${1}/{r}$ cusp, which we have replaced by the smoother profile
of\citetads{2008PhRvD..77f3527D} within 0.1~kpc of the Galactic centre.

\vskip 0.1cm
In the case of pulsars, which can be modelled as point-like sources in space and time, the time-dependent transport
equation~(\ref{eq:CR_prop_1}) can be solved with the Green function method. The positron propagator in that case
describes the probability that a particle released at position $\vec{x}_{S}$ and time $t_{S}$ with energy $E_{S}$
is observed at position $\vec{x}$ and time $t$ with energy $E$. It can be easily related to the steady state propagator
through
\beq
G_{e^{+}}(\vec{x} , t , E \leftarrow {\vec{x}_{S}} , t_{S} , E_{S}) = \frac{b(E_{\star})}{b(E)} \,
\tilde{G}(\vec{x} \leftarrow {\vec{x}_{S}} ; \lambda_{D}) \, \delta(E_{S} - E_{\star}) \, ,
\eeq
where $E_{\star}$ denotes the energy at which a positron needs to be injected to be detected
with energy $E$ after a laps of time $t - t_{S}$. This initial energy is related to the age $t_{\star}$
of the source by
\beq
t_{\star} \equiv t - t_{S} = {\displaystyle \int_{E}^{E_{\star}}} \frac{dE'}{b(E')} \, .
\label{eq:age_E_star}
\eeq
We assume that pulsars release instantaneously positrons. A source located at position $\vec{x}_{\star}$
with age $t_{\star}$ contributes then at the Earth a flux
\beq
\Phi_{e^{+}}^{\rm psr}(\odot , E) = \frac{c}{4 \pi} \, \frac{b(E_{\star})}{b(E)} \,
\tilde{G} \left\{ \vec{x} \leftarrow {\vec{x}_{\star}} ; \lambda_{D}(E,E_{\star}) \right\} g(E_{\star}) \, ,
\label{eq:flux_PSR_1}
\eeq
should a value of the injection energy $E_{\star}$ exist that satisfies the age relation~(\ref{eq:age_E_star}).
In that respect, the positron spectrum exhibits a high-energy cut-off at the Earth arising from energy losses. Even if
the injection energy $E_{\star}$ is infinite, the positron energy $E$ after a time $t_{\star}$ cannot exceed a maximal
bound. The positron spectrum at the source is parameterised by
\beq
g(E) = Q_{0} \, \left( \frac{E_{0}}{E} \right)^{\gamma} \, \exp(- {E}/{E_{C}}) \, .
\label{eq:PSR_injection_spectrum}
\eeq
The normalisation constant $Q_{0}$ is determined by requiring that the total energy provided by the pulsar
to the positrons above an energy $E_{\rm min}$ is a fraction $f$ of the initial spinning
energy $W_{0}$. This leads to
\beq
{\displaystyle \int_{E_{\rm min}}^{+ \infty}} E_{S} \, g(E_{S}) \,  dE_{S} = f W_{0} \, .
\eeq
When pulsars form, they initially rotate with a period as small as a few milliseconds. The initial kinetic energy of
a 3~ms pulsar is of the order of $10^{51}$ ergs, or equivalently $10^{54}$ GeV, which sets the natural unit in
which we will express in Sec.~\ref{sec:single_pulsar} the energy $f W_{0}$ carried out by positrons.
The energy $E_{C}$ in relation~(\ref{eq:PSR_injection_spectrum}) is a cut-off in the injection spectrum.
It has been set equal to 1~TeV throughout our analysis. The exact value does not matter much, since the
high-energy cut-off of the positron spectrum at the Earth comes from the age $t_{\star}$ of the pulsar and not
from a cut-off at the source\citepads{2009PhRvD..80f3005M}.

\vskip 0.1cm
We then compute the total positron flux at the Earth $\Phi_{e^{+}} = \Phi_{e^{+}}^{\rm sec} + \Phi_{e^{+}}^{\rm prim}$,
where the primary component is produced either by DM particles or by pulsars. The calculation is performed consistently
with the same cosmic ray propagation model for both components. In most of this work, we have used the MED configuration,
which best fits the boron to carbon ratio B/C\citepads[see]{2004PhRvD..69f3501D}. In Sec.~\ref{sec:CR_uncertainty}, we
study how changing the transport parameters affects the DM and pulsar results, and gauge the effects of cosmic ray propagation
uncertainties.
Since the positron excess appears at high energy, we have concentrated our analyses above 10~GeV, where
we can safely ignore diffusive reacceleration and convection\citepads{2009A&A...501..821D}.
The former mechanism originates from the motion, with Alfv\'en velocity $V_A$, of the magnetic diffusion centres in the Galactic frame and induces a diffusion
in energy and a reacceleration of cosmic rays. This process can sweep particles out of the MH along the vertical direction with a convection velocity $V_c$.
Above 10~GeV, solar modulation is a subdominant effect. We have taken that process into account by modelling it with the force-field
approximation\citepads{1971JGR....76..221F}, with a potential $\phi_{\rm F}$ of 600 MV.
%

%
\begin{figure*}[ht!]
\begin{center}
\includegraphics[width=0.75\textwidth]{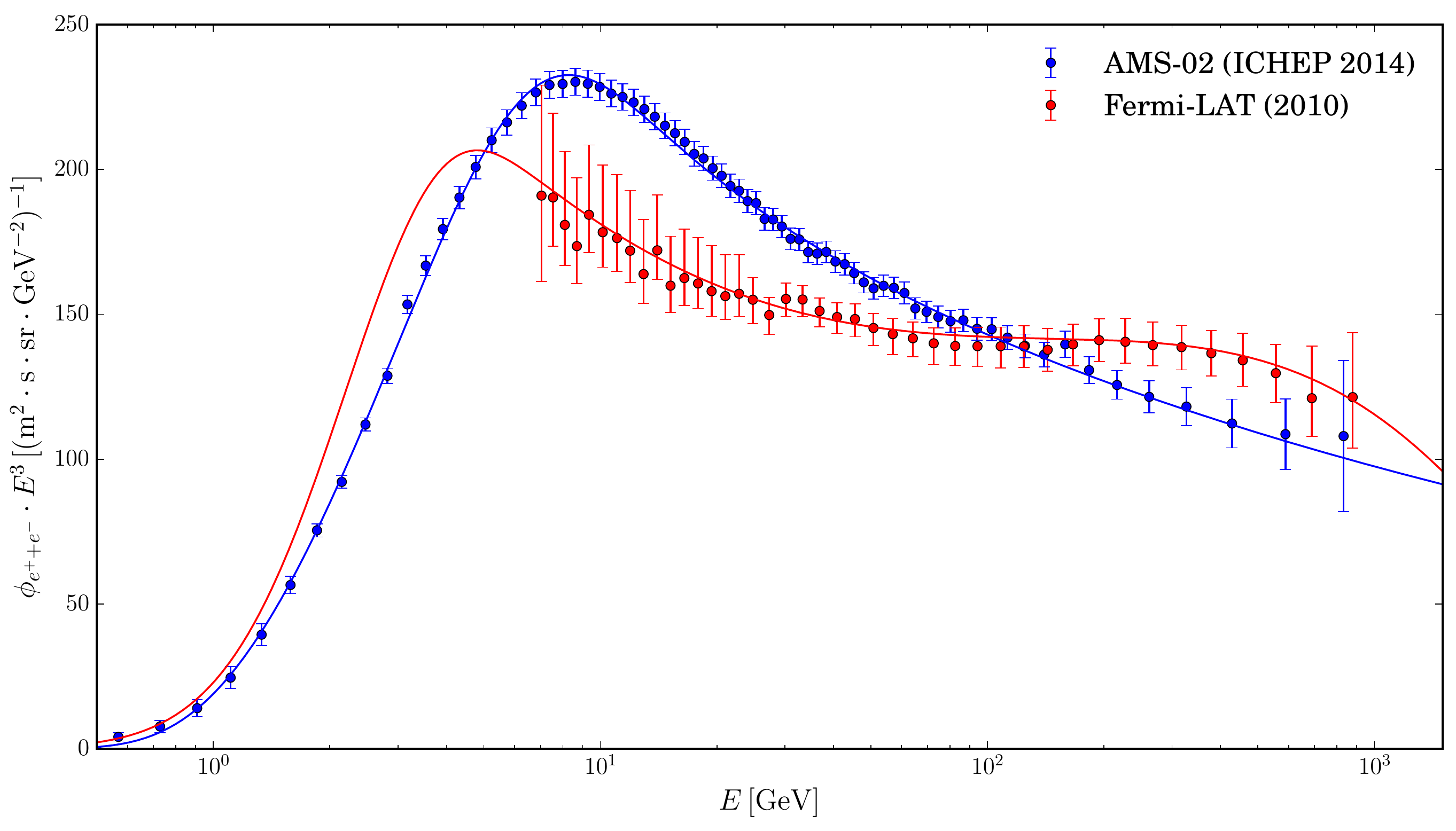}
\caption{
The lepton flux $\Phi_{L} = \Phi_{e^{+}} + \Phi_{e^{-}}$ is plotted as a function of the energy $E$. We included
a rescaling factor of $E^{3}$ to make clear the power-law behaviour of the flux at high energy. The red points correspond
to the Fermi-LAT data\citepads{2010PhRvD..82i2004A} whereas the blue ones stand for the more recent measurements
by AMS-02 presented at the ICHEP 2014 conference \citep{leptons_ICHEP_2014}. The lines are fits to the data.}
\label{fig:lepton_fluxes}
\end{center}
\end{figure*}
%

\vskip 0.1cm
The positron fraction is defined as ${\rm PF} = {\Phi_{e^{+}}}/{\Phi_{L}}$, i.e. the ratio between the positron flux and the lepton flux
$\Phi_{L} = \Phi_{e^{+}} + \Phi_{e^{-}}$. Usually, the electron flux
is derived theoretically to get $\Phi_{L}$. However, contrary to positrons, the astrophysical background of electrons
has a strong contribution, which is accelerated with nuclear species in supernova shock waves. This primary component is very
model dependent and relies, in particular, on the actual three-dimensional distribution of sources in the Galactic spiral
arms\citepads{2013PhRvL.111b1102G,2014PhRvD..89h3007G}.
We have therefore used measurements of the lepton flux $\Phi_{L}$ to derive the positron fraction more accurately.
The most recent measurements of the total lepton flux come from the space-borne experiments Fermi-LAT\citepads{2010PhRvD..82i2004A},
PAMELA\citepads{2011PhRvL.106t1101A,2013PhRvL.111h1102A}, and AMS-02 \citep{leptons_ICHEP_2014}. Whereas the data published
by the PAMELA and AMS-02 collaborations are consistent within their respective uncertainties, both are in tension with the Fermi-LAT result (see Fig.1).
Our analysis strongly depends on the lepton flux. We henceforth anticipate significant variations in our results, as will be illustrated in Section~\ref{sec:DM_analysis}.
Using the experimental lepton flux implies an additional error $\sigma_{\rm PF}^{\rm th} = {\rm PF} \times ({\sigma_{\Phi_{\rm L}}}/{\Phi_{\rm L}})$
on the positron fraction, arising from the error bars ${\sigma_{\Phi_{\rm L}}}$ of the lepton flux data. This uncertainty is partially
anti-correlated with the published errors $\sigma_{\rm PF}^{\rm mes}$ on the positron fraction measurements. To be conservative,
we have added $\sigma_{\rm PF}^{\rm mes}$ and $\sigma_{\rm PF}^{\rm th}$ in quadrature to get the so-called corrected errors
$\sigma_{\rm PF}$, which we have included in our $\chi^{2}$ analyses. The measured $\sigma_{\rm PF}^{\rm mes}$ and corrected
$\sigma_{\rm PF}$ errors are very close to each other when the AMS-02 ICHEP 2014 lepton flux is used. This is not the case for Fermi-LAT
where ${\sigma_{\Phi_{\rm L}}}/{\Phi_{\rm L}}$ can be substantial, as featured in Fig.~\ref{fig:lepton_fluxes}.

The use of the positron fraction data from AMS-02~\citep{Accardo:2014lma} instead of the positron flux \citepads{2014PhRvL.113l1102A} results in a more precise estimation of the extra component parameters. This comes from the combination of less systematic uncertainties on the fraction and the lepton flux as well as higher statistics on the lepton flux. The published AMS-02 positron flux has been reconstructed using more severe cuts insuring the purity of the data sample, but leading to larger error bars. Since this published flux is compatible with the estimated flux obtained by multiplying the published positron fraction with the lepton flux, we expect similar results in both cases.
Finally, our DM and pulsar studies are based on the minimisation of the $\chi^{2}$
\beq
\chi^{2} = \sum_{i} \left\{ \frac{{\rm PF}^{\rm \, mes}(i) - {\rm PF}^{\rm \, th}(i)}{\sigma_{\rm PF}(i)} \right\}^{2} \, ,
\label{eq:chi_2_def}
\eeq
where the sum runs on the data points $i$ whose energies exceed 10~GeV. We have checked that increasing this threshold
from 10 to 15 or 20~GeV does not affect our results. The reduced-$\chi^{2}$ ($\chi^{2}_{\rm dof}$) is obtained by dividing
the result of equation~(\ref{eq:chi_2_def}) by the number of degrees of freedom, {\ie} the number of data points minus the
number of parameters over which the fit is performed.
%
In a forthcoming work, we will use the positron flux to investigate the robustness of our DM and pulsar results. 

%
\section{Dark matter analysis}
\label{sec:DM_analysis}

As a first interpretation of the AMS-02 results, we investigate the possibility that the excess of positrons at high energies originates
from DM annihilation. Contrary to recent studies which consider specific DM models (see for example \citeads{2013arXiv1307.6204B}), we make no assumptions about the underlying DM model and explore different annihilation channels.
The positron flux resulting from DM annihilation is computed with
micrOMEGAs$\_{3.6}$\citepads{2011CoPhC.182..842B,2014CoPhC.185..960B} for the MED set of propagation parameters and
the positron fraction is obtained using the lepton spectrum measured by AMS-02 \citep{leptons_ICHEP_2014}.

%
\begin{table*}[ht!]
\centering
\caption{Best fits for specific DM annihilation channels assuming the MED propagation parameters. The recently published
positron fraction \citep{Accardo:2014lma} and the AMS-02 lepton spectrum \citep{leptons_ICHEP_2014} are used to
derive the $\chi^{2}$, as in formula~(\ref{eq:chi_2_def}). The {\it p}-value, indicated in the last column, is defined
in formula~(\ref{eq:p-value}).}
\label{tab:MED_onechannel_PF14_ICHEP14}
\vspace{.1cm}
\begin{tabular}{|c|c|c|c|c|c|c|}
\hline
{Channel} &  {$m_\chi \, [{\rm TeV}]$} & {$\langle \sigma v \rangle \, [{\rm cm^{3} \, s^{-1}}]$} & {$\chi^2$} & {$\chi^{2}_{\rm dof_{}}$} & $p$ \\
\hline
e & $0.350 \pm  0.004$ & $(2.31 \pm 0.02) \cdot 10^{-24}$ & 1489 & 37.2 & 0 \\
$\mu$ & $0.350 \pm 0.003$ & $(3.40 \pm 0.03) \cdot 10^{-24}$ & 346 & 8.44 & 0  \\
$\tau$ & $0.894 \pm 0.040$ & $(2.25 \pm 0.15) \cdot 10^{-23}$ & 93.0 & 2.27 & $4.2 \cdot 10^{-6}$ \\
$u$ &  $31.5 \pm 2.9$ & $(1.43 \pm 0.20) \cdot 10^{-21}$ & 25.2 & 0.61 & 0.97 \\
$b$ &  $27.0 \pm 2.2$ & $(1.00 \pm 0.12) \cdot 10^{-21}$ & 26.5 & 0.65 & 0.95 \\
$t$ &  $42.5 \pm 3.3$ & $(1.81 \pm 0.21) \cdot 10^{-21}$ & 29.4 & 0.72 & 0.89 \\
$Z$ &  $14.2 \pm 0.9$ &$(6.02 \pm 0.58) \cdot 10^{-22}$ & 43.8 & 1.07 & 0.31 \\
$W$ &  $12.2 \pm 0.08$ & $(5.10 \pm 0.48) \cdot 10^{-22}$ & 41.1 & 1.00 & 0.42 \\
$H$ &  $23.2 \pm 1.5$ & $(8.17 \pm 0.77) \cdot 10^{-22}$ & 39.1 & 0.95 & 0.51 \\
$\phi \rightarrow e$ &  $0.350 \pm 0.0008$ & $(1.56 \pm 0.01) \cdot 10^{-24}$ & 534 & 13.0 & 0 \\
$\phi \rightarrow \mu$ & $0.590 \pm 0.022$ & $(5.87 \pm 0.36) \cdot 10^{-24}$ & 175 & 4.27 & 0 \\
$\phi \rightarrow \tau$ & $1.76 \pm 0.08$ & $(4.51 \pm  0.32) \cdot 10^{-23}$ & 83.5 & 2.04  & $7.7 \cdot 10^{-5}$ \\
\hline
\end{tabular}
\end{table*}
%

%
\begin{table*}[ht!]
\centering
\caption{Same as in Table~\ref{tab:MED_onechannel_PF14_ICHEP14} with the first measurements of the positron fraction
released by AMS-02 \citep{Aguilar:2013qda} and the Fermi lepton flux\citepads{2010PhRvD..82i2004A}. The DM mass is
systematically larger than for the previous analysis based on more recent data. The large error bars of the Fermi lepton flux translate
into small values for the $\chi^{2}$.}
\label{tab:MED_onechannel_PF13_Fermi}
\vspace{.1cm}
\begin{tabular}{|c|c|c|c|c|c|c|}
\hline
{Channel} &  {$m_\chi \, [{\rm TeV}]$} & {$\langle \sigma v \rangle \, [{\rm cm^{3} \, s^{-1}}]$} & {$\chi^2$} & {$\chi^{2}_{\rm dof_{}}$} & $p$ \\
\hline
$e$ & $0.260 \pm 0.046$ & $(9.35 \pm 0.18) \cdot 10^{-25}$ & 119 & 2.97& $0$ \\
$\mu$ & $0.621 \pm 0.049$ & $(6.71 \pm 0.91) \cdot 10^{-24}$ & 11.8 & 0.29 & $0.99$ \\
$\tau$ & $2.49 \pm 0.32$ & $(9.07 \pm 1.95)\cdot10^{-23}$ & 7.68 & 0.19 & $0.99$\\
$u$ & $227 \pm 57$ & $(2.56 \pm 1.07) \cdot 10^{-20}$ & 12.4 & 0.31 & $0.99$ \\
$b$ & $186 \pm 52$ & $(1.68 \pm 0.80) \cdot 10^{-20}$ & 12.7 & 0.32 & $0.99$ \\
$t$ & $237 \pm 55$ & $(2.16 \pm 0.84) \cdot 10^{-20}$ & 11.4 & 0.28 & $0.99$ \\
$Z$ & $55.2 \pm 9.3$ & $(4.08 \pm 1.16) \cdot 10^{-21}$ & 8.87 & 0.22 & $0.99$ \\
$W$ & $49.3 \pm 8.0$ & $(3.66 \pm 1.00) \cdot 10^{-21}$ & 9.00 & 0.22 & $0.99$ \\
$H$ & $98.0 \pm 17.5$  & $(6.32 \pm 1.89) \cdot 10^{-21}$ & 9.62 & 0.24 & $0.99$ \\
$\phi \rightarrow e$ & $0.447 \pm 0.322$ & $(1.78 \pm 0.24) \cdot 10^{-24}$ & 15.2 & 0.38 & $0.99$ \\
$\phi \rightarrow \mu$ & $1.31 \pm 0.15$ & $(1.65 \pm 0.29) \cdot 10^{-23}$ & 8.00 & 0.20 & $0.99$ \\
$\phi \rightarrow \tau$ & $5.07 \pm 0.71$ & $(1.92 \pm  0.45) \cdot 10^{-22}$ & 7.85 & 0.19 & $0.99$ \\
\hline
\end{tabular}
\end{table*}
%

%
\subsection{Single annihilation channel analysis}
\label{subsec:single_channel}

Assuming a specific DM annihilation channel, we scan over two free parameters, the annihilation cross section $\langle \sigma v \rangle$
and the mass $m_{\chi}$ of the DM species. A fit to the AMS-02 measurements of the positron fraction is performed using MINUIT to
determine the minimum value of the $\chi^{2}$ defined in relation~(\ref{eq:chi_2_def}). We find that the data can be fitted very well, i.e. with
$\chi^{2}_{\rm dof} \le 1$, for annihilation channels into quark and boson final states as featured in
Table~\ref{tab:MED_onechannel_PF14_ICHEP14}.
In each case, the preferred DM mass is above 10~TeV and the annihilation cross section is at least a factor $10^{4}$ larger than the canonical cross section.
Leptophilic DM candidates, in particular for the channels $e^{+}e^{-}$ and $\mu^{+}\mu^{-}$, feature a sharp drop in the positron spectrum at
the DM mass. The favourite DM mass is therefore much lower than for hadronic channels. Fitting both the low and high energy part of the spectrum
with only two free parameters is in both cases difficult, leading to a poor overall $\chi^{2}$. The situation is better for the $\tau^{+}\tau^{-}$ channel,
however the best fit corresponds only to $\chi^{2}_{\rm dof} \approx 2.3$ for a DM mass near $900~{\rm GeV}$.
The case where DM annihilates into four leptons, for example through the annihilation into a pair of new scalar (or vector) particles that decay
into lepton pairs, provides an interesting alternative. Each four-lepton channel leads to a better fit than the corresponding two-lepton channel.
Nevertheless the best fit for the $4\tau$ channel is near $\chi^{2}_{\rm dof}=2.04$ with a preferred mass of $m_{\chi} = 1.76$~TeV.
The spectrum for the positron fraction corresponding to the best fit for the $b \bar{b}$ and $4\tau$ channels is compared with the AMS-02 data in
Fig.~\ref{fig:bb_4tau_bestfit}.
%
\begin{figure*}[ht!]
\begin{center}
\includegraphics[width=0.49\textwidth]{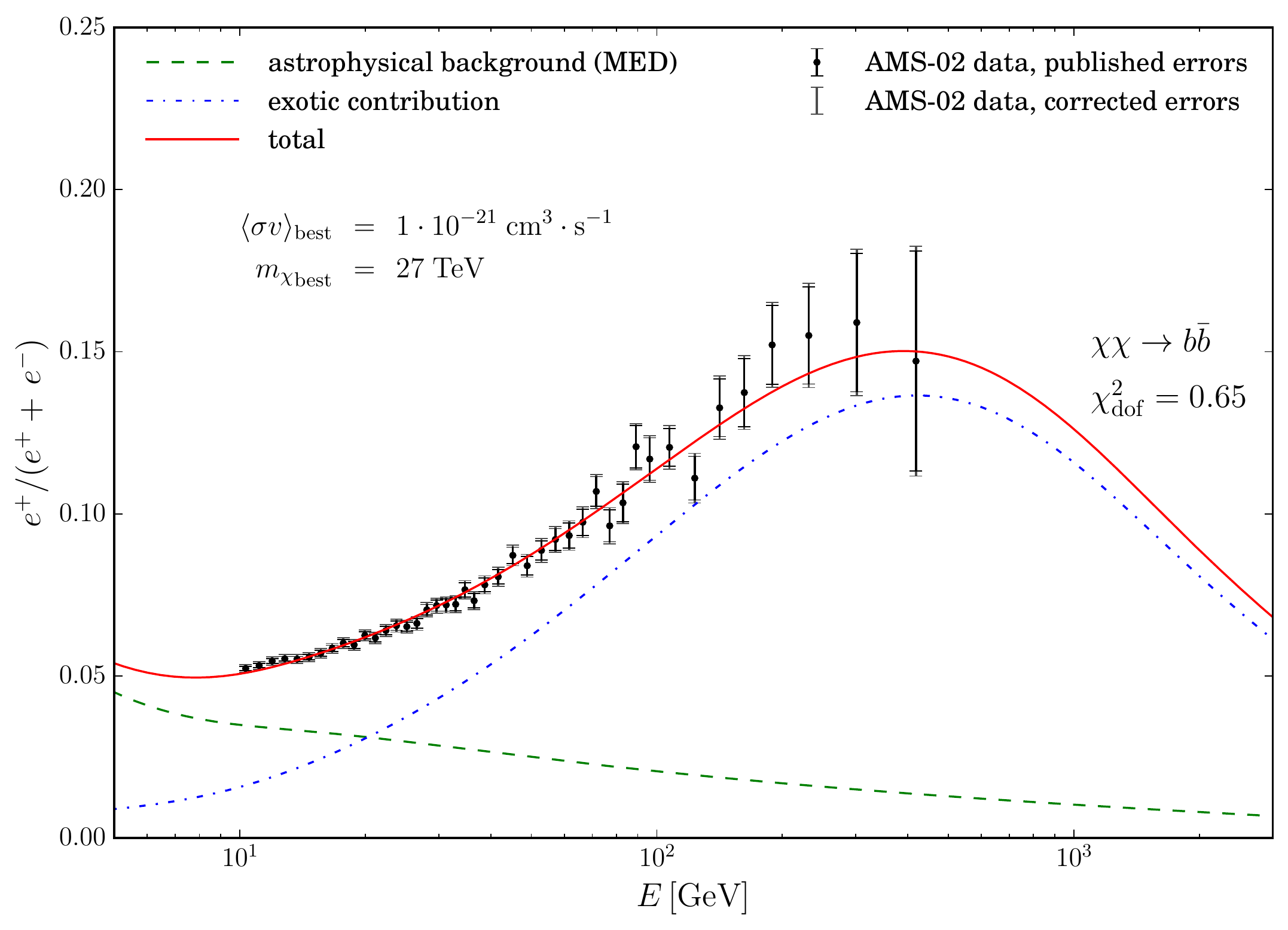}
\includegraphics[width=0.49\textwidth]{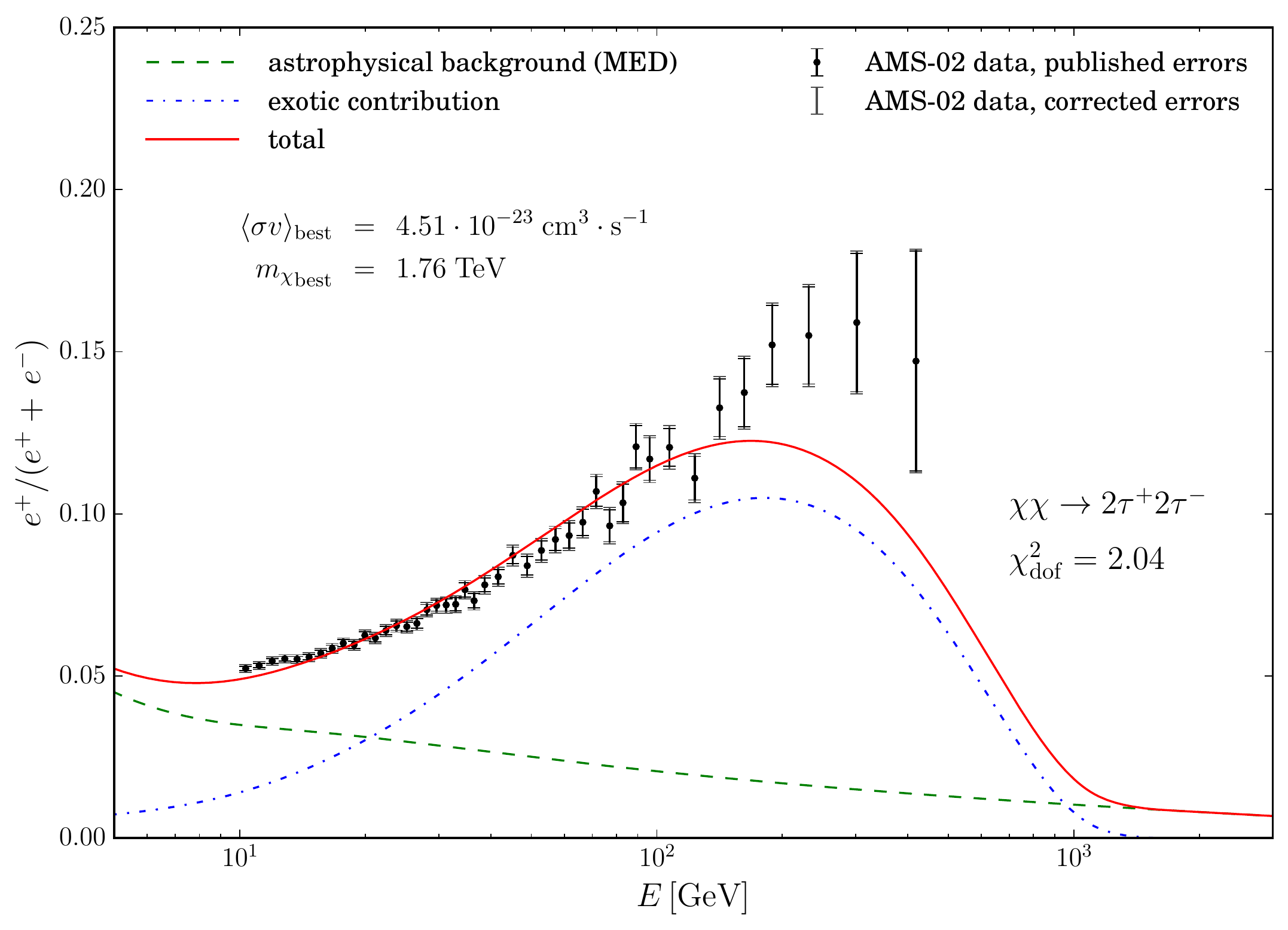}
\caption{
Positron fraction as a function of the positron energy corresponding to the best-fit value of $\langle \sigma v \rangle$ and DM mass
$m_{\chi}$ for $b \bar{b}$ (left) and $4\tau$ annihilation channels (right), compared with AMS data~\citep{Accardo:2014lma}. The propagation
parameters correspond to the MED model. The AMS-02 lepton spectrum \citep{leptons_ICHEP_2014} is used to derive the $\chi^{2}$.}
\label{fig:bb_4tau_bestfit}
\end{center}
\end{figure*}
%
%
\begin{figure*}[ht!]
\begin{center}
\includegraphics[width=0.49\textwidth]{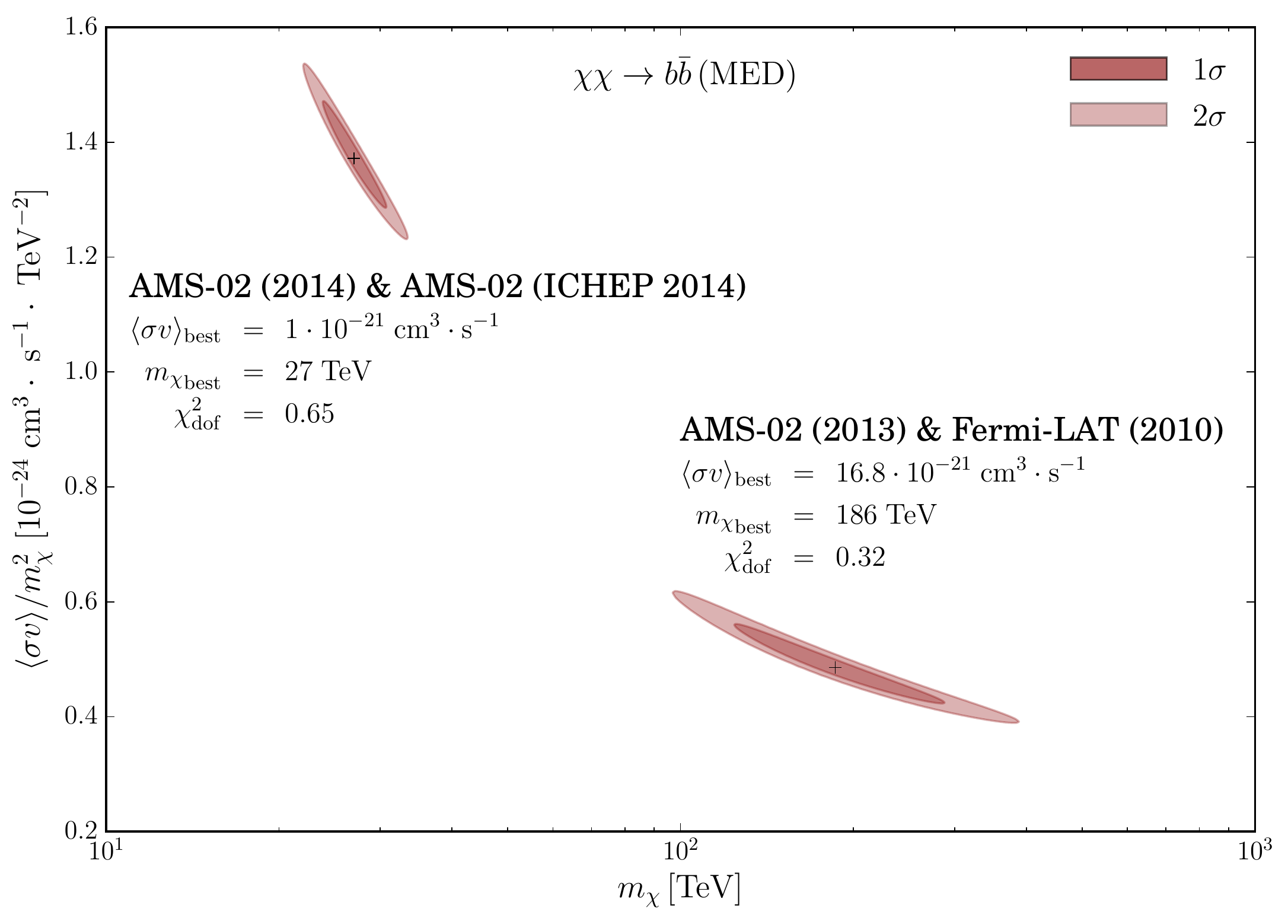}
\includegraphics[width=0.49\textwidth]{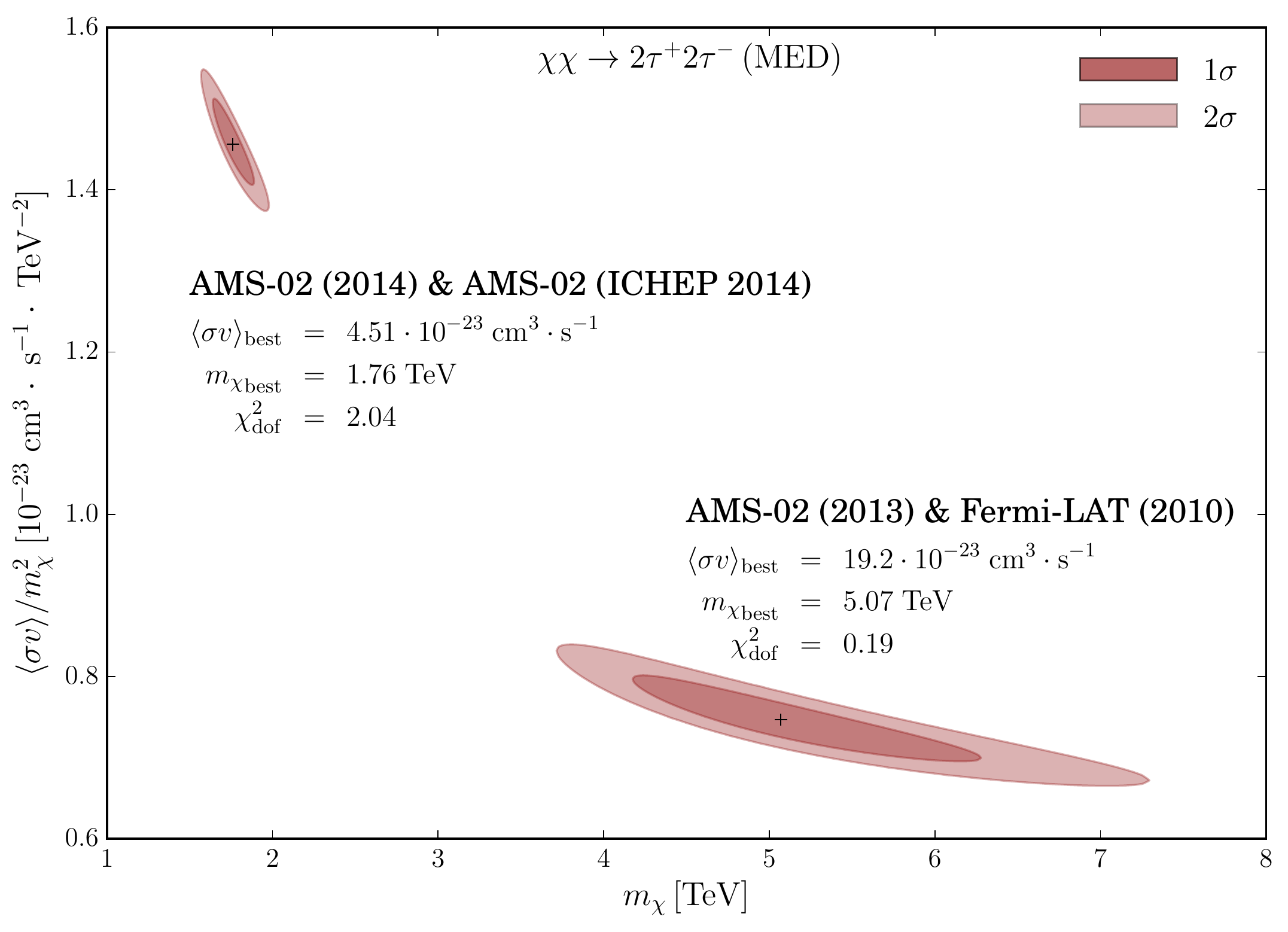}
\caption{
Comparison of the best fits in the  ${\langle \sigma v \rangle}{/m_{\chi}^{2}}$ and $m_\chi$ plane using the lepton spectrum
measured by AMS-02 \citep{leptons_ICHEP_2014} and Fermi\citepads{2010PhRvD..82i2004A}. We assume DM annihilating into
$b \bar{b}$ pairs (left) and $4\tau$ (right) while the propagation parameters correspond to the MED model. The $1\sigma$ and
$2\sigma$ regions are defined by a difference $\chi^{2} - \chi_{\rm min}^{2}$ of 2.30 and 6.18, which
are typical of a two-dimensional fit.}
\label{fig:bb_4tau_contour}
\end{center}
\end{figure*}
%

\vskip 0.1cm
The measurement of the lepton flux performed by AMS-02 has a significant impact on the DM interpretation of the positron fraction.
In particular these results systematically point towards lighter DM candidates and with lower annihilation cross sections than those
obtained using the flux of Fermi as is clear from a comparison between Tables~\ref{tab:MED_onechannel_PF14_ICHEP14} and
\ref{tab:MED_onechannel_PF13_Fermi}. The shift in the $2\sigma$ allowed region in the $\langle \sigma v \rangle$ - $m_\chi$ plane
using the AMS-02 \citep{leptons_ICHEP_2014} or the Fermi\citepads{2010PhRvD..82i2004A} lepton flux is clearly displayed in
Fig.~\ref{fig:bb_4tau_contour} for both the $b \bar{b}$ and $4\tau$ channels.

\vskip 0.1cm
However because of the much smaller error bars of the AMS-02 lepton flux with respect to those of Fermi, the overall $\chi^{2}$ values
displayed in Table~\ref{tab:MED_onechannel_PF14_ICHEP14} are not as good as those of Table~\ref{tab:MED_onechannel_PF13_Fermi}.
As the precision of measurements improves, the goodness of fit lessens. To better illustrate this point, we calculate the {\it p}-value from the
$\chi^{2}_{n}$ test statistic with $n$ degrees of freedom obtained from each fit
\beq
p = 1 - \frac{\gamma (n/2 , \chi^{2}_{n}/2)}{\Gamma(n/2)} \, ,
\eeq
where $\gamma$ and $\Gamma$ are the lower incomplete and complete gamma functions, respectively. We furthermore define two critical
{\it p}-values for which we accept the resulting fit based on a 1 ($p > 0.3173$) and 2 ($p > 0.0455$) standard deviation ($\sigma$) significance
level for a normal distribution
\beq
p = 1 - \Phi ({\cal N} \sigma) = 1 - \text{erf} ({\cal N} \sigma/\sqrt{2}),
\label{eq:p-value}
\eeq
where ${\cal N} \sigma$ is the number of standard deviations and $\Phi$ is the cumulative distribution function of the Gaussian distribution.
We readily conclude from Table~\ref{tab:MED_onechannel_PF13_Fermi} that analyses based on the Fermi lepton flux cannot discriminate
among the various annihilation channels as the corresponding {\it p}-values are larger than 0.99, except in the case of the $e^{+}e^{-}$ pair
production, which always provides a bad fit. 
Taking the recent measurements of the lepton flux by AMS-02 into account \citep{leptons_ICHEP_2014}
strongly favors quark and boson channels. The {\it p}-values quoted in
Table~\ref{tab:MED_onechannel_PF14_ICHEP14} are larger than $0.31$ for these channels whereas they are vanishingly small for the
leptonic channels. For consistency in the data sets, we restrict our study in the following to the AMS-02 lepton flux for the positron fraction calculation.

%
\subsection{Combination of channels}
\label{subsec:combi_channel}

The description of DM annihilation into a single channel may be too simplistic.  Indeed, in most models annihilation proceeds through a combination
of channels. Here we consider this possibility. To avoid introducing many free parameters and since the spectra are rather similar for different types of
quarks, we only use the $b \bar{b}$ flux to describe quark final states. To a certain extent, spectra are also similar  for gauge and Higgs bosons since both
decay dominantly into hadrons. Since the spectra show a dependence on the lepton flavour, we allow non-universal lepton contributions. For each case
study, we use the fitting procedure described above, adding the branching fractions into specific channels as free parameters and scanning over the
DM mass $m_{\chi}$.

As a first example, we consider the leptophilic case corresponding to the favoured DM candidate that originally explained the PAMELA positron excess without
impacting the antiproton spectrum, as pointed out by\citetads{2009NuPhB.813....1C} and also by\citetads{2009PhRvL.102g1301D}. We find a
good fit, {\ie} with $\chi^{2}_{\rm dof}<1$, only for a DM mass near 500 GeV with a strong dominance of the $\tau^{+} \tau^{-}$ channel and
only $10\%$ of direct annihilation into $e^{+}e^{-}$. This induces a sharper drop of the spectra near the last data point of AMS-02.
%
Our results are qualitatively in agreement with what\citetads{2014arXiv1409.7317C} have recently found. The branching ratios are similar in both analyses,
with a large admixture of tau leptons, although our DM mass is much more constrained.

%
\begin{figure*}[ht!]
\begin{center}
\includegraphics[width=0.75\textwidth]{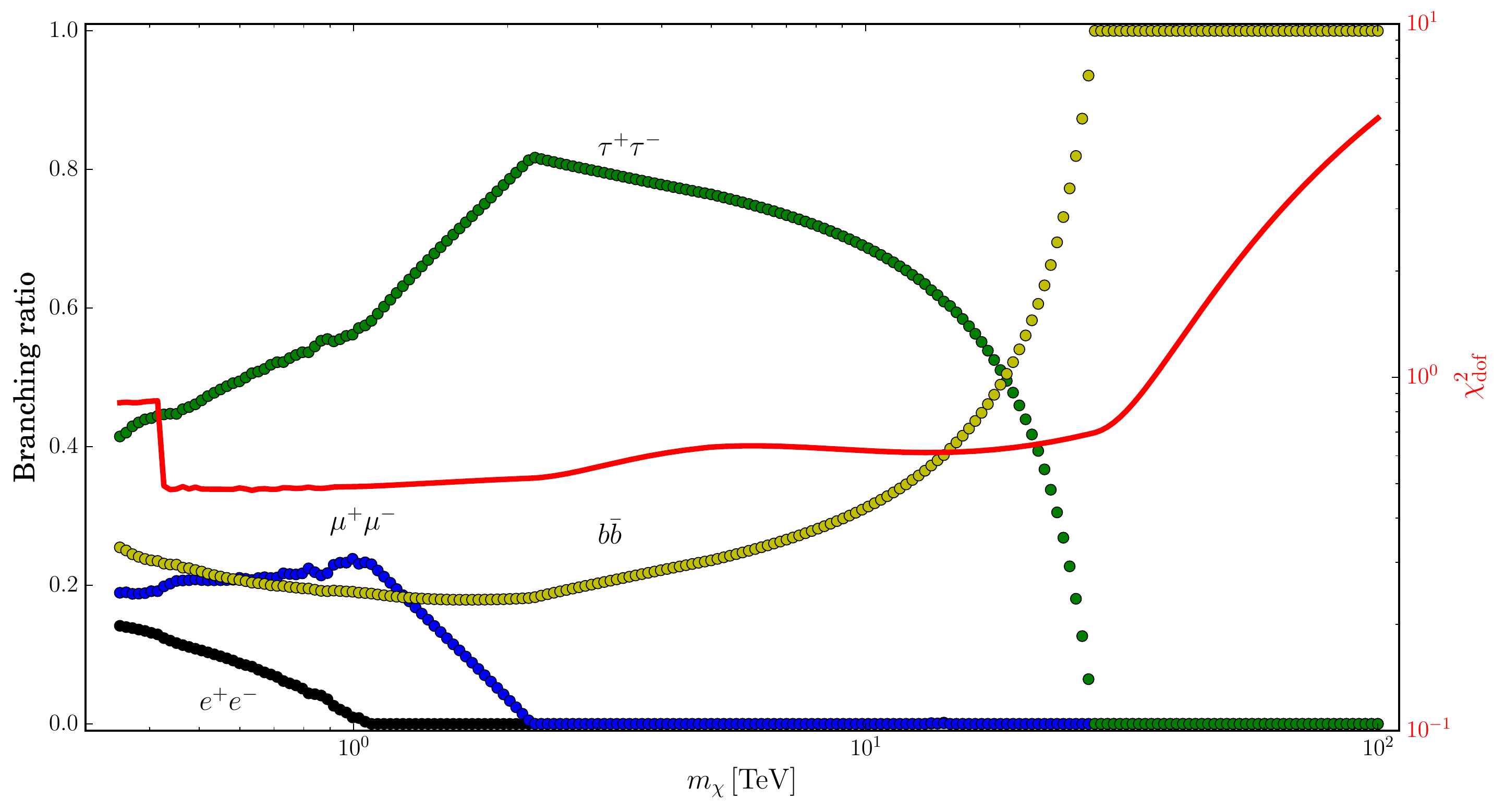}
\caption{
Best fit for the branching fractions for each DM mass $m_{\chi}$ assuming only annihilations into lepton and $b \bar{b}$ pairs.
The red line indicates on the right vertical axis the best $\chi^{2}_{\rm dof}$ value. As the DM mass exceeds the energy of the last data point,
the fit improves discontinuously and the $\chi^{2}_{\rm dof}$ drops sharply, hence the kink in the red curve above 400~GeV.
Between 20 and 30~TeV, the three lepton channels yield similar spectra in the range of positron energies that come into play in the
fit. The branching ratio into $\tau^{+} \tau^{-}$ should be understood as a sum over the three lepton families.}
\label{fig:BR_leptons_b}
\end{center}
\end{figure*}
%

%
\begin{figure*}[ht!]
\begin{center}
\includegraphics[width=0.75\textwidth]{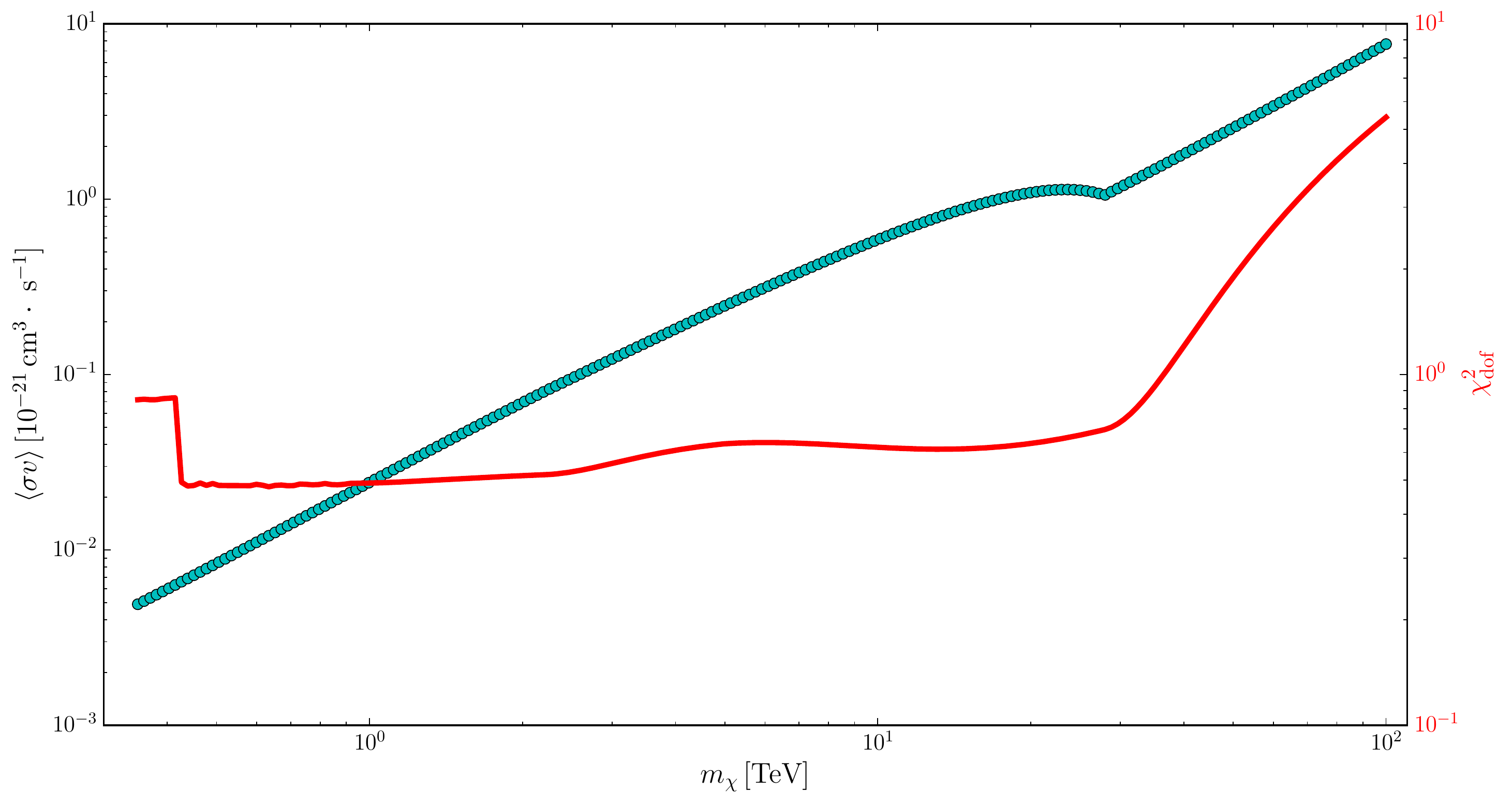}
\caption{
Best fit for the annihilation cross sections $\langle \sigma v \rangle$ for each DM mass $m_{\chi}$ assuming only annihilations into lepton and $b \bar{b}$ pairs.
The red line indicates on the right vertical axis the best $\chi^{2}_{\rm dof}$ value.}
\label{fig:SigV_leptons_b}
\end{center}
\end{figure*}
%

\vskip 0.1cm
It is much easier to find excellent fits with $\chi^{2}_{\rm dof}<1$ when allowing for some hadronic channel and this for any DM mass in
the range between 0.5 and 40~TeV. The preferred cross sections range from $10^{-23}$ cm$^{3}$ s$^{-1}$ for $m_{\chi} = 500$~GeV to
$10^{-21}$ cm$^{3}$ s$^{-1}$ for $m_{\chi} = 40$~TeV. The preferred branching fractions for the range of masses considered are displayed
in Fig.~\ref{fig:BR_leptons_b}. Not surprisingly the leptonic contribution strongly dominates below the TeV scale while the $b \bar{b}$
component increases with the DM mass.
The corresponding annihilation cross sections $\langle \sigma v \rangle$ are represented in Fig.~\ref{fig:SigV_leptons_b}.
 Figure~\ref{fig:leptons_b_best_fit} shows the positron fraction corresponding to the best fit for the cross
section and the branching fractions for the two sample masses of 600~GeV (left) and 20~TeV (right). The contributions of the various channels
to the DM signal are also indicated.

%
\begin{figure*}[ht!]
\begin{center}
\includegraphics[width=0.49\textwidth]{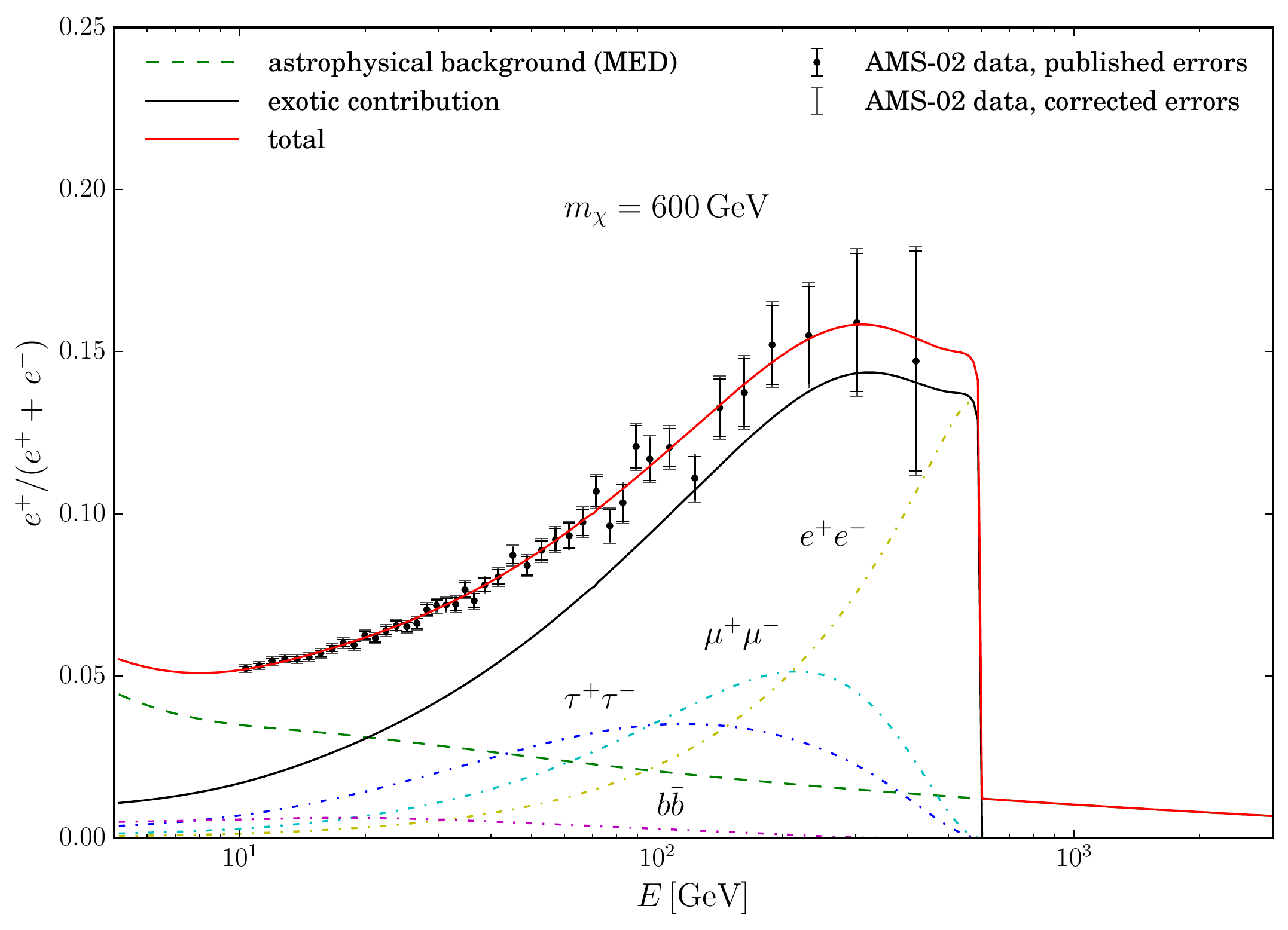}
\includegraphics[width=0.49\textwidth]{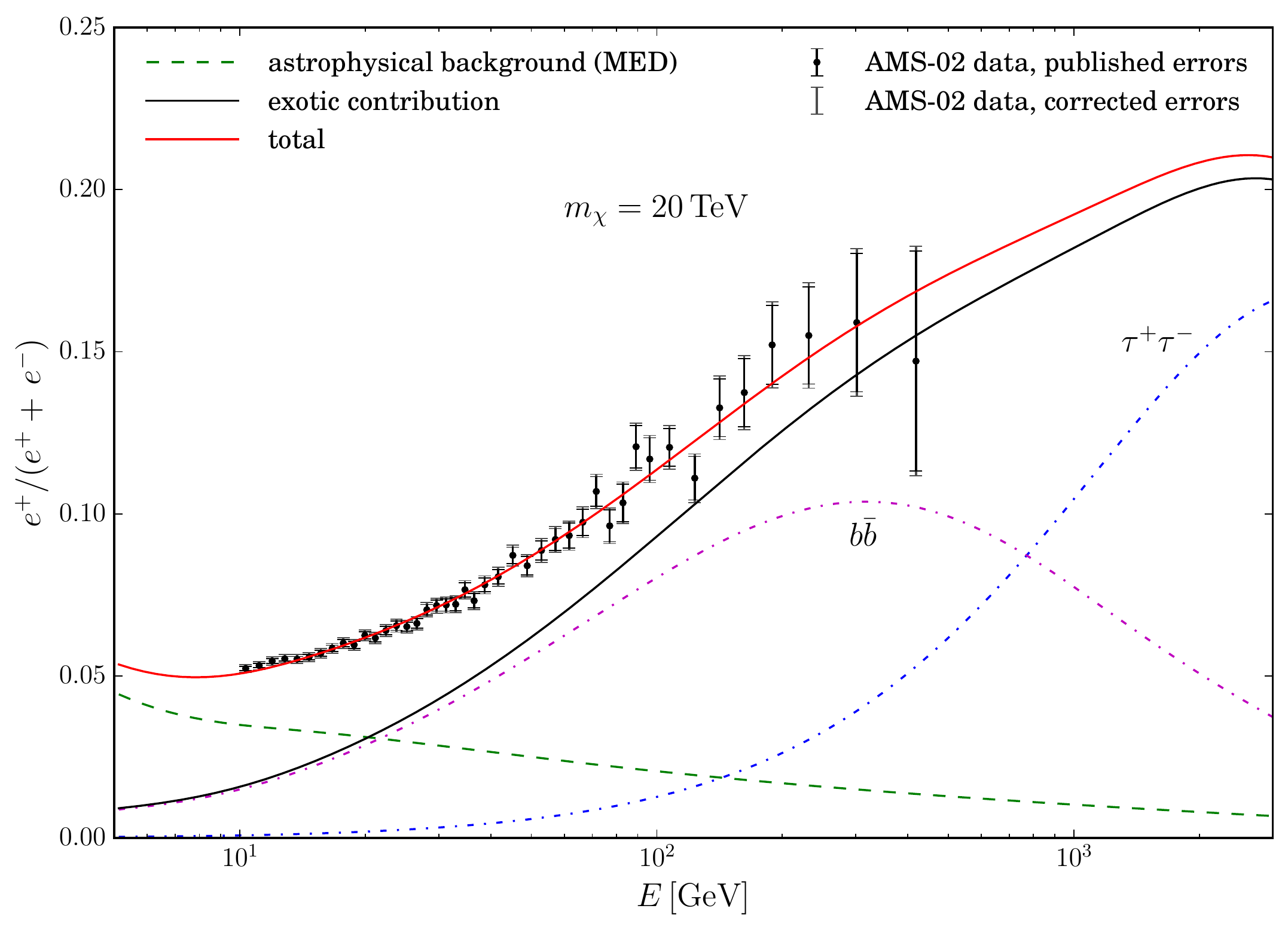}
\caption{Positron fraction as a function of the positron energy, for a DM mass m$_{\chi}$ = 600 GeV  (left panel)
and m$_{\chi}$ = 20 TeV (right panel), compared to AMS-02 data~\citep{Accardo:2014lma}. These values
correspond to the cross section $\langle \sigma v \rangle$
and branching ratios into lepton and $b \bar{b}$ pairs, which best fit the positron fraction, as from Fig.~\ref{fig:BR_leptons_b} and \ref{fig:SigV_leptons_b}. Both values give excellent fits with $\chi^{2}_{\rm dof}$ of 0.5 (left)
and 0.6 (right). The branching ratio into $\tau^{+} \tau^{-}$ amounts to 50\% whereas the quark contribution increases from 20\% (left)
to 50\% (right). The $e^{+}e^{-}$ and $\mu^{+}\mu^{-}$ channels disappear above 1 and 2~TeV, respectively.
The cross section is equal to $\langle \sigma v \rangle = 1.11 \cdot 10^{-23}$ cm$^{3}$ s$^{-1}$ (left) and
$1.09 \cdot 10^{-21}$ cm$^{3}$ s$^{-1}$  (right).
The contribution of each channel to the positron fraction is also indicated.
}
\label{fig:leptons_b_best_fit}
\end{center}
\end{figure*}
%

\vskip 0.1cm
Imposing the condition that the branching fractions into leptons are universal, while allowing for quark channels, deteriorates the fits somewhat.
Nevertheless, excellent agreement is found for masses above 5~TeV and branching fractions around 20\% in each lepton flavour as displayed
in the left panel of Fig.~\ref{fig:UED_case}.
These branching fractions are typical of the minimal universal extra dimension model (mUED) although  the preferred mass is larger
than expected in that model from the relic density constraint\citepads{2011JCAP...02..009B}. The corresponding annihilation cross sections $\langle \sigma v \rangle$ are represented in the right panel of Fig.~\ref{fig:UED_case}.
In the left panel of Fig.~\ref{fig:best_fit_UED_4leptons} the positron fraction has been plotted for $m_{\chi} = 600$~GeV (dashed-dotted lines) and 20~TeV (solid lines), corresponding to
$\langle \sigma v \rangle = 1.05 \cdot 10^{-23}$ cm$^{3}$ s$^{-1}$ and $1.12 \cdot 10^{-21}$ cm$^{3}$ s$^{-1}$, and compared to AMS-02 data~\citep{Accardo:2014lma}.
The 600~GeV DM species does not provide a good fit. The corresponding reduced $\chi^{2}$ is of the order of 2.
On the contrary, the 20~TeV WIMP reproduces the observations with a $\chi^{2}_{\rm dof}$ value of 0.6 but induces
a sharp increase of the positron fraction above 1~TeV.

%
\begin{figure*}[ht!]
\begin{center}
\includegraphics[height=0.33\textwidth]{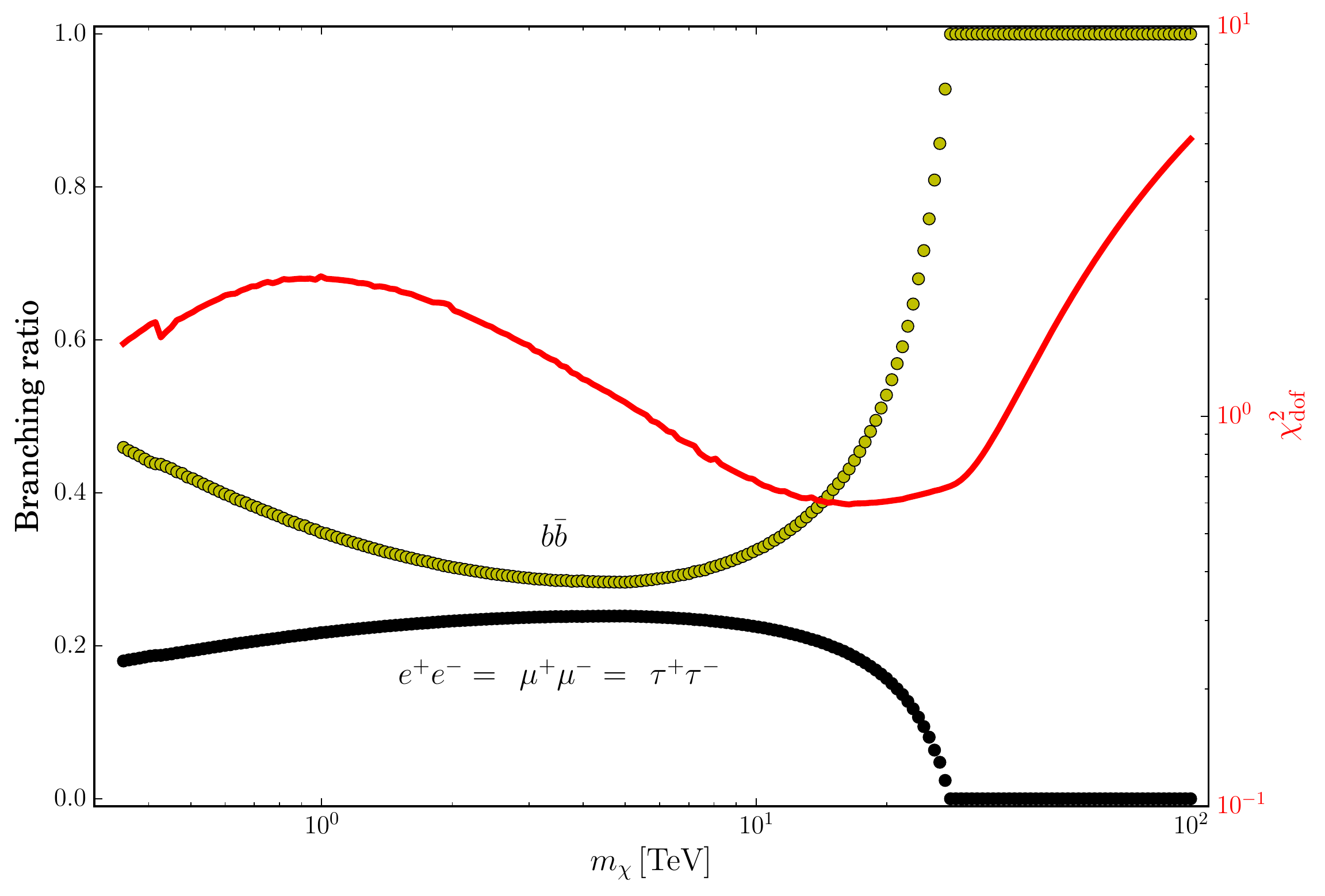}
\includegraphics[height=0.33\textwidth]{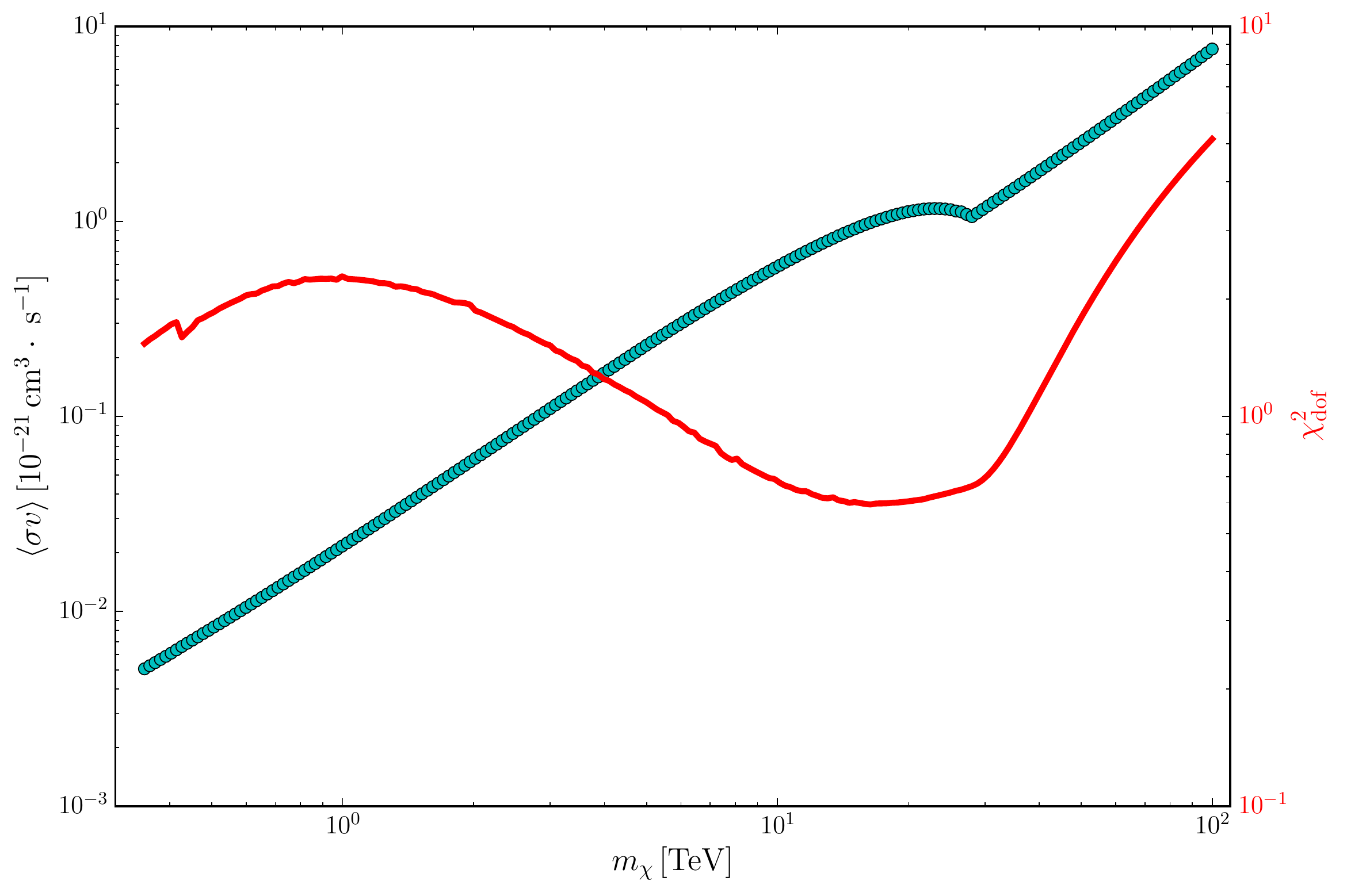}
\caption{
The left (right) panel shows the best fit for the branching ratios (annihilation cross section $\langle \sigma v \rangle$) as a function of the DM mass m$_{\chi}$, assuming annihilation into leptons and $b \bar{b}$.
The red line indicates on the right vertical axis the best $\chi^{2}_{\rm dof}$ value.
The plot is similar to Fig.~\ref{fig:BR_leptons_b} (Fig.~\ref{fig:SigV_leptons_b}), with the additional assumption of universal branching fractions into lepton pairs.}
\label{fig:UED_case}
\end{center}
\end{figure*}
%

\vskip 0.1cm
Finally allowing for any combination of the four-lepton channels allows for a very good fit to the data but only for a DM mass  between 0.5 and 1~TeV.
Annihilation into $4\tau$ is by far dominant -- at least 70\% as featured in left panel of Fig.~\ref{fig:4leptons_case}. Note that the $4e$ channel is
subdominant and that the $4\mu$ channel is strongly disfavoured.
The positron fraction for DM masses of 600~GeV (20~TeV) and cross sections $\langle \sigma v \rangle = 7.37 \cdot 10^{-24}$ cm$^{3}$ s$^{-1}$
($2.72 \cdot 10^{-21}$ cm$^{3}$ s$^{-1}$) with branching fraction into $4\tau$ of 75\% (100\%) is shown in the right panel of
Fig.~\ref{fig:best_fit_UED_4leptons}. For these parameters, the reduced $\chi^{2}$ is respectively equal to 0.8 and 3.
%
\begin{figure*}[ht!]
\begin{center}
\includegraphics[height=0.33\textwidth]{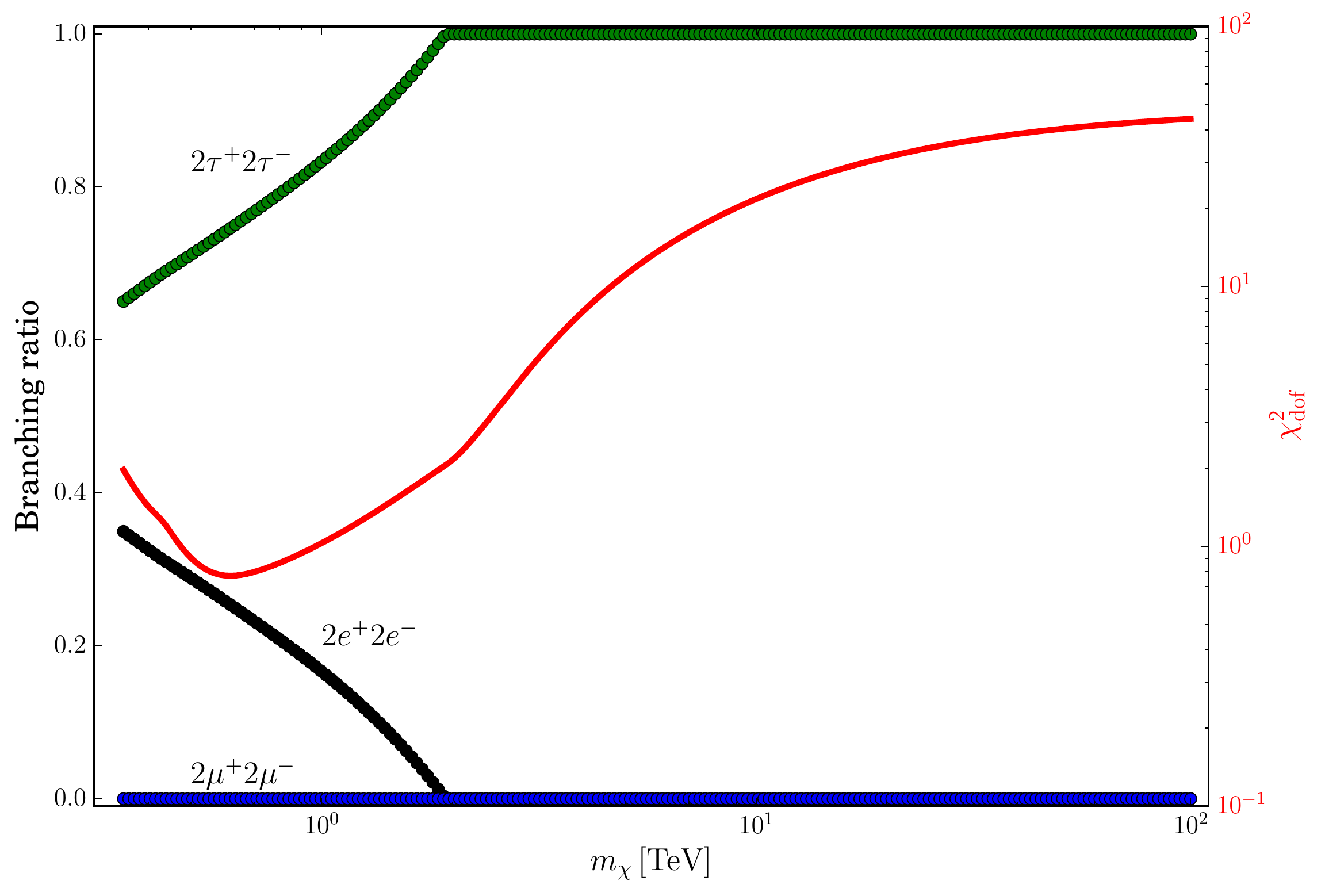}
\includegraphics[height=0.33\textwidth]{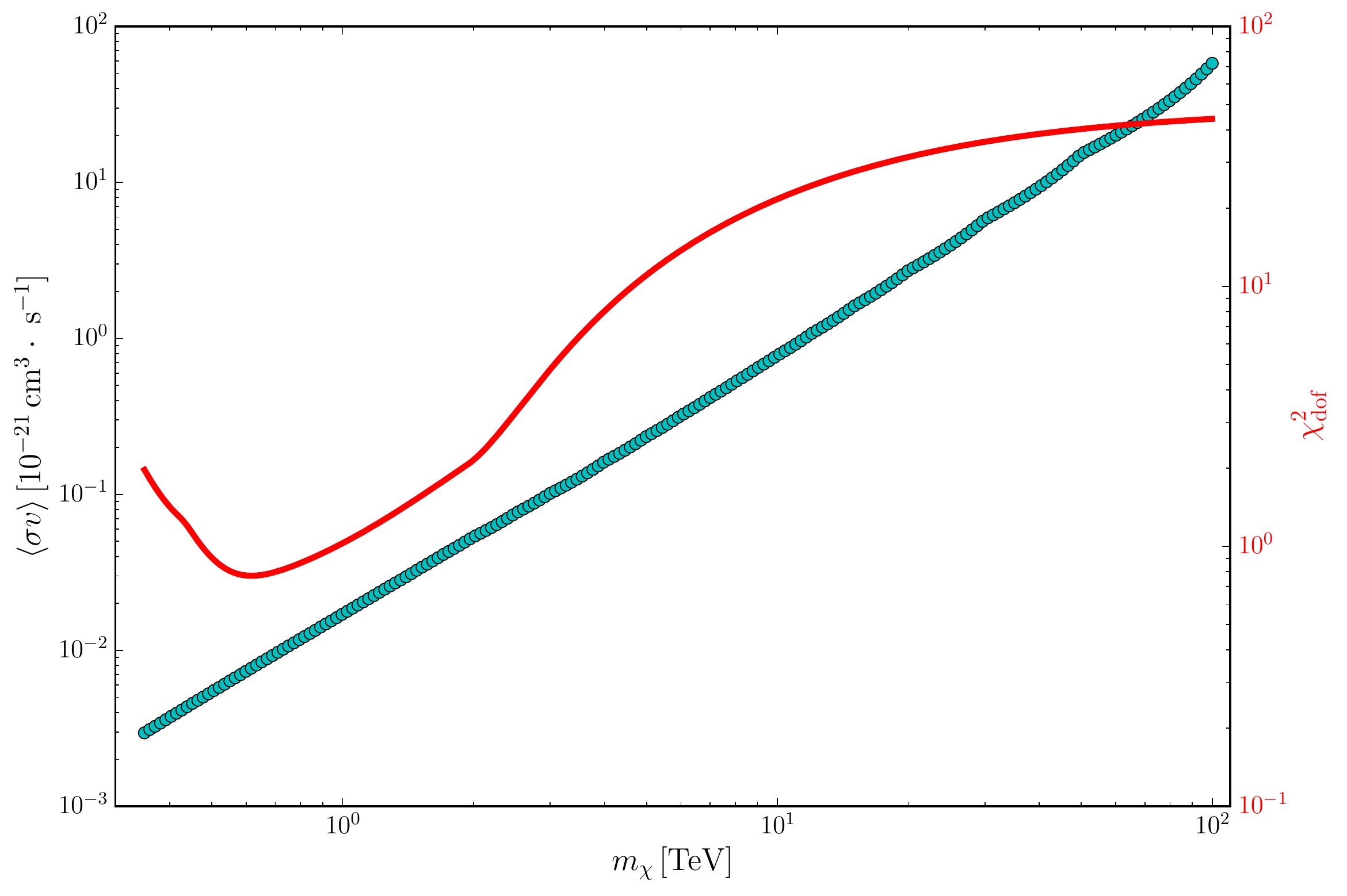}
\caption{
The left (right) panel shows the best fit for the branching ratios (annihilation cross section $\langle \sigma v \rangle$) as a function of the DM mass m$_{\chi}$, assuming annihilation
into four leptons. The red line indicates the best $\chi^{2}_{\rm dof}$ value on the right vertical axis.}
\label{fig:4leptons_case}
\end{center}
\end{figure*}
%

%
\begin{figure*}[ht!]
\begin{center}
\includegraphics[height=0.345\textwidth]{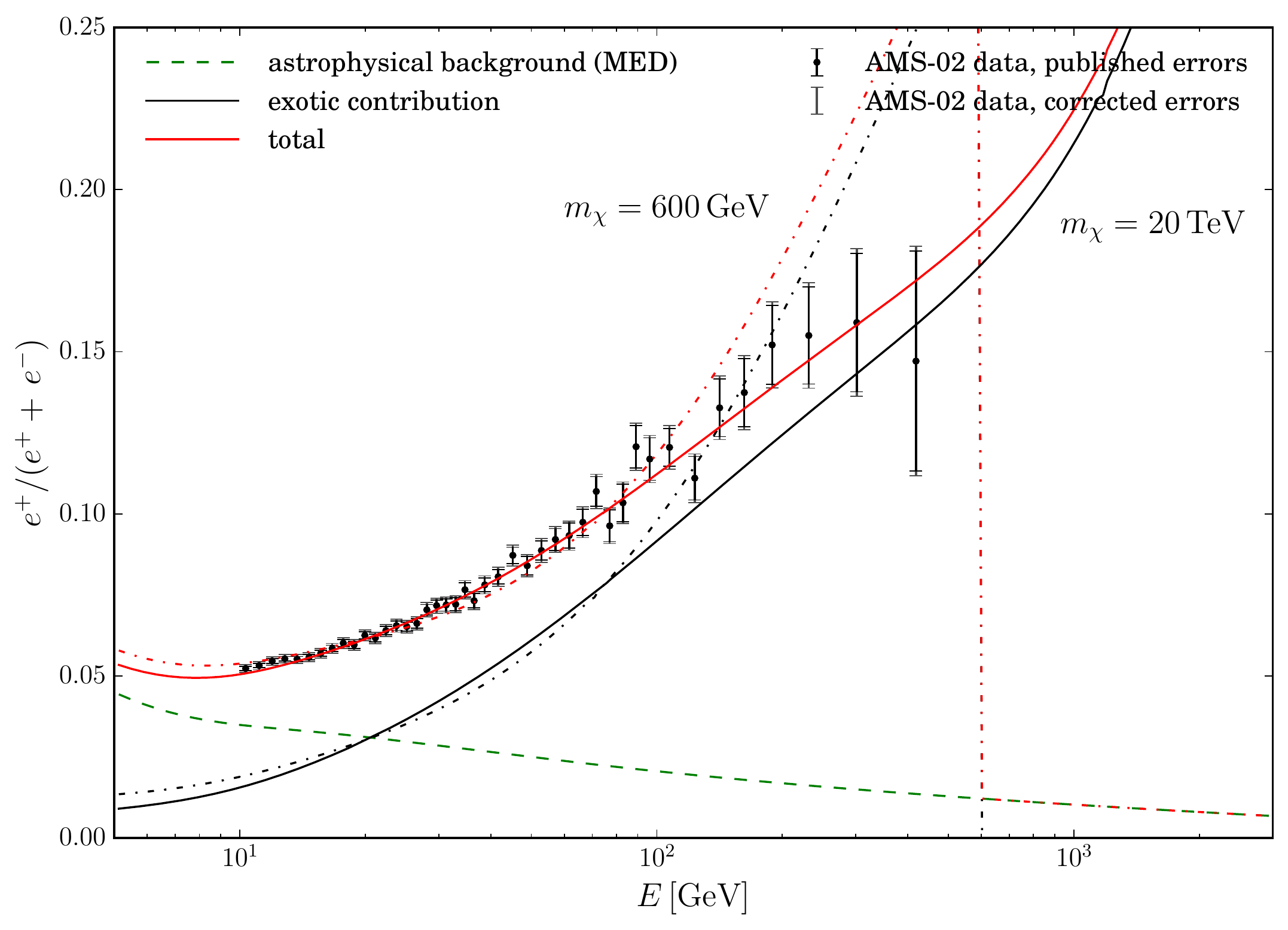}
\includegraphics[height=0.345\textwidth]{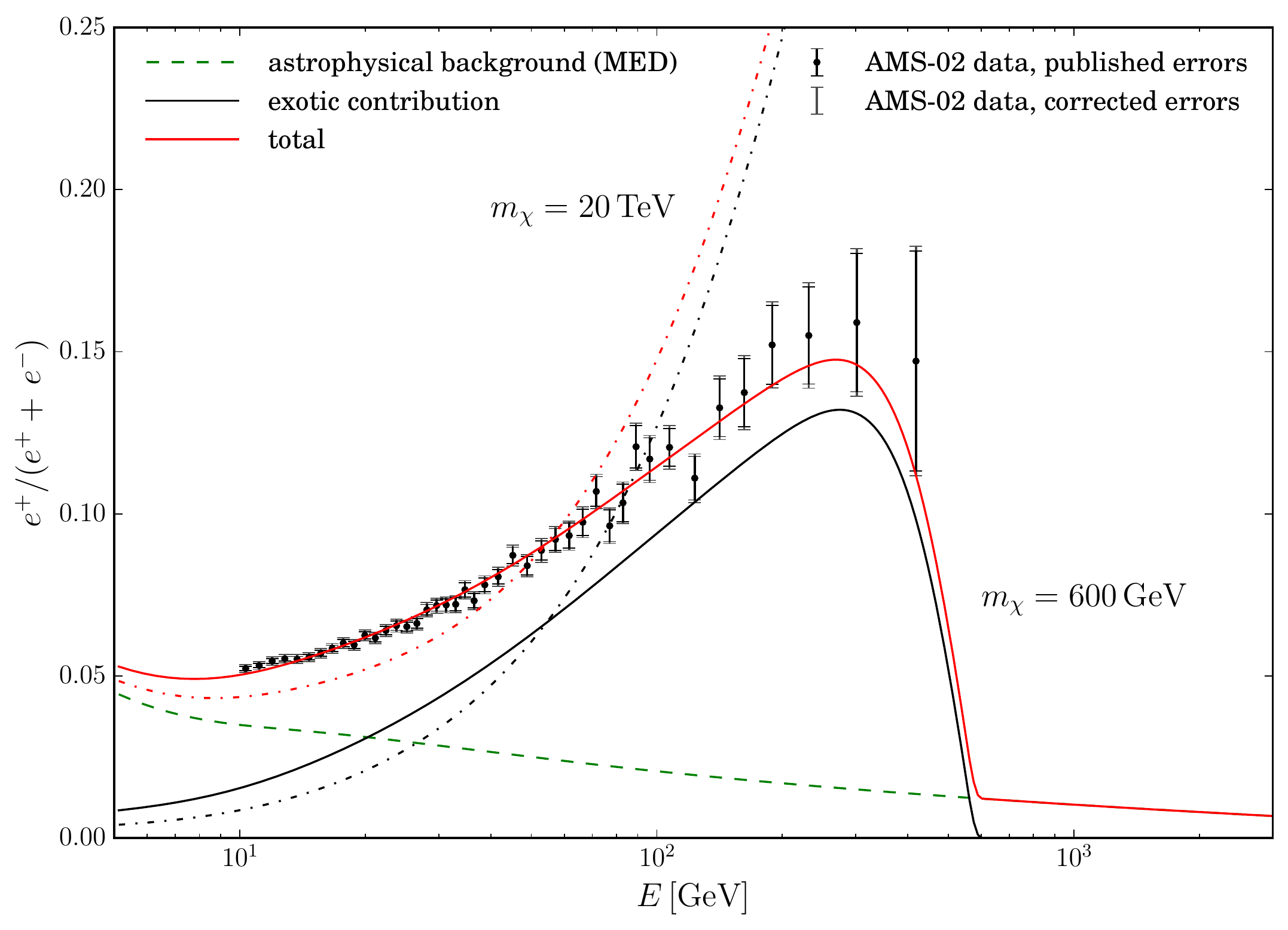}
\caption{
The left panel corresponds to the leptons and $b\bar{b}$ channels with the additional assumption of universal branching fractions into lepton pairs.
The positron fraction has been plotted for $m_{\chi} = 600$~GeV (dashed-dotted lines) and 20~TeV (solid lines), corresponding to
$\langle \sigma v \rangle = 1.05 \cdot 10^{-23}$ cm$^{3}$ s$^{-1}$ and $1.12 \cdot 10^{-21}$ cm$^{3}$ s$^{-1}$, and compared to AMS-02
data~\citep{Accardo:2014lma}.
The 600~GeV DM species does not provide a good fit. The corresponding reduced $\chi^{2}$ is of the order of 2.
On the contrary, the 20~TeV WIMP reproduces the observations with a $\chi^{2}_{\rm dof}$ value of 0.6 but induces
a sharp increase of the positron fraction above 1~TeV.
The right panel corresponds to the four-lepton channel. The positron fraction has been plotted for $m_{\chi} = 600$~GeV (solid lines) and 20~TeV
(dashed-dotted lines).
The 600~GeV case provides a good fit with a $\chi^{2}_{\rm dof}$ value of 0.8 and a cross section of $7.37 \cdot 10^{-24}$ cm$^{3}$ s$^{-1}$.
The 20~TeV WIMP mass scenario, on the contrary, barely fits the data below 50~GeV, with a reduced $\chi^{2}$ larger than 3 and
$\langle \sigma v \rangle = 2.72 \cdot 10^{-21}$ cm$^{3}$ s$^{-1}$.
}
\label{fig:best_fit_UED_4leptons}
\end{center}
\end{figure*}
%

%
\section{The single pulsar hypothesis reinvestigated}
\label{sec:single_pulsar}

Following\citetads{2013ApJ...772...18L} and\citetads{2013PhRvD..88b3013C}, the aim of this section is to investigate if the rise of the positron
fraction measured by AMS-02 can be explained by a single pulsar contribution.
Assuming a pulsar origin for the rise of the positron fraction leads to a cumulative contribution from all detected and yet undiscovered pulsars. Nevertheless, demonstrating that the positron fraction can be explained by a unique pulsar contribution, provides us with a valid alternative to the DM explanation of this anomaly. If the single pulsar hypothesis is viable, the whole of pulsars is capable of reproducing the experimental data. Indeed, as there is only an upper limit on the injection normalisation $fW_0$, adjusting $fW_0$ for each individual pulsar will result in even better fits when more pulsars are added.
An extensive analysis has been recently performed by\citetads{2014JCAP...04..006D} along the same direction, with a global fit over all
available electron and positron observables. However, these may be correlated with each other, and we have taken the standpoint of concentrating
only on the positron fraction.
We also assess the goodness of our fits using the $p$-value, which is an absolute statistical estimator.

\subsection{Selection of possible pulsars: the five survivors of the ATNF catalogue}
\label{subsec:five_samourais}

The contribution of a single pulsar is calculated using the injection spectrum given in Sec.~\ref{sec:CR_transport} Eq.~\ref{eq:PSR_injection_spectrum}.
The free parameters are the spectral index $\gamma$ and  the energy released by the pulsar through positrons $fW_0$, which are related to the spectral shape and normalisation, respectively. In our analysis, we assume a fictional source placed
at a distance \textit{d} from the Earth and of age $t_\star$. We then estimate the parameters $\gamma$ and $fW_0$, which give the best fit to the positron fraction. We allow the spectral index $\gamma$ to vary from 1 to 3 and we fix the upper limit of $fW_0$ to $10^{54} \; \rm{GeV}$ (see Sec.~\ref{sec:CR_transport}).
Since only close and relatively young single pulsars reproduce the experimental data well, we repeat this procedure for 2500 couples of $(d,\,t_{\star})$
with $d<1\,\rm{kpc}$ and $t_{\star}< 1\,\rm{Myr}$. We perform our analysis with the benchmark set of propagation parameters MED.

%
\begin{figure*}[ht!]
\begin{center}
\includegraphics[height=0.427\textwidth]{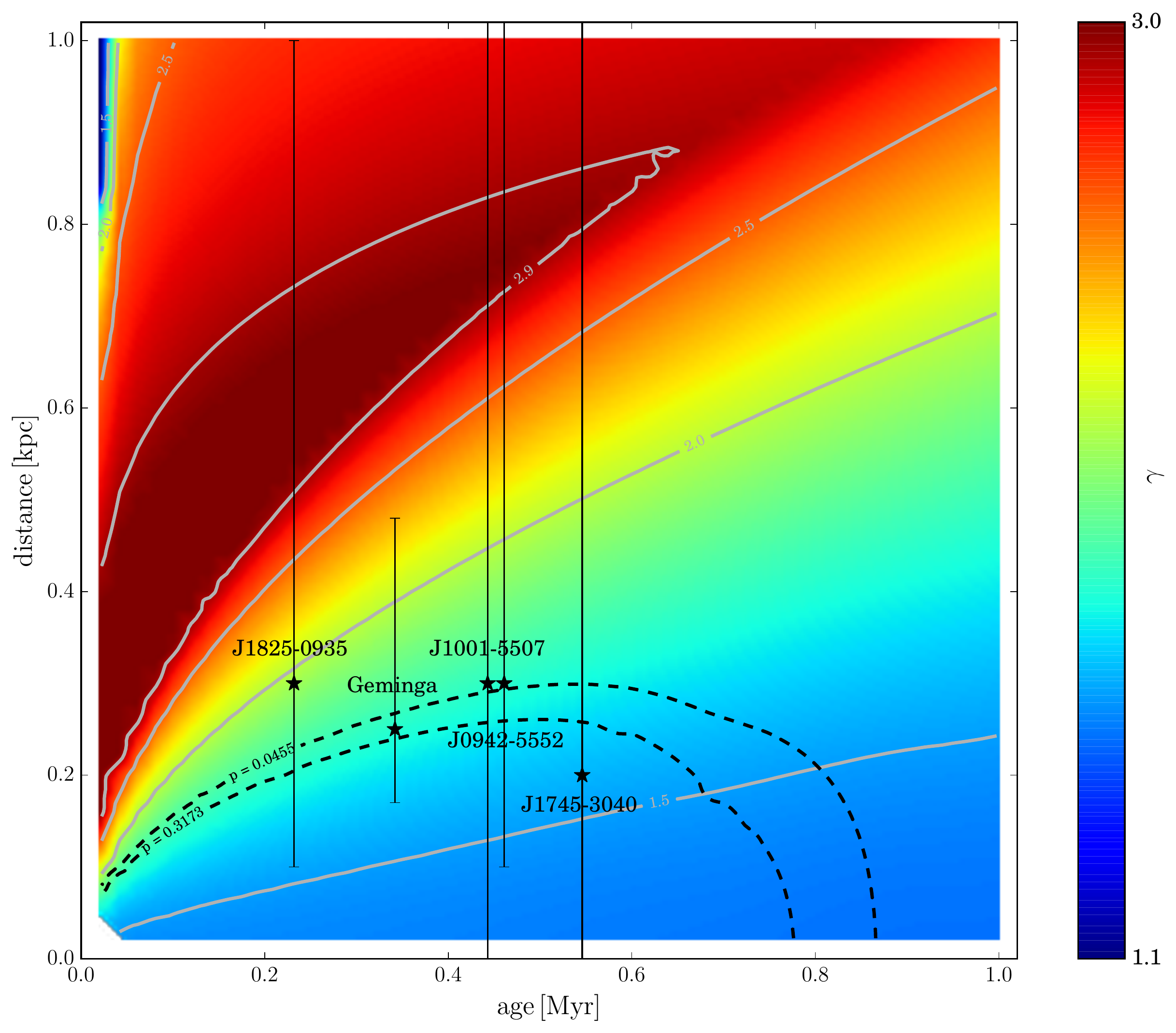}
\includegraphics[height=0.427\textwidth]{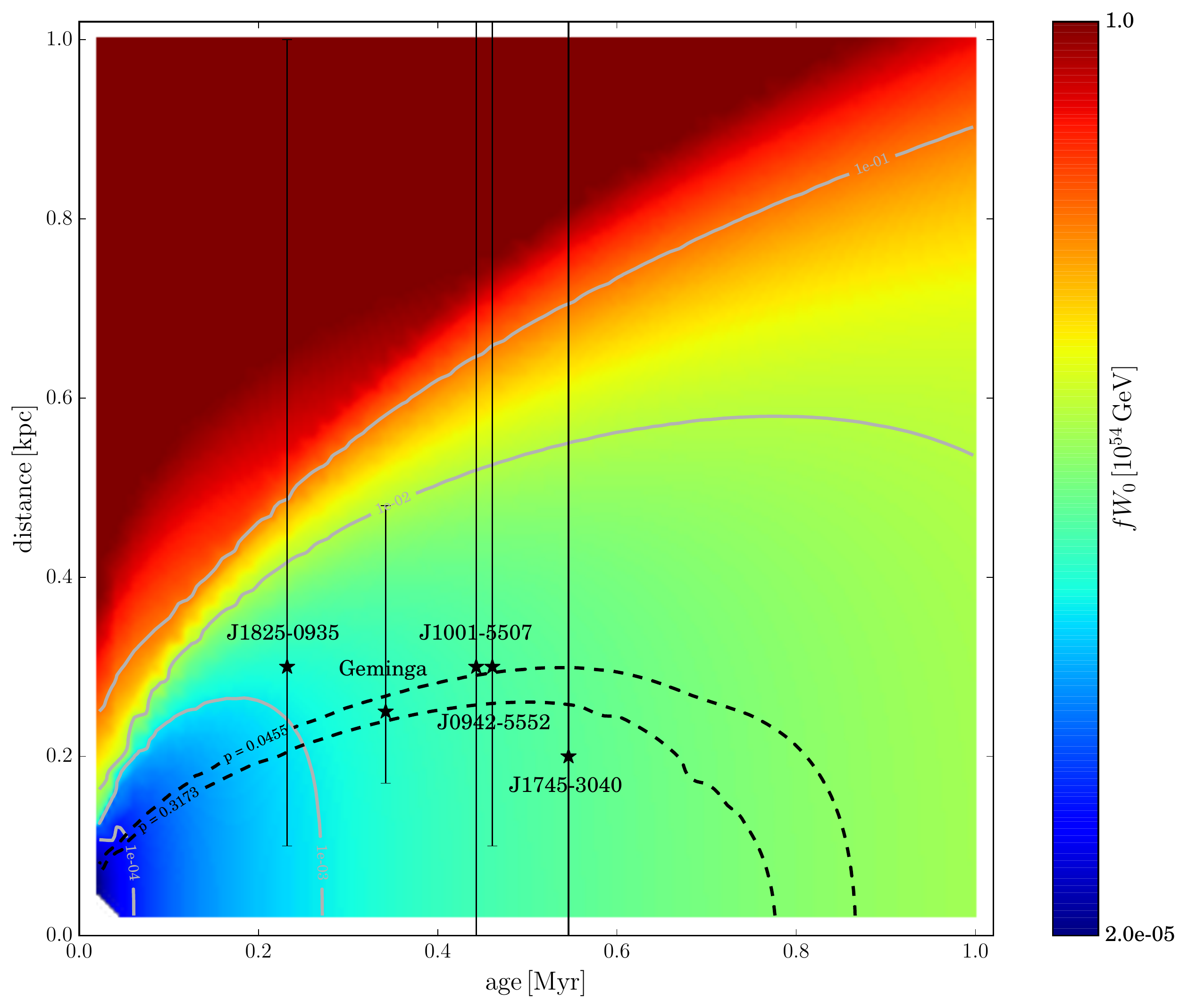}
\caption{Best-fit values of the spectral index $\gamma$ (left panel) and the total energy carried by positrons $fW_0$ (right panel) for each point of the plane (age, distance) with the benchmark propagation model MED. The grey lines display the iso-contours for given values of $\gamma$ (left) and $fW0$ (right). The black dashed lines represent the iso-contours of the critical {\it p}-values. The five selected pulsars with their associated uncertainty on their distance are indicated by the black stars.}
\label{fig:pulsar_gamma_fW_0}
\end{center}
\end{figure*}
%

\vskip 0.1cm
The results are shown in Fig.~\ref{fig:pulsar_gamma_fW_0} where the colour scale indicates the value of  $\gamma$ (left panel) and $fW_0$ (right panel). The grey lines highlight the iso-contours for given values of $\gamma$ and $fW_0$. We observe a positive (negative) correlation between the distance (age) of the pulsar and its injection spectral index $\gamma$. This can be explained by the fact that the free parameters of the pulsar ($\gamma$, $fW_0$) are predominantly determined by the well-measured low-energy shape of the positron fraction. Indeed, the positron flux between 10 and $\sim100$\,GeV can be approximated by $ \Phi_{e+}(E) \propto \exp ( - {d^2}/{\lambda_{D}^2})$, with the positron sphere radius $\lambda_{D}^2 \simeq 4K_0 t_{\star} \left( {E}/{E_0} \right)^{\delta}$. We can hence define a lower energy limit $E_{\rm{min}} = E_0 \, ({d^2}/{4K_0 t_{\star}})^{1/ \delta}$ below which the positron flux becomes negligible since the positrons have not had enough time to reach the Earth. Given a pulsar age, lengthening the distance implies, on the one hand, an increase of $E_{\rm min}$, i.e. the spectrum becomes harder and the best-fit value of $\gamma$ larger. On the other hand, the positron flux decreases exponentially and the value of $fW_0$ increases consequentially. In the same way, for a fixed pulsar distance, an older source yields at the Earth positrons at lower energies and needs a smaller $\gamma$ and $fW_0$ to reproduce the experimental data. In the special case of a very close pulsar ($d  \lesssim 0.3$\,kpc), the shape of the injected positron flux mildly depends on the pulsar distance and varies like $\Phi_{e+}(E) \propto \lambda_{D}^{-3} \propto {t_{\star}}^{-3/2}$. In this situation, $fW_0$ and the age are positively correlated.

\vskip 0.1cm
In the same figures, the two iso-contours of the critical {\it p}-values (black dashed lines) as defined in Sec.~\ref{subsec:single_channel} are represented. Those define the good-fit region with $\gamma \lesssim 2 $ and $fW_0$ within the range of $[10^{49}, 10^{52}] \; \rm{GeV}$. These value ranges are consistent with previous studies\citepads{2009JCAP...01..025H,2009PhRvD..80f3005M,2010A&A...524A..51D,2013ApJ...772...18L,2014JCAP...04..006D}. We select the pulsars from the ATNF catalogue that fall into this good-fit region. The pulsar distance suffers from large uncertainties, which are taken into account for the pulsar selection. The uncertainty on the pulsar age is negligible due to a precise measurement of its spin and spin-down. Only five pulsars from the ATNF catalogue fulfil the goodness-of-fit criteria. The chosen pulsars and their distance uncertainties are indicated in Fig.~\ref{fig:pulsar_gamma_fW_0} by black stars with error bars.

\subsection{Results for the five pulsars}

For each of these five selected pulsars we estimate the values of $\gamma$ and $fW_0$ that best reproduce the experimental data. The results are listed in Table~\ref{tab:table_pulsar_best_fits} with the corresponding $\chi^2$ and {\it p}-values. The nominal age and distance (bold line) are taken from the ATNF catalogue. We also perform this procedure for their minimal (first line) and maximal distances (third line) according to the experimental uncertainty, which is not taken into account in the minimisation procedure. A further study will include this uncertainty, but it is beyond the scope of this paper. Finally, we study the contribution to the positron fraction of the well-known pulsars Monogem and Vela, and present these results in the Table.

\vskip 0.1cm
As can be seen in Fig.~\ref{fig:pulsar_gamma_fW_0}, for their nominal distances, the pulsar J1745$-$3040 (J1825$-$0935) reproduces best (worst) the AMS-02 positron fraction. This is well reflected in their respective {\it p}-values. In contrast, Monogem and Vela cannot adjust the data. Because of their very young age, they are not able to contribute to the low-energy positron fraction between 10 and 50\,GeV where the error bars are the smallest. For all studied pulsars, the {\it p}-values increase with decreasing distance. This can be explained by the above mentioned low-energy cut-off $E_{\rm min}$, which is significantly lowered and allows hence the pulsar to cover a larger part of the positron fraction. An example is given in Fig.~\ref{fig:PF_Geminga_distance} where the contribution of Geminga is studied for its nominal (left) and minimal (right) ATNF distance. In the case of most pulsars the fit does not converge for the maximal distance and reaches the defined limits of the free parameters. The associated $\chi^2$ and {\it p}-values are hence not meaningful.
The resulting positron fractions of the pulsars J1745$-$3040 (solid line), Geminga (dashed-dotted line), and Monogem (dotted line) are shown in Fig.~\ref{fig:PF_J1745_Geminga_Monogem} for their nominal distances. Because of the large error bars at high energies the contribution of J1745$-$3040 reproduces well the experimental data, reflected by the good {\it p}-value, even though it does not reach the highest energy data points.
As mentioned in Sec.~\ref{sec:CR_transport}, increasing $E_C$ neither changes our conclusions nor modifies our list of selected pulsars.

%
\begin{table*}[ht!]
\centering
\caption{Results for the pulsar parameters $fW_0$ and $\gamma$ for the best fits in the single pulsar approach. Only pulsars with a {\it p}-value $> 0.0455$, taking their distance uncertainty into account, are listed, besides the well-known pulsars Monogem and Vela. The bold lines correspond to the nominal distance value.}
\label{tab:table_pulsar_best_fits}
\vspace{.1cm}
\begin{tabular}{|c|c|c|c|c|c|c|c|}
\hline
{Name} & {Age [kyr]} & {Distance [kpc]} & {$fW_0 \; [{10^{54} \, \rm GeV]}$} & {$\gamma$} & {$\chi^{2}$} & {$\chi^{2}_{\rm dof}$} & $p$ \\
\hline
 				    & &  0 & $ (2.95 \pm 0.07) \cdot 10^{-3}  $ & $ 1.45 \pm 0.02$ & 23.4 & 0.57 & $ 0.99 $ \\
J1745$-$3040	& 546 & \textbf{0.20} & $ \mathbf{(3.03 \pm 0.06) \cdot 10^{-3}} $ & $ \mathbf{1.54 \pm 0.02} $ & \textbf{33.6} & \textbf{0.82} & $ \textbf{0.79} $ \\
				   & & 1.3 & $ 1 $ & $ 2.54 $ & 9902 & 241 & $ 0 $ \\
\hline
 			 &	     & 0.17 & $ (1.48 \pm 0.03) \cdot 10^{-3}  $ & $ 1.56 \pm 0.02$ & 26.8 & 0.65 & $ 0.96 $ \\
J0633+1746 & 342 & \textbf{0.25} & $ \mathbf{(1.63 \pm 0.02) \cdot 10^{-3}} $ & $ \mathbf{1.68 \pm 0.02} $ & \textbf{49.6} & \textbf{1.21} & $ \textbf{0.17} $ \\
\textit{Geminga}	&         & 0.48 & $ (1.01 \pm 0.06) \cdot 10^{-2}  $ & $ 2.29 \pm 0.02$ & 332 & 8.10 & $ 0 $ \\
\hline
			&        & 0.10 & $ (2.28 \pm 0.05) \cdot 10^{-3}  $ & $ 1.48 \pm 0.02$ & 21.7 & 0.53 & $ 0.99 $ \\
J0942$-$5552 & 461 & \textbf{0.30} & $ \mathbf{(2.61 \pm 0.04) \cdot 10^{-3}} $ & $ \mathbf{1.69 \pm 0.02}$ & \textbf{61.0} & \textbf{1.49} & $ \textbf{0.02} $ \\
			&        & 1.1 & $ 1 $ & $ 2.65$ & 7747 & 189 & $ 0 $ \\
\hline
			&	    & 0 & $ (2.13 \pm 0.05) \cdot 10^{-3}  $ & $ 1.46 \pm 0.02$ & 19.8 & 0.48 & $ 0.99 $ \\
J1001$-$5507 & 443 & \textbf{0.30} & $ \mathbf{(2.49 \pm 0.03) \cdot 10^{-3}} $ & $\mathbf{ 1.70 \pm 0.02} $ & \textbf{62.4} & \textbf{1.52} & $ \textbf{0.02} $ \\
			&	    & 1.4 & $ 1 $ & $ 2.46$ & 13202 & 322 & $ 0 $ \\
\hline
			&	    & 0.1 & $ (0.80 \pm 0.02) \cdot 10^{-3}  $ & $ 1.52 \pm 0.02$ & 21.0 & 0.51 & $ 0.99 $ \\
J1825$-$0935 & 232 & \textbf{0.30} & $ \mathbf{(1.45 \pm 0.03) \cdot 10^{-3}} $ & $ \mathbf{1.94 \pm 0.02} $ & \textbf{126} & \textbf{3.07} & $ \textbf{0} $ \\
			&	    & 1.0 & $ 1 $ & $ 2.64 $ & 12776 & 312 & $ 0 $ \\
\hline \hline
		 	&	    & 0.25 & $ (1.06 \pm 0.05) \cdot 10^{-3}  $ & $ 2.18 \pm 0.02$ & 216 & 5.27 & $ 0 $ \\
J0659+1414 & 111 & \textbf{0.28} & $\mathbf{ (2.53 \pm 0.16) \cdot 10^{-3}} $ & $ \mathbf{2.37 \pm 0.02} $ & \textbf{316} & \textbf{7.71} & $ \textbf{0} $ \\
\textit{Monogem}	&	    & 0.31 & $ (7.96 \pm 0.61) \cdot 10^{-3} $ & $ 2.58 \pm 0.02$ & 444 & 10.8 & $ 0 $ \\
\hline
		 	 &	     & 0.26 & $ (2.53 \pm 0.08) \cdot 10^{-1}  $ & $ 3$ & 14316 & 349 & $ 0 $ \\
J0835+4510 & 11.3 & \textbf{0.28} & $ \mathbf{(3.90 \pm 0.14) \cdot 10^{-1}} $ & $ \textbf{3} $ & \textbf{14982} & \textbf{365} & $ \textbf{0} $ \\
\textit{Vela}		&	     & 0.3 & $ (6.00 \pm 0.26) \cdot 10^{-1} $ & $ 3$ & 15446 & 377 & $ 0 $ \\
\hline
\end{tabular}
\end{table*}
%

%
\begin{figure*}[ht!]
\begin{center}
\includegraphics[width=0.49\textwidth]{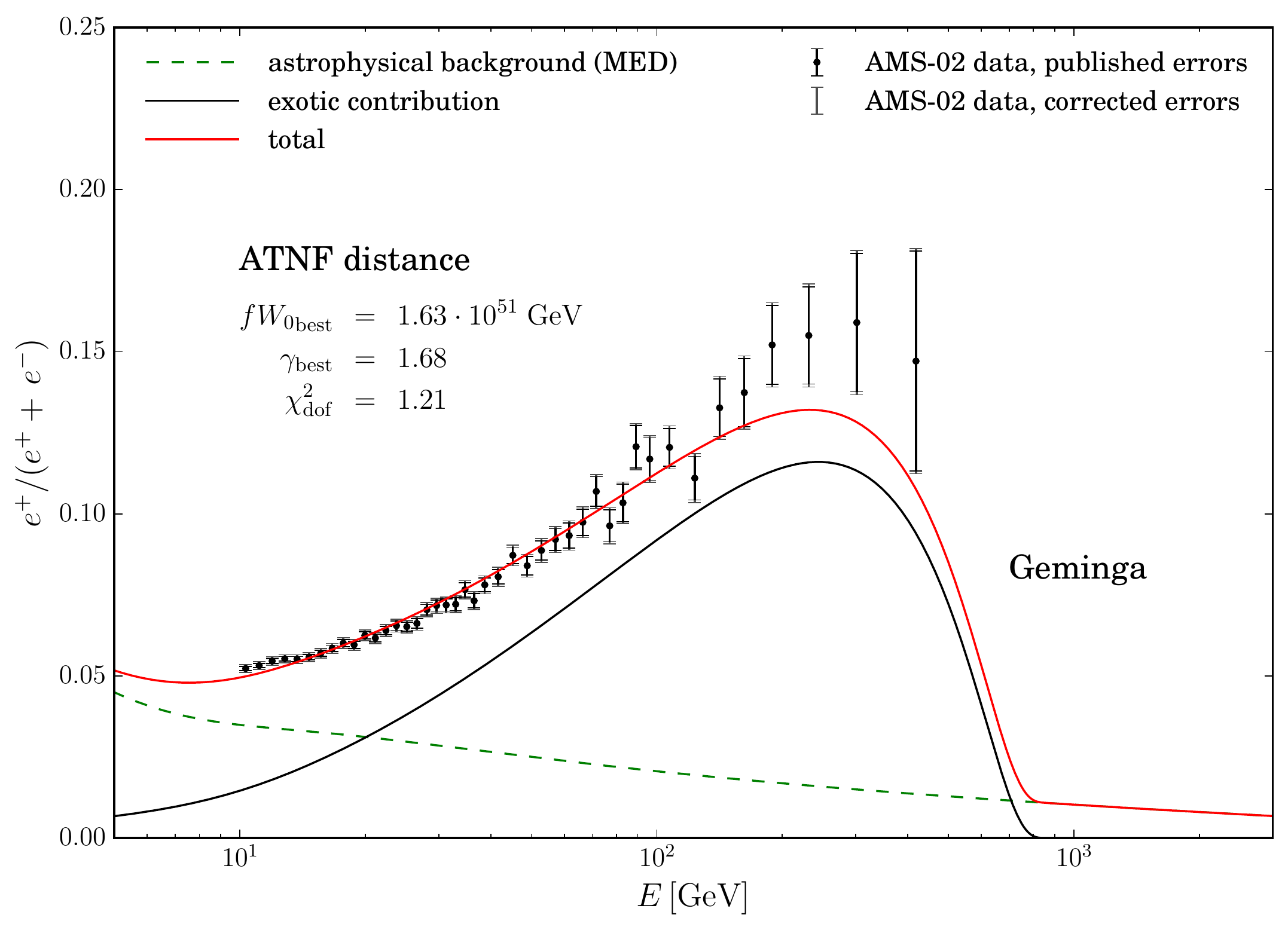}
\includegraphics[width=0.49\textwidth]{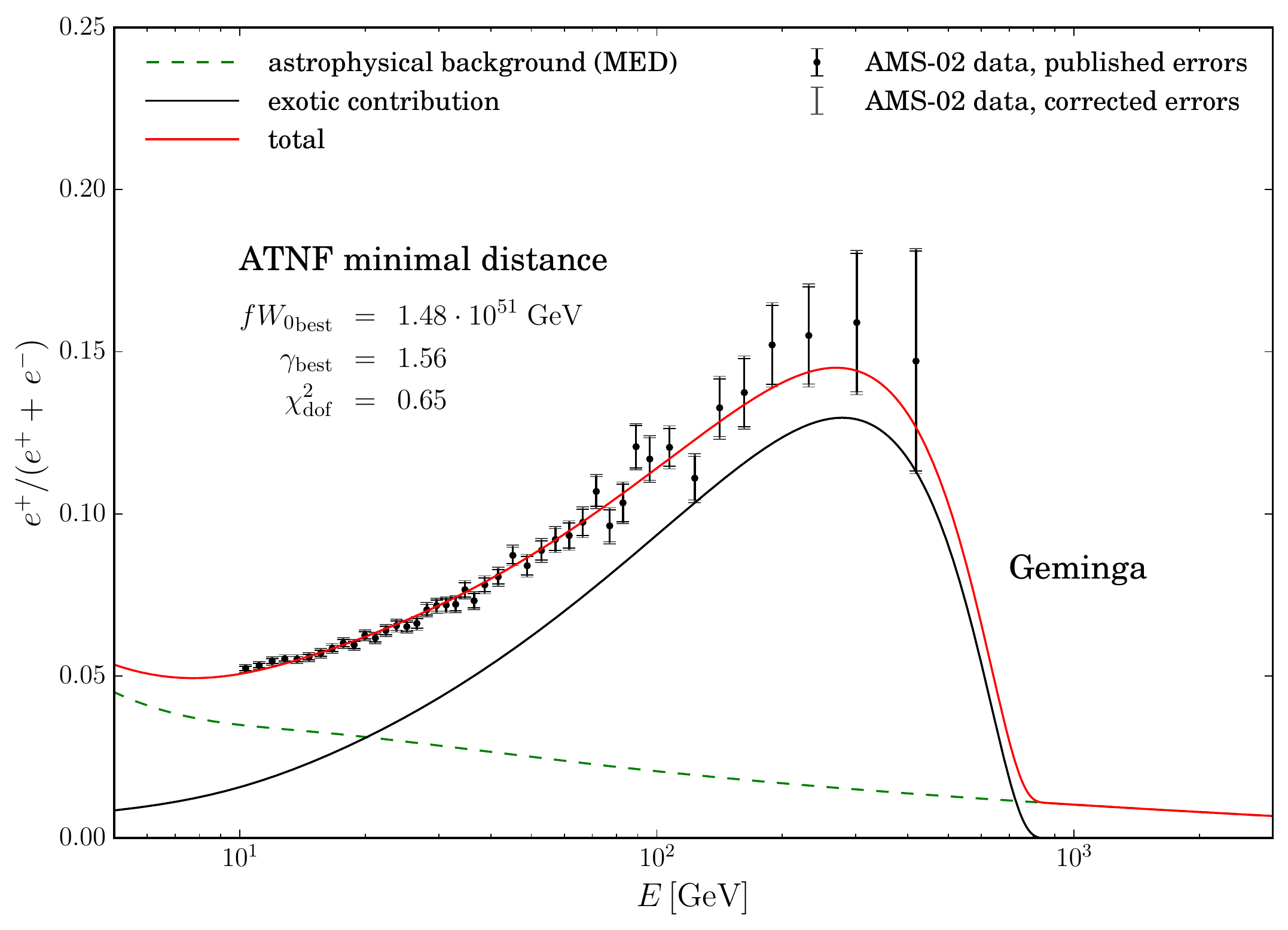}
\caption{
Positron fraction for the best fits for the pulsar Geminga considering the nominal (left panel) and minimal (right panel) distances.
The spectral index $\gamma$ at the source decreases with the pulsar distance. The positron flux becomes harder and better fits
the highest-energy data points.
}
\label{fig:PF_Geminga_distance}
\end{center}
\end{figure*}
%

%
\begin{figure}[ht!]
\begin{center}
\includegraphics[width=0.49\textwidth]{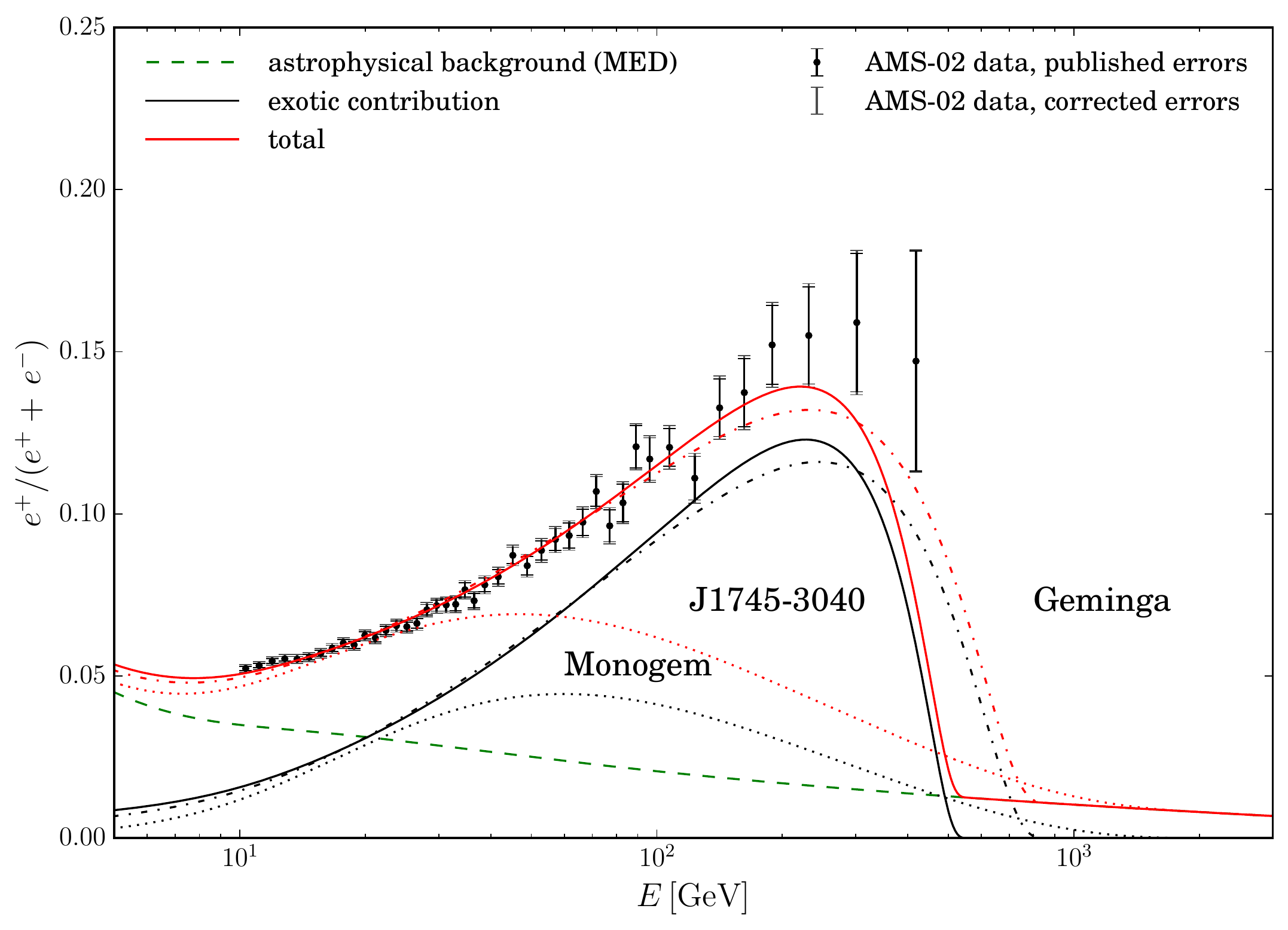}
\caption{Positron fraction for the best fits for the pulsars J1745-3040 (solid line), Geminga (dashed-dotted line), and Monogem (dotted line) with the propagation model MED.}
\label{fig:PF_J1745_Geminga_Monogem}
\end{center}
\end{figure}
%

\subsection{What happens when we get more statistics?}

We can now investigate how the list of selected pulsars would change if AMS-02 publishes a positron fraction in ten years with more statistics. To estimate the new error bars, we assume that the number of events follows a Gaussian distribution in each bin. This is a reasonable assumption since the last bin already contains 72 positrons. Therefore, the statistical uncertainty $\sigma_{\rm stat}$ decreases with time $t$ as $\sigma_{\rm stat} \propto 1/\sqrt{t}$. The systematic uncertainty $\sigma_{\rm syst}$ is here assumed to be constant with time. The uncertainty on the lepton flux is expected to follow the same variation with time as that on the positron fraction. Accordingly, the total uncertainty in each energy bin is multiplied by the reduction factor $RF(t)$ defined as:
\begin{equation}
RF(t) = \sqrt{\frac{\sigma_{\rm stat, \, AMS}^2 {\displaystyle \frac{t_0}{t}} + \sigma_{\rm syst, \, AMS}^2}{\sigma_{\rm stat, \, AMS}^2 + \sigma_{\rm syst, \, AMS}^2}},
\end{equation}
where $\sigma_{\rm stat, \, AMS}$ and $\sigma_{\rm syst, \, AMS}$ are the statistical and systematic uncertainties, while $t_{0} = 2.47$ yr stands for
the data taking time of the published AMS-02 data, to be compared to the time $t$ of the assumed data collection (10 years).

\vskip 0.1cm
In Fig.~\ref{fig:pulsar_gamma_fW_0_10yr} the same analysis as in Sec.~\ref{subsec:five_samourais} and Fig.~\ref{fig:pulsar_gamma_fW_0} is performed. Since the mean value of the positron fraction does not change, the colour variations of Fig.~\ref{fig:pulsar_gamma_fW_0} and \ref{fig:pulsar_gamma_fW_0_10yr} are the same. However, the good-fit regions defined by the iso-contours of the {\it p}-values drastically shrink. Thus, if the tendency of the positron fraction remains similar, the single pulsar hypothesis would be excluded by our criterion. The currently allowed five pulsars benefit from the large statistical uncertainties of the last bins.
%
\begin{figure*}[ht!]
\begin{center}
\includegraphics[height=0.427\textwidth]{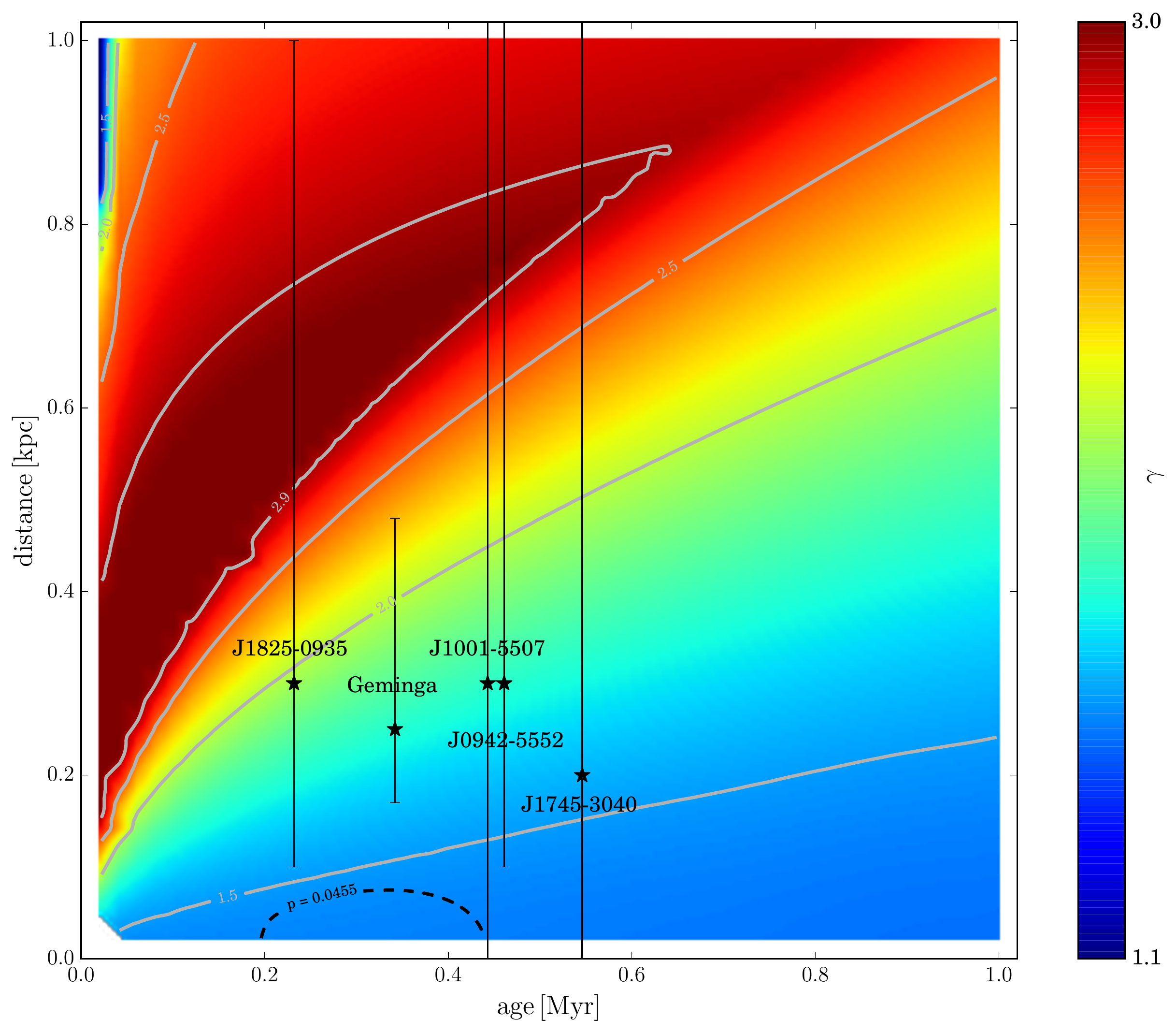}
\includegraphics[height=0.427\textwidth]{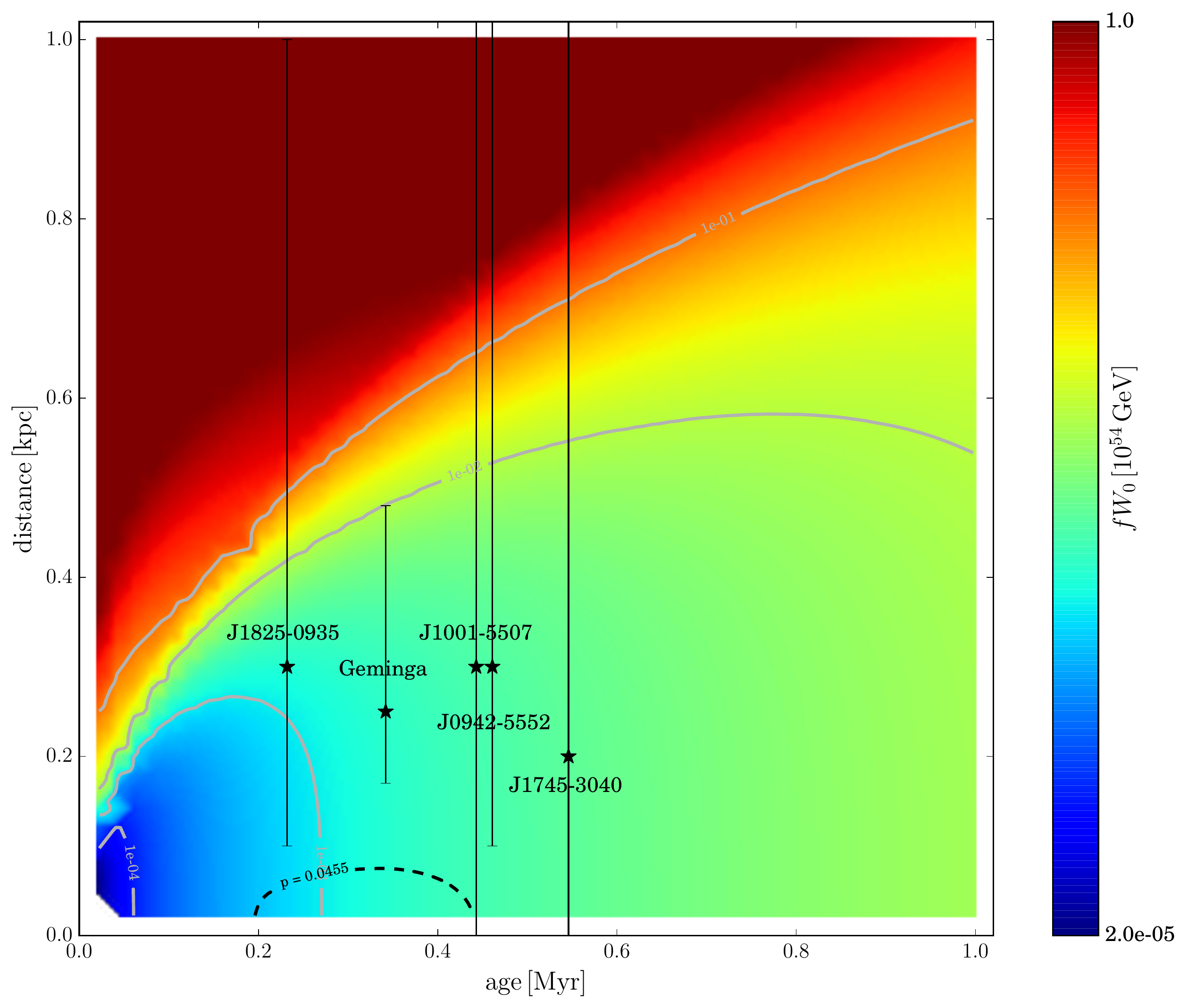}
\caption{
Same as Fig.~\ref{fig:pulsar_gamma_fW_0} but with ten years of measurements of the positron fraction by AMS-02.}
\label{fig:pulsar_gamma_fW_0_10yr}
\end{center}
\end{figure*}
%

%
\section{The effect of cosmic ray propagation uncertainties}
\label{sec:CR_uncertainty}

In Secs.~\ref{sec:DM_analysis} and \ref{sec:single_pulsar} we studied the constraints on an additional contribution of DM or a single pulsar to the positron fraction measured by the AMS-02 experiment above 10\,GeV. These constraints have been obtained by modelling the expected positron flux with the cosmic ray diffusion benchmark model MED defined in  \citet{2004PhRvD..69f3501D}. However, the transport mechanisms of Galactic cosmic rays are still poorly understood. The uncertainties on cosmic ray transport parameters are not negligible and have a major impact on searches for new physics. To take these uncertainties into account and to study their effect on modelling the positron fraction with an additional contribution, we use a set of 1623 combinations of the transport parameters $\{\delta, K_0, L, V_c, V_A \}$. These parameter sets result from a secondary-to-primary ratio analysis \citep{2001ApJ...555..585M} where 26 data points of the boron-to-carbon (B/C) ratio were fitted over the energy range from 0.1 to 35\,GeV/n leading to a $\chi^2$ less than 40. The advantage of choosing this study over more recent
studies\citepads{2010A&A...516A..66P, 2012A&A...539A..88C, 2011ApJ...729..106T} are the wider and more conservative ranges of the transport parameters. In addition, the benchmark models MIN, MED, and MAX of\citetads{2004PhRvD..69f3501D}, widely used in the DM literature, are based on the parameters found in \citet{2001ApJ...555..585M}.

\vskip 0.1cm
In the following, we extrapolate these models to higher energies without taking any contribution from secondaries accelerated in nearby sources into account \citep{2009PhRvL.103e1104B,2014PhRvD..90f1301M}. We furthermore marginalise over $V_c$ and $V_A$ since the reaccelerating and convection processes are negligible at higher energies and are not taken into account in our positron flux calculation. Finally, we only show the $\chi \chi \rightarrow b\bar{b}$ channel and the pulsar J1745-3040 as an example to highlight the correlations between the transport parameters and the parameters necessary to model the additional exotic contribution to the positron fraction at higher energies.

\begin{figure*}[ht!]
	\centering
	\includegraphics[height=0.37\textwidth]{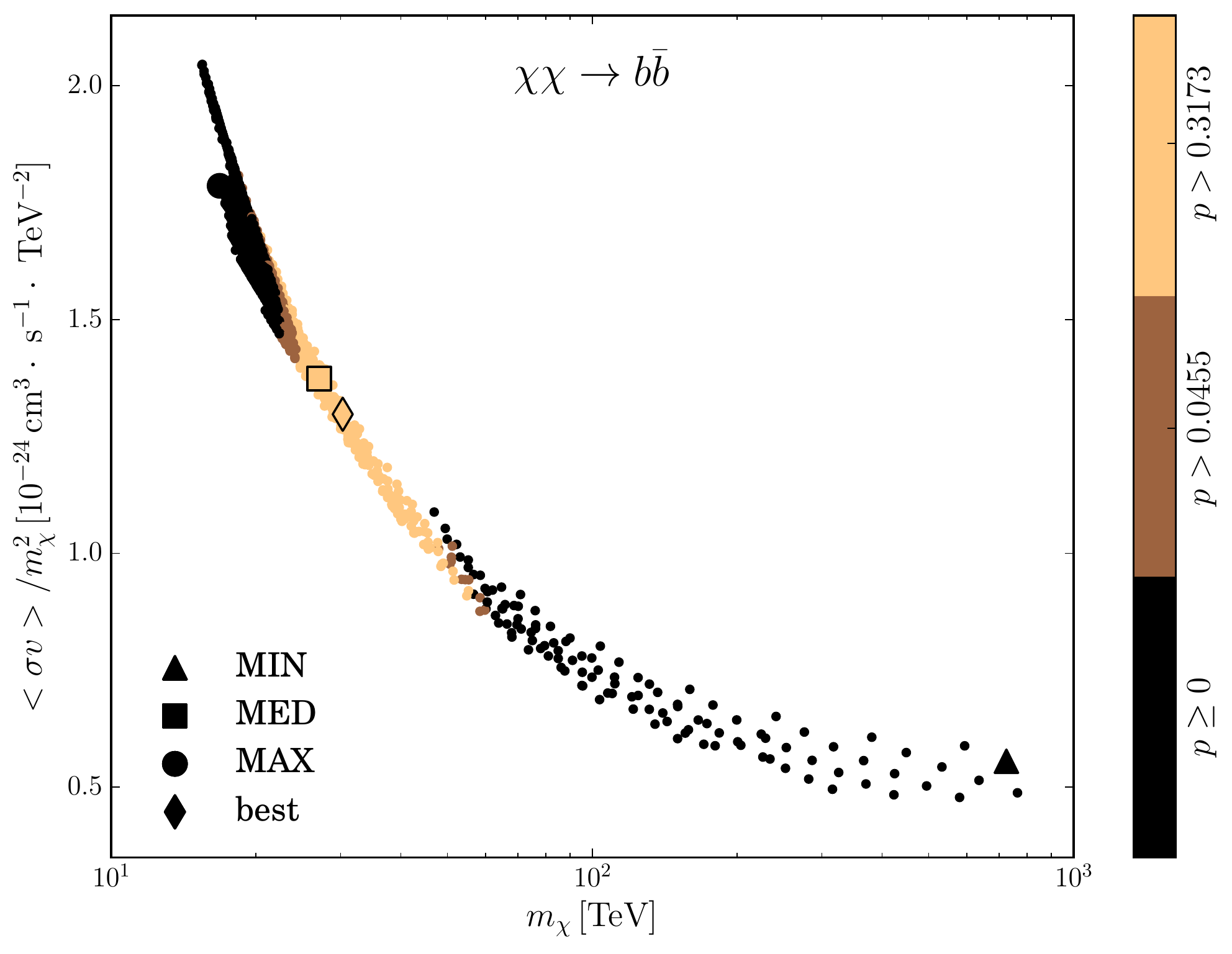}
	\includegraphics[height=0.37\textwidth]{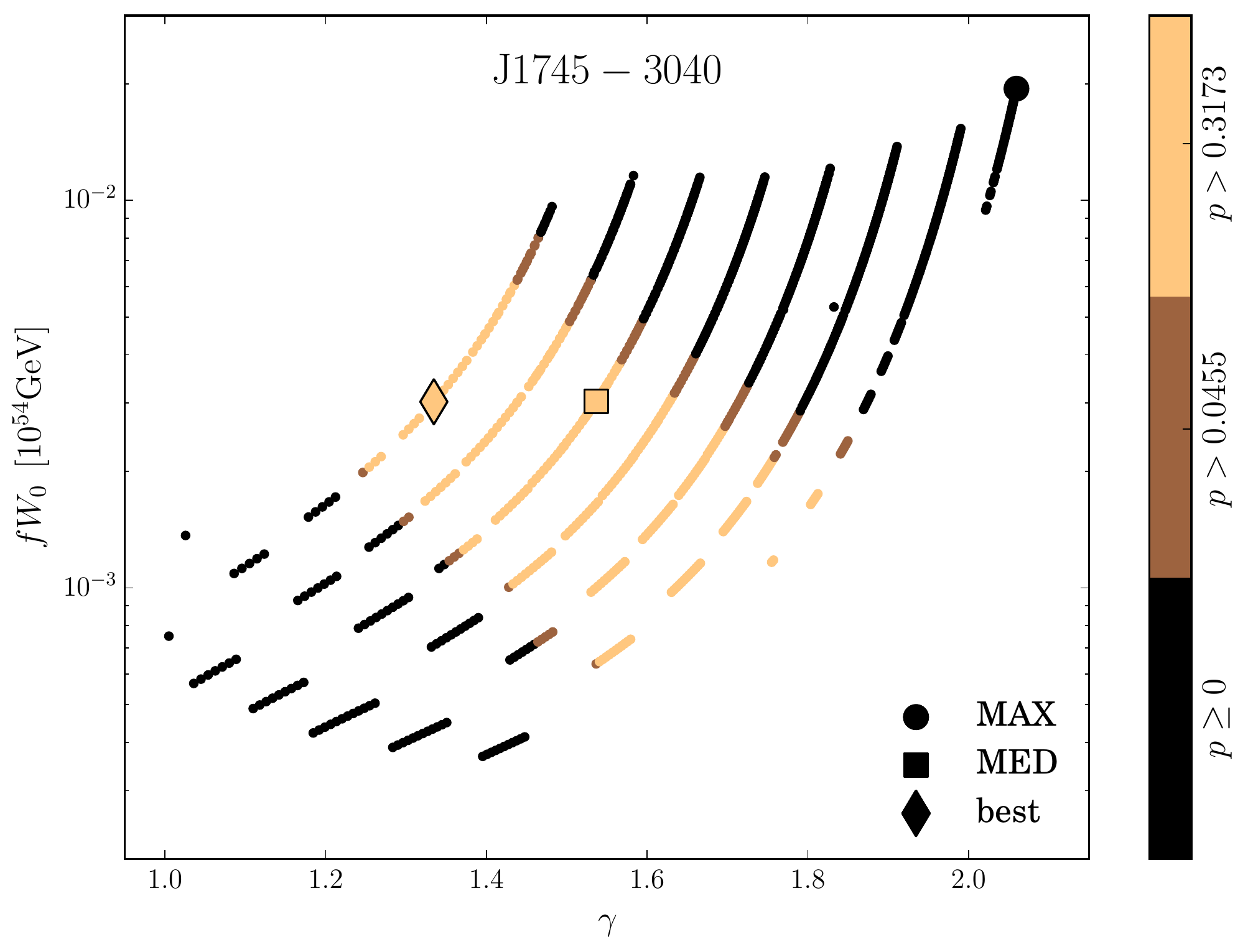}
	\caption{{\it p}-value distributions of the 1623 transport parameter sets for the DM $\{m_\chi, \langle \sigma v \rangle/ m_\chi^2 \}$ (left plot) and the pulsar $\{ \gamma, fW0 \}$ (right plot) parameters. The colour coding represents the increasing {\it p}-value from darker to lighter colours. The benchmark models MIN, MED, and MAX are represented with a triangle, square, and circle symbol, respectively. In addition, the best transport parameter set is highlighted with a diamond symbol.}
	\label{fig:pvalues}
\end{figure*}

\subsection{Which transport parameters give a good fit?}
For each set of transport parameters, we fit the positron fraction to find the best combination of $\{\langle \sigma v \rangle, \, m_{\chi}\}$ or $\{fW_0, \, \gamma\}$ for the DM and pulsar contributions, respectively. We calculate the {\it p}-value given in formula~\ref{eq:p-value} to determine for which transport parameter set the modelled sum of secondary and exotic contributions reproduce well the positron fraction measured by AMS-02.
In Fig.~\ref{fig:pvalues}, the {\it p}-value distributions of the 1623 transport parameter sets for the DM $\{m_\chi, \, \langle \sigma v \rangle/ m_\chi^2 \}$ (left plot) and the pulsar $\{ \gamma, \,fW0 \}$ (right plot) parameters are shown. The colour coding represents the increasing {\it p}-value from darker to lighter colours, which is binned into the three defined {\it p}-value ranges:
\begin{enumerate}
\item$p > 0.3173$: the modelled positron fraction reproduces the experimental data very well (yellow  dots),
\item$0.0455 < p \leq 0.3173$: the modelled positron fraction reproduces the experimental data well enough  (brown dots),
\item$p \leq 0.0455$: the modelled positron fraction reproduces the experimental data (black dots) badly. It is excluded for the final results.
\end{enumerate}

\vskip 0.1cm
The benchmark models MIN, MED, and MAX are represented with a triangle, square, and circle symbol, respectively. In addition, the best transport parameter set is highlighted with a diamond symbol. For the pulsar J1745-3040 some transport models resulting in a very low and unphysical $\gamma$ values were excluded from the analysis, including the benchmark model MIN. The criterion of goodness-of-fit defined above reduces the number of transport parameter sets considered from 1623 to a few hundred. In general, the benchmark models MIN (triangle) and MAX (filled circle) are disfavoured by the experimental data. We observe on these figures that the transport parameters are strongly correlated with the DM and pulsar parameters. We discuss these correlations in the following section.

\subsection{Correlations between parameters}

\begin{figure*}[ht!]
	\centering
	\includegraphics[height=0.365\textwidth]{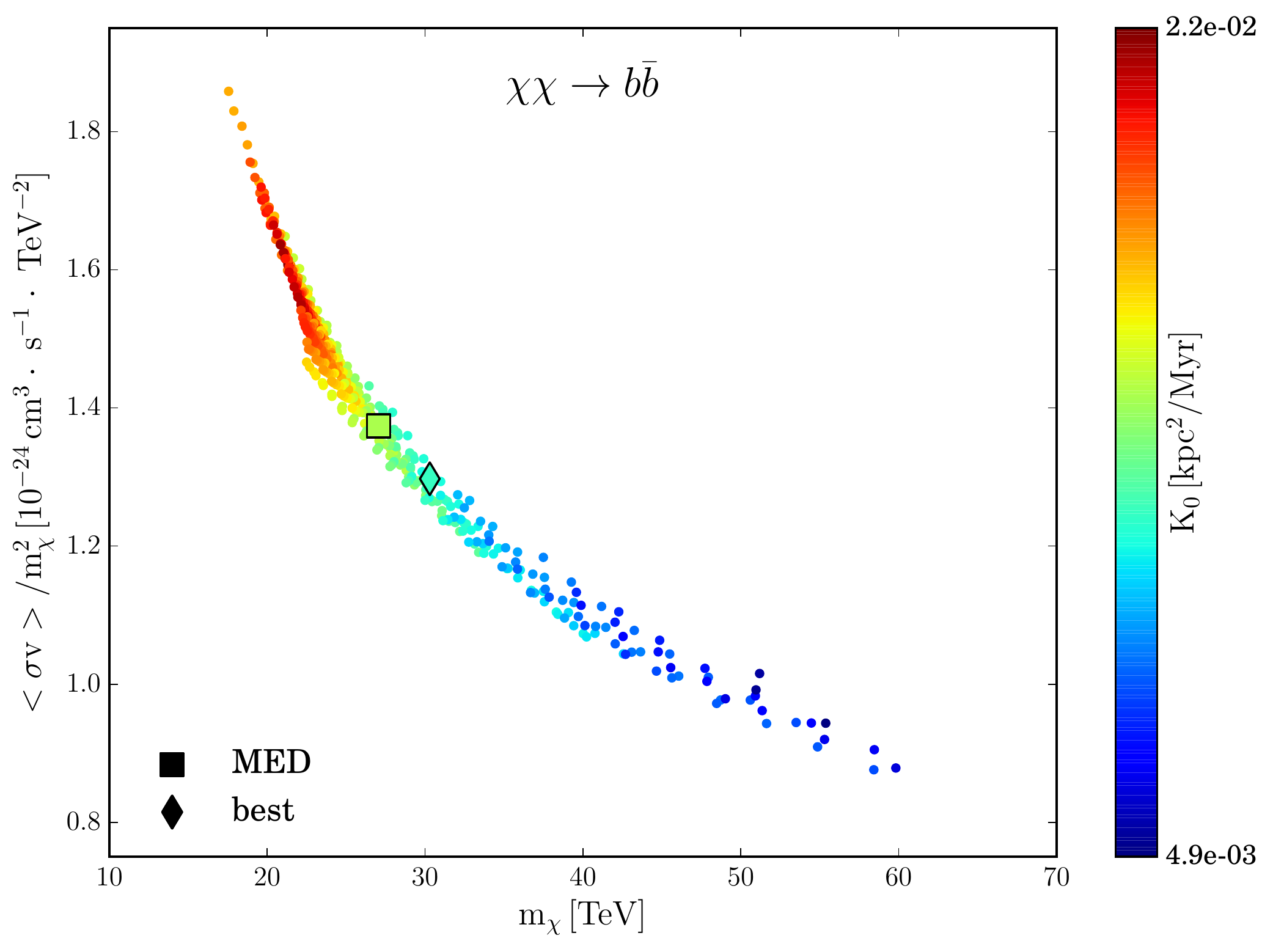}
	\includegraphics[height=0.365\textwidth]{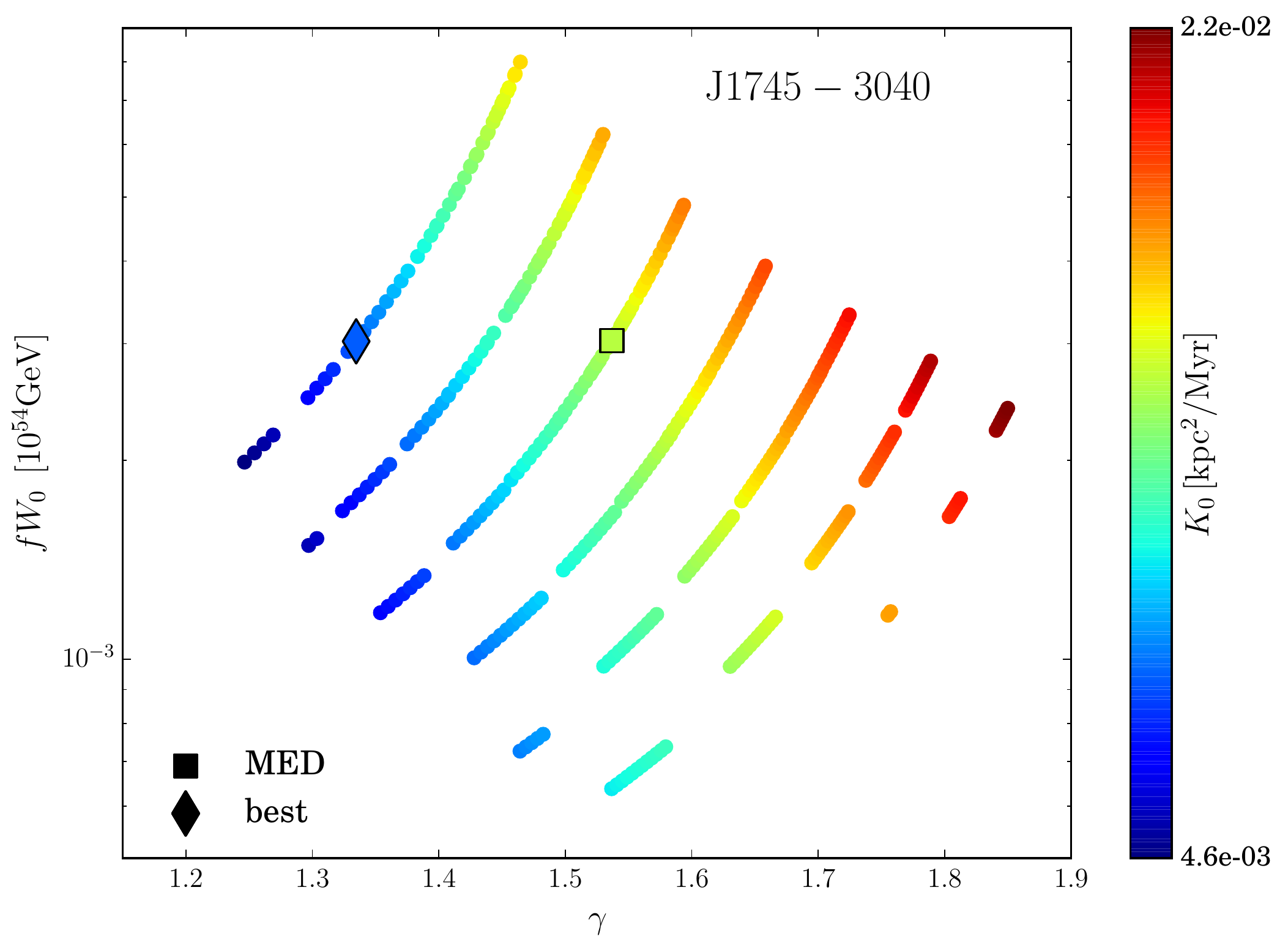}
	\caption{Surviving transport parameter sets' $K_0$ distributions for the DM $\{m_\chi, \langle \sigma v \rangle/ m_\chi^2 \}$ (left plot) and the pulsar $\{ \gamma, fW_0 \}$ (right plot) parameters. The colour coding represents the increasing $K_0$ value from blue to red. The benchmark model MED is represented with a square. In addition, the best transport parameter set is highlighted with a diamond symbol.}
	\label{fig:K0L}
\end{figure*}

\begin{figure*}[ht!]
	\centering
	\includegraphics[height=0.375\textwidth]{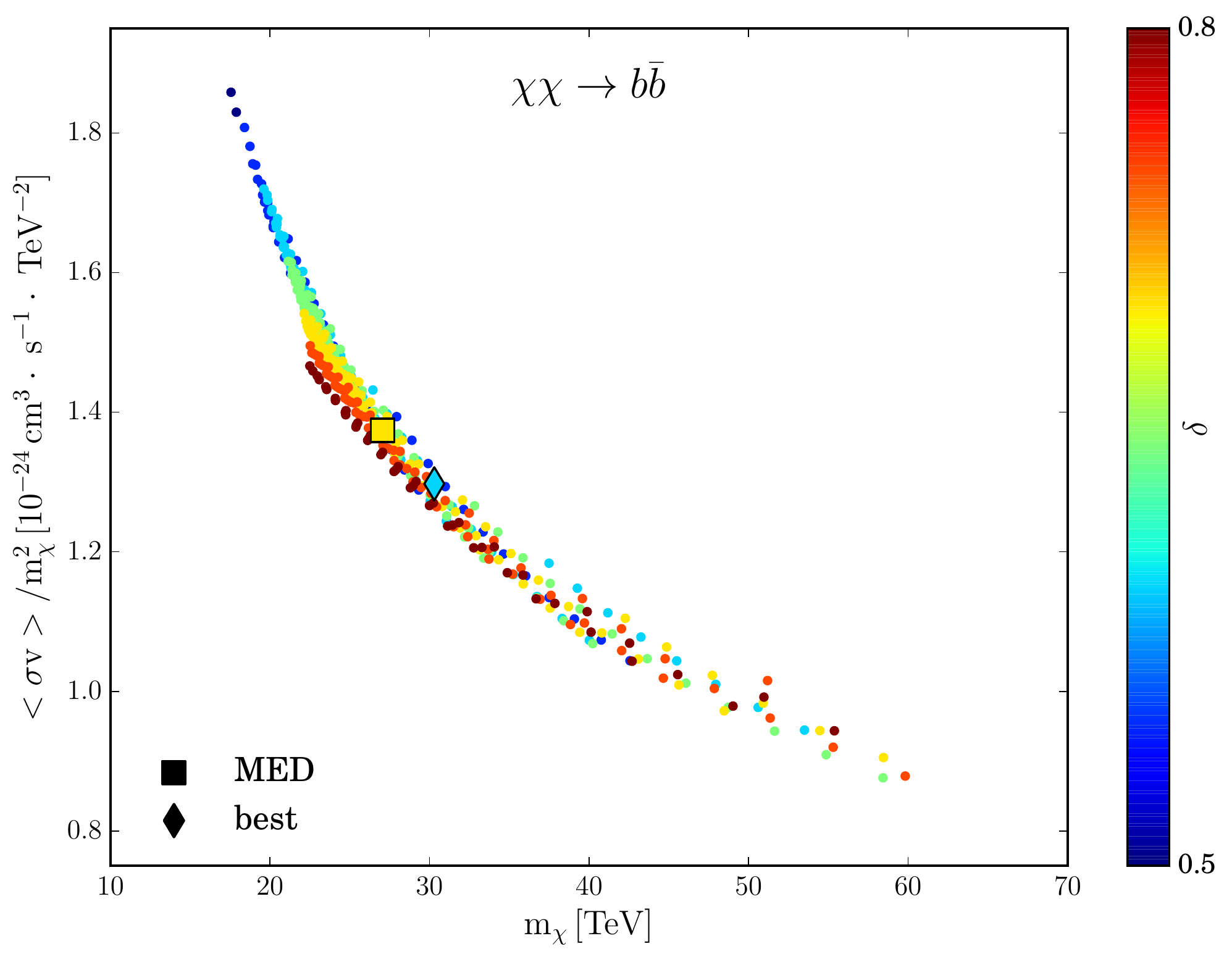}	
	\includegraphics[height=0.375\textwidth]{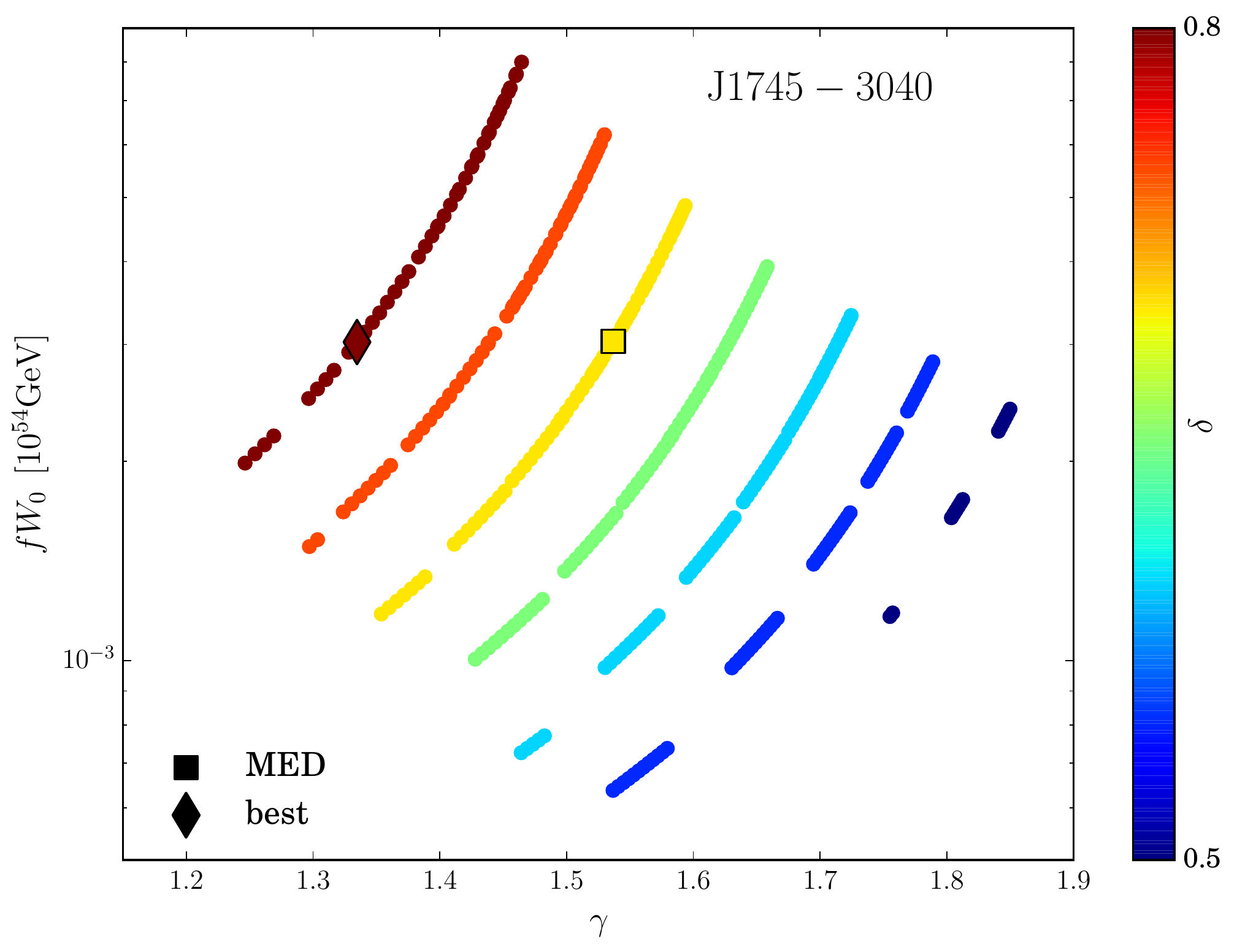}	
	\caption{Surviving transport parameter sets' $\delta$ distributions for the DM $\{m_\chi, \langle \sigma v \rangle/ m_\chi^2 \}$ (left plot) and the pulsar $\{ \gamma, fW_0 \}$ (right plot) parameters. The colour coding represents the increasing $\delta$ value from blue to red. The benchmark model MED is represented with a square. In addition, the best transport parameter set is highlighted with a diamond symbol.}
	\label{fig:delta}
\end{figure*}

Figures~\ref{fig:K0L} and \ref{fig:delta} show the best fit for each transport parameter set (coloured dot) in the $\langle \sigma v \rangle/m_\chi^2$ -- $m_{\chi}$ (left plots) and $\gamma$ -- $fW_0$ (right plots) planes, representing the correlations between the transport parameters and DM or pulsar parameters. Plotting $\langle \sigma v \rangle/m_{\chi}^{2}$ instead of $\langle \sigma v \rangle$ allows us to clearly see the correlations. The colour indicates the value of a given transport parameter ($K_0$ in Fig.~\ref{fig:K0L}, $\delta$ in Fig.~\ref{fig:delta}) from lower (blue) to higher (red) values. One can clearly see a strong correlation between the transport and DM or pulsar parameters showing a huge impact on the best-fit values for the considered free parameters.

\vskip 0.1cm
The main correlations are due to the normalisation and the shape of the fluxes. Indeed, $\langle \sigma v \rangle/m_\chi^2$ and $fW_0$ are related
to the number of positrons injected in the MH, whereas the amount of produced secondary particles is inversely proportional to the diffusion
length $\lambda_{D}$ (see Eq.~\ref{eq:lambdaD}) and is hence negatively correlated with $K_0$.
If enough secondary particles are created, we need fewer particles through DM or pulsar injection and vice-versa. For the DM candidate,
$\langle \sigma v \rangle/m_\chi^2$ increases with $K_0$ as shown in the left panel of Fig.~\ref{fig:K0L}. In the case of the J1745$-$3040
pulsar, that trend is reinforced by the fact that given the small distance of the source, the positron flux scales as $\lambda_{D}^{-3} \propto K_0^{-3/2}$.
We observe hence a strong positive correlation between $f W_0$ and $K_0$ in the right panel of Fig.~\ref{fig:K0L}.
The spectral index of the diffusion coefficient $\delta$ modifies the high energy shape of the secondary positron flux.
A lower (higher) value of $\delta < 0.5$ ($\delta > 0.5$) has a harder (softer) spectrum and therefore allows for a DM induced contribution
at smaller (higher) DM masses $m_{\chi}$. This trend is clearly visible in the left panel of Fig.~\ref{fig:delta}. The correlation between
$\langle \sigma v \rangle/m_\chi^2$ and $m_{\chi}$ is due to the form of the DM spectrum and varies from channel to channel.
For pulsars, the primary positron flux behaves as $f W_0 / E^{(\gamma + 3 \delta /2)}$.
This scaling accounts for the positive (negative) correlation between $f W_0$
($\gamma$) and the spectral index $\delta$ observed in the right panel of Fig.~\ref{fig:delta}.


\vskip 0.1cm
In each of the above figures, the best transport parameter set (highest $p$-value) is shown (diamond symbol) for the $\chi\chi \rightarrow b \bar{b}$ channel and the pulsar J1745$-$3040.
In the same way, we also extract the best set of parameters for the other studied DM annihilation channels and pulsars. The results are summarised in Tab.~\ref{tab:onechannel_PF14_ICHEP14_with_best_propagation} and \ref{tab:table_pulsar_best_fits_with_best_propagation}.
In each case, independent of the primary positron source, we can find a set of parameters that better describes the experimental data than the benchmark model MED. 
Moreover, in the framework of our analysis, the experimental data favour small halo sizes ($L \lesssim 3.5$\,kpc). Eventually, taking the uncertainties into account of the propagation parameters does not change the discrepancy between the AMS-02 data and the modelled positron fraction, neither for the electron DM annihilation channel nor for the Monogem and Vela pulsars.

%
\begin{table*}[ht!]
\centering
\caption{Best fits for specific DM annihilation channels assuming the best propagation parameter set. The recently published
positron fraction \citep{Accardo:2014lma} and the AMS-02 lepton spectrum \citep{leptons_ICHEP_2014} are used to
derive the $\chi^{2}$, as in formula~(\ref{eq:chi_2_def}). The {\it p}-value, indicated in the last column, is defined
in formula~(\ref{eq:p-value}).}
\label{tab:onechannel_PF14_ICHEP14_with_best_propagation}
\vspace{.1cm}
\begin{tabular}{|c|c|c|c|c|c|c|c|c|}
\hline
{Channel} &  {$m_\chi $} & {$<\sigma v>$  } 	   & {$K_0$ } 			    & {$L$ } & {$\delta$} & {$\chi^2$} & {$\chi^{2}_{\rm dof}$} & $p$ \\ 
		   & [\rm TeV] 	& $[\rm cm^{3} \, s^{-1}]$ & $ [\rm kpc^2 \, Myr^{-1}]$ & 	$[\rm kpc]$			&			&			&					    &  \\
\hline
$e$ & 0.350 $\pm$  0.001 & $(1.88 \pm 0.03) \cdot 10^{-24}$ & 0.00405 & 1.5 & 0.70 & 682 & 16.6 & 0 \\
$\mu$ & 0.422 $\pm$ 0.018 & $(3.99 \pm 0.30) \cdot 10^{-24}$ & 0.00540 & 2.0 & 0.70 & 99.2 & 2.42 & $9.8 \cdot 10^{-7}$\\
$\tau$ & 1.26 $\pm$ 0.07 & $(3.47 \pm 0.31) \cdot 10^{-23}$ & 0.00670 & 2.0 & 0.65 & 24.1 & 0.59  & 0.98 \\
$u$ &  33.7 $\pm$ 3.3 & $(1.61 \pm 0.24) \cdot 10^{-21}$ & 0.00930 & 2.0 & 0.60 & 18.1 & 0.44  & 0.99 \\
$b$ &  30.3 $\pm$ 2.7 & $(1.19 \pm 0.16) \cdot 10^{-21}$ & 0.00910 & 2.0 & 0.60 & 17.7 & 0.43 & 0.99 \\
$t$ &  50.8 $\pm$ 4.1 & $(2.39 \pm 0.30) \cdot 10^{-21}$ & 0.00775 & 2.5 & 0.70 & 17.7 & 0.43 & 0.99 \\
$Z$ &  20.0 $\pm$ 1.3 & $(9.65 \pm 1.01) \cdot 10^{-22}$ & 0.00675 & 3.0 & 0.75 & 17.7 & 0.43 & 0.99 \\
$W$ &  17.5 $\pm$ 0.1 & $(8.40 \pm 0.90) \cdot 10^{-22}$ & 0.00675 & 3.0 & 0.75 & 17.6 & 0.43 & 0.99 \\
$H$ &  31.7 $\pm$ 2.2 & $(1.26 \pm 0.14) \cdot 10^{-21}$ & 0.00690 & 3.0 & 0.75 & 17.6 & 0.43 & 0.99 \\
$\phi \rightarrow e$ &  $0.350 \pm 0.005$ & $(1.31 \pm 0.02) \cdot 10^{-24}$ &	0.00440  & 2.0	    &	 0.65 & 143 & 3.48 & 0 \\
$\phi \rightarrow \mu$ & $ 0.763 \pm 0.034$ & $(7.77 \pm 0.56) \cdot 10^{-24}$ & 0.00660	&	2.0    &	0.65 & 44.7 & 1.09 & 0.32 \\
$\phi \rightarrow \tau$ & $2.43 \pm 0.13$ & $(6.80 \pm  0.61) \cdot 10^{-23}$ &	0.00680 	&	2.0   &	0.65 & 22.9 & 0.56  & 0.99 \\ 
\hline 
\end{tabular}
\end{table*}
%
\vskip 0.1cm
%
\begin{table*}[ht!]
\centering
\caption{Best fits for the five pulsars selected in Sec.~\ref{subsec:five_samourais} as well as Monogem and Vela in the single pulsar approach, assuming the best propagation parameter set.
 The pulsar ages and distances are the nominal values taken from the ATNF catalogue. The recently published
positron fraction \citep{Accardo:2014lma} and the AMS-02 lepton spectrum \citep{leptons_ICHEP_2014} are used to
derive the $\chi^{2}$, as in formula~(\ref{eq:chi_2_def}). The {\it p}-value, indicated in the last column, is defined
in formula~(\ref{eq:p-value}).}
\label{tab:table_pulsar_best_fits_with_best_propagation}
\vspace{.1cm}
\begin{tabular}{|c|c|c|c|c|c|c|c|c|c|c|}
\hline
{Name} & {Age } & {Distance} & {fW0}                             & {$\gamma$} & {$K_0$ }                           & {$L$}         & {$\delta$} & {$\chi^{2}$} & {$\chi^{2}_{\rm dof}$} & $p$ \\ 
	      & [kyr]    &  [kpc]        &  $[10^{54} \, \rm GeV]$  & 			     & $ [\rm kpc^2 \, Myr^{-1}]$ & $[\rm kpc]$ &                  &                   &                                    &    \\
\hline
J1745$-$3040	& 546 & 0.20 & $ (3.02 \pm 0.08) \cdot 10^{-3} $ & $ 1.33 \pm 0.01 $ &  0.00647         &   3.5     &  0.80        & 21.6 & 0.53 & $ 0.99 $ \\
\hline
J0633$+$1746 & 342 & 0.25 & $ (6.29 \pm 0.08) \cdot 10^{-4} $ & $ 1.74 \pm 0.02 $ & 0.00850 & 2.0 & 0.60 & 18.0 & 0.44 & 0.99 \\
\textit{Geminga}	&         &  &  &  &  &  & & & & \\
\hline
J0942$-$5552 & 461 & 0.30 & $ (3.02 \pm 0.08) \cdot 10^{-3} $ & $ 1.34 \pm 0.01$ & 0.00647 & 3.5 & 0.80 & 21.6 & 0.53 & $ 0.99 $ \\ 
\hline
J1001$-$5507 & 443 & 0.30 & $ (3.02 \pm 0.08) \cdot 10^{-3} $ & $ 1.34 \pm 0.01$ & 0.00647 & 3.5 & 0.80 & 21.6 & 0.53 & $ 0.99 $ \\ 
\hline
J1825$-$0935 & 232 & 0.30 & $ (1.08 \pm 0.07) \cdot 10^{-3} $ & $ 2.18 \pm 0.02 $ & 0.00802 & 1.5 & 0.55 & 23.0 & 0.56 & $ 0.99 $ \\
\hline
\hline
J0659$+$1414 & 111 & 0.28 & $ (6.20 \pm 0.01) \cdot 10^{-2} $ &  3.00  & 0.00590 & 1.0 & 0.55 & 63.7 & 1.6 & $ 0.01$ \\
\textit{Monogem}	&	    &  &  &  &  &  & & & &  \\
\hline
J0835$+$4510 & 11.3 & 0.28 & 1.00 & $ 2.49 \pm 0.04 $ & 0.00335 & 1.0 & 0.65 & 3453 & 84 & $ 0 $ \\
\textit{Vela}		&	     &  & &  &  &  &  & & &\\
\hline
\end{tabular}
\end{table*}
%
\vskip 0.1cm
\subsection{Comparison of systematic and statistical uncertainties}

\begin{figure*}[ht!]
	\centering
	\includegraphics[height=0.35\textwidth]{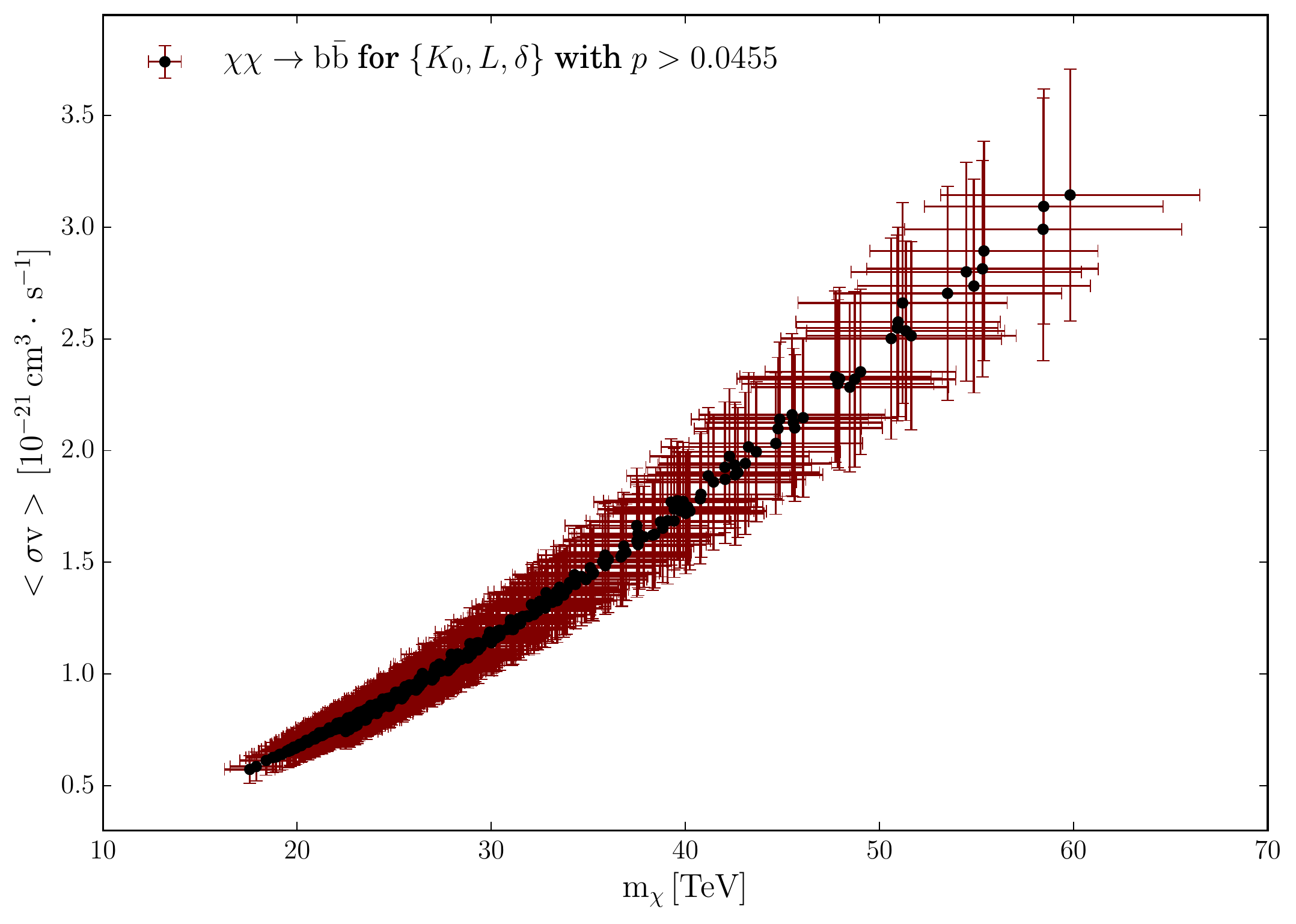}
	\includegraphics[height=0.35\textwidth]{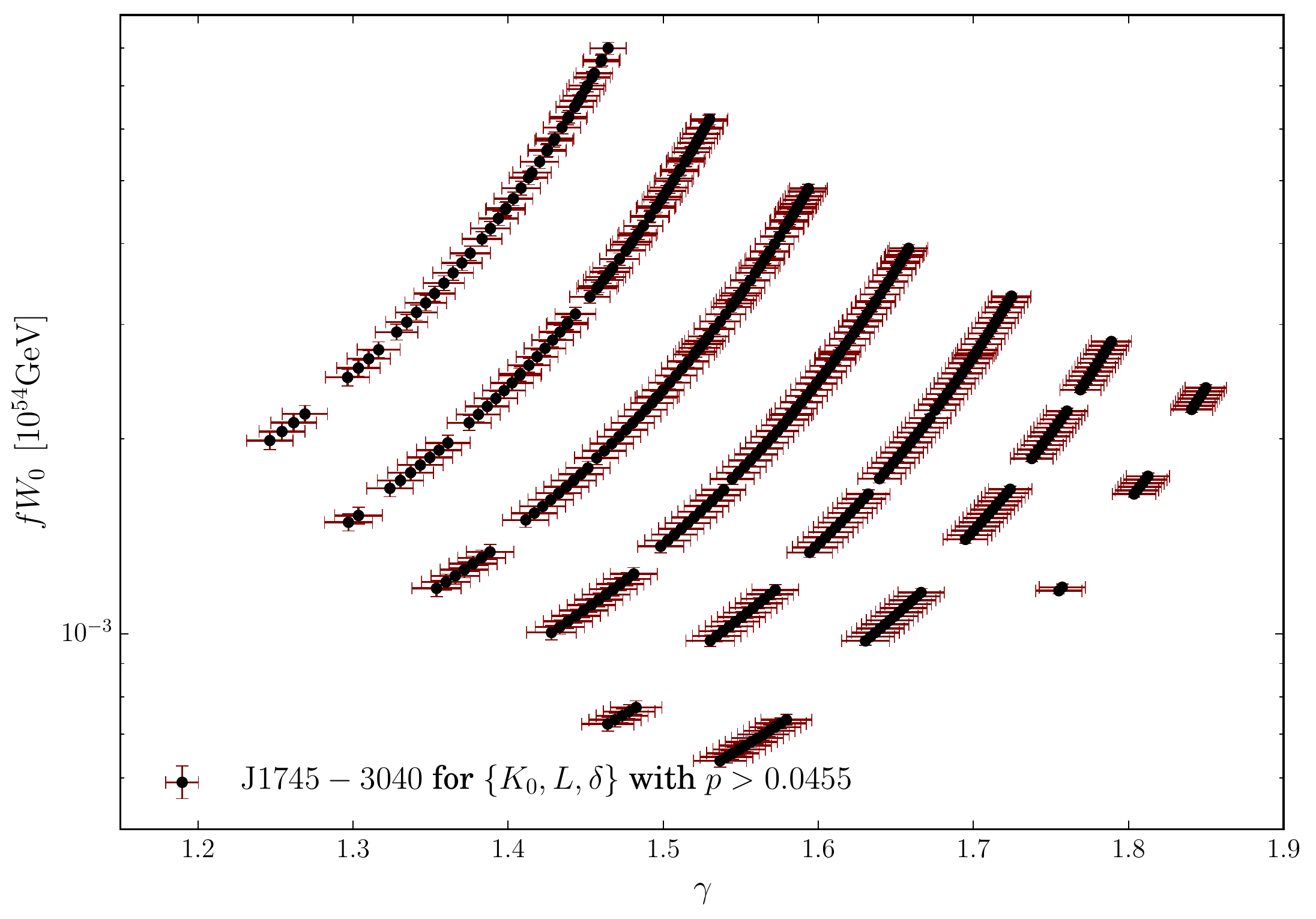}
	\caption{Best-fit values of the surviving transport parameter sets for the DM $\{m_\chi, \langle \sigma v \rangle \}$ (left plot) and the pulsar $\{ \gamma, fW0 \}$ (right plot) parameters. The error bars represent the errors on the fit parameter resulting from the statistical uncertainty on the experimental data.}
	\label{fig:errorbars}
\end{figure*}

The goodness-of-fit criterion ($p > 0.0455$) allows us to select the parameter sets that describe the experimental data reasonably well. The spread of these parameter sets shown in Fig.~\ref{fig:errorbars} (black dots) in the $\langle \sigma v \rangle$ -- $m_{\chi}$ and $\gamma$ -- $fW_0$ planes therefore represents the systematic uncertainty of the determination of $\langle \sigma v \rangle$ and $m_{\chi}$ as well as $\gamma$ and $fW_0$. Compared to the statistical uncertainties of these parameters because of the errors of the experimental data (red crosses), the systematic uncertainties dominate completely their determination. A perfect knowledge of the distance and age of the pulsar is assumed: in reality, this is not the case and would lead to larger uncertainties. The inclusion of the uncertainties on the pulsar distance is beyond the scope of this paper and will be considered in a follow-up study. To better estimate the transport parameters and reduce their impact on the study of an additional contribution to the positron fraction, more precise measurements of secondary-to-primary ratios over a large energy range are needed.

\subsection{How can the positron fraction constrain the diffusive halo size?}

\begin{figure*}[ht!]
	\centering
	\includegraphics[height=0.38\textwidth]{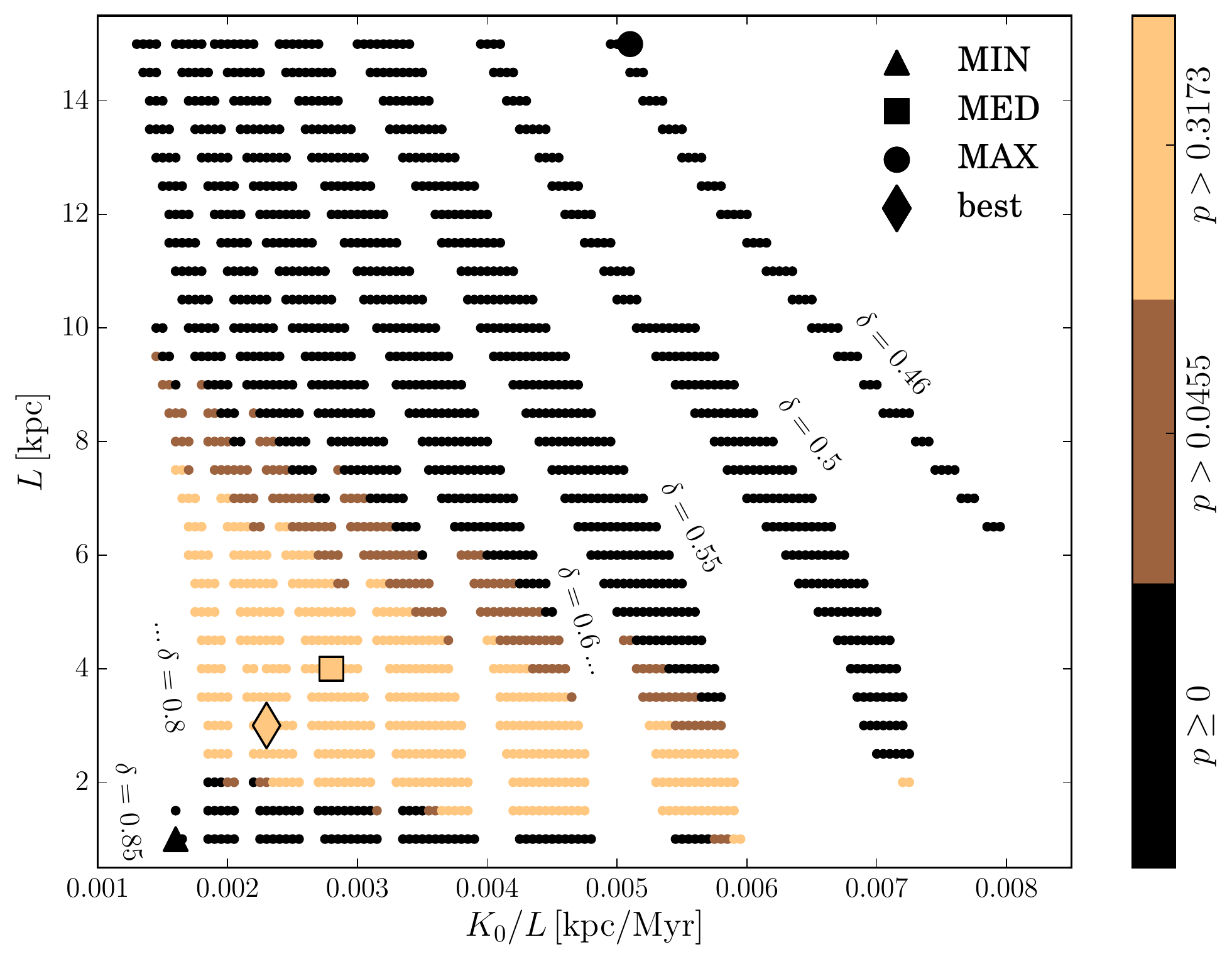}
	\includegraphics[height=0.38\textwidth]{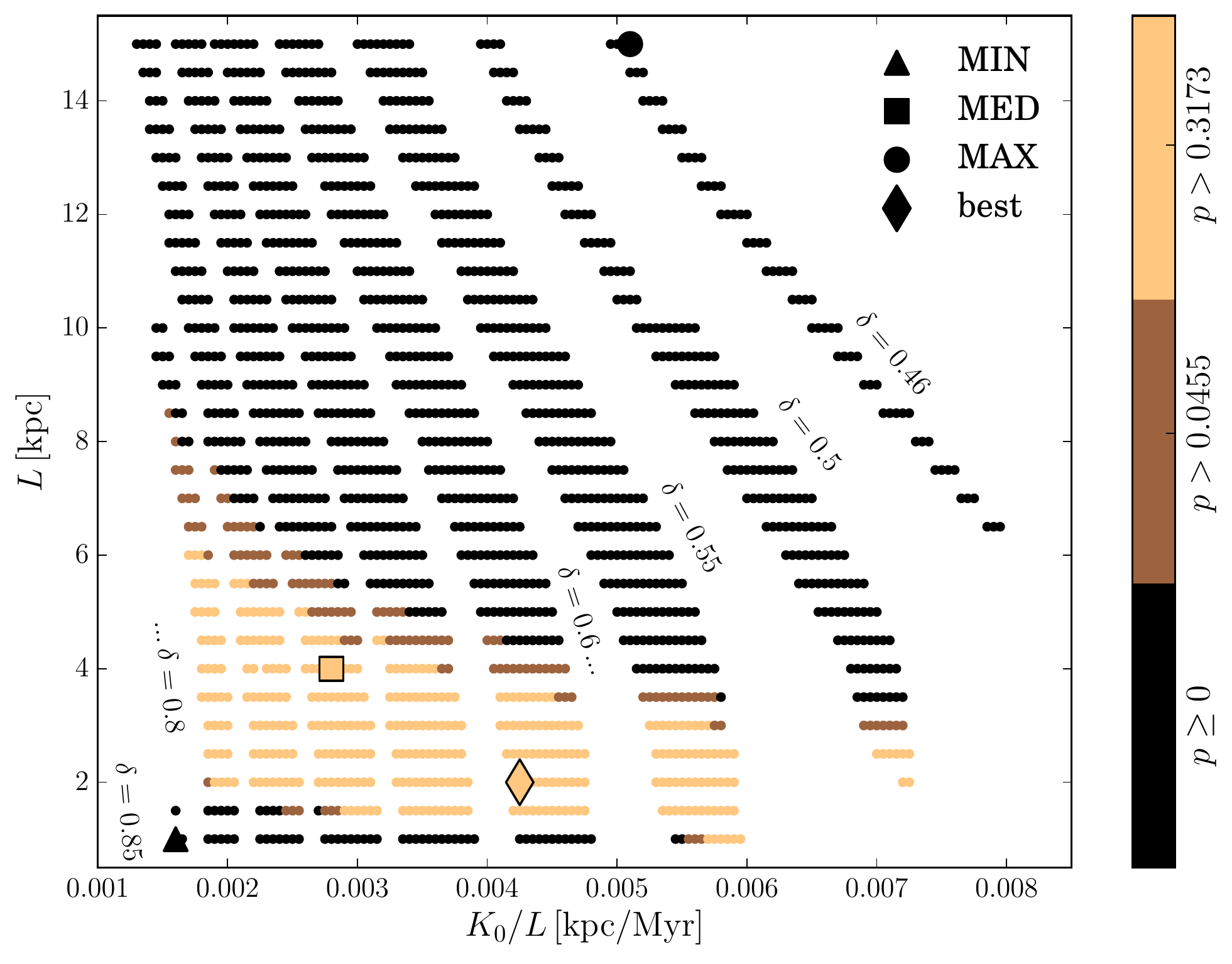}
	\caption{For each transport parameter set, represented in the $K_0/L$ -- $L$ plane, the best {\it p}-value distributions of all the 12 channels (five pulsars) are shown in the left (right) plot. The colour coding represents the increasing {\it p}-value from darker to lighter colours. The benchmark models MIN, MED, and MAX are represented with a triangle, square, and circle symbol, respectively. In addition, the best transport parameter set is highlighted with a diamond symbol.}
	\label{fig:K0L_constraints}
\end{figure*}

The B/C ratio is sensitive to the matter density in the Galactic disc, which is related to $L/K_0$. This degeneracy can be broken by an observable, which is sensitive to only one of the two parameters. In general, one uses radioactive secondary-to-stable secondary ratios, such as $^{10}$Be/$^{9}$Be. The radioactive secondaries decay before they can reach the edge of the Galaxy and escape. Their modelling is hence independent of the Galactic diffusive halo size $L$. Because of a lack of precise measurements over a sufficient large energy range up to now, the halo size is still not estimated well. Recently, \citet{2014PhRvD..90h1301L} demonstrated that low-energetic secondary positrons can directly constrain diffusion models with small haloes and large spectral indices due to very high energy losses and hence small diffusion lengths. Besides the availability of very precise positron data over a large energy range, this method is less sensitive to the modelling of the local interstellar medium compared to the standard approach. In this study, only a secondary positron spectrum was used. We propose to extend the analysis by considering here an additional contribution to the positron fraction from DM annihilation or pulsars as well as taking their different spectral shapes into account.

\vskip 0.1cm
Figure~\ref{fig:K0L_constraints} shows the 1623 transport parameter sets in the ($K_0/L,\, L$) plane. As before, we divide the sets into three different bins of {\it p}-values obtained by a fit with a  contribution from either a given DM channel or a single pulsar. For each transport parameter set, we choose the best {\it p}-value from all the 12 (five) channels for the DM (single pulsar) contribution considered in this analysis. In both cases, very small ($L \lesssim 2$\,kpc) and very big halo sizes ($L \gtrsim 7$\,kpc) as well as small diffusion slopes ($\delta \lesssim 0.5$) are disfavoured by the experimental data due to the different spectral features at high energies of the additional contribution. In our analysis, the benchmark models MIN and MAX are largely disfavoured by the experimental data. However these constraints are model dependent since they are sensitive to the shape of the additional contribution.

%
\section{Discussion and conclusions}
\label{sec:discussion_conclusion}

This analysis aimed at testing the DM and pulsar explanations of the cosmic ray positron anomaly with the most recent
available AMS-02 data. A first very important observation that we made is the sensitivity of the results to the lepton
flux $\Phi_{L}$ used in the derivation of the positron fraction. This is particularly obvious from the comparison between
Tables~\ref{tab:MED_onechannel_PF14_ICHEP14} and \ref{tab:MED_onechannel_PF13_Fermi} where the DM
hypothesis is investigated. The improved accuracy of AMS-02 on the lepton flux now excludes channels previously
allowed when small $\chi^{2}$ values were still easily obtained.

\vskip 0.1cm
As regards the DM analysis, we first performed a $\chi^{2}$ minimization analysis for each single channel and found that
leptons are strongly disfavoured by the recent AMS-02 data. Lepton pairs never provide a good fit, although the situation slightly
improves for secluded DM candidates whose annihilations into four leptons proceed through light vector or scalar mediators.
On the contrary, the measurements are well explained in the case of quarks and gauge or Higgs bosons, with a preferred
DM mass between 10 and 40~TeV, and a large annihilation cross section of the order of $10^{-21}$ cm$^{3}$ s$^{-1}$.
These large DM annihilation rates, however, also yield gamma rays, antiprotons, and neutrinos for which no evidence has been
found so far. In particular, the upper limits on the DM annihilation cross section for a given mass and annihilation channel
obtained from the observations of dwarf spheroidal galaxies challenge the DM interpretation of the positron anomaly.
For example, the best fit values given in Table~\ref{tab:MED_onechannel_PF14_ICHEP14} for the $\tau^{+} \tau^{-}$,
$b \bar{b}$, and $W^{+}W^{-}$ channels assuming the MED propagation parameters are all excluded by one order of
magnitude by Fermi-LAT combined analysis of dwarf galaxies\citepads{2013arXiv1310.0828T},
VERITAS\citepads{2012PhRvD..85f2001A}, and MAGIC\citepads{2014JCAP...02..008A} observations of Segue 1, as well as
by the recent results of the H.E.S.S. collaboration \citepads{2014arXiv1410.2589H} on Sagittarius and other dwarf galaxies.
As mentioned in the introduction, measurements of the CMB temperature and polarisation allows us to put constraints on the annihilation
cross section of DM. For example, for $m_{\chi} = 1$~TeV, cross sections larger than roughly $5 \times 10^{-24}$ cm$^3$ s$^{-1}$
are excluded\citepads{2012JCAP...12..008G} for a value of the energy injected typical of DM annihilation into
$\tau^{+} \tau^{-}$\citepads{2009PhRvD..80d3526S}, thus constraining our best-fit single-channel scenarios. The Higgs and the
four-tau channels, for which dwarf galaxies do not provide any information, are excluded by\citetads{2013JCAP...03..044C} from
their CMB analysis.
The IceCube neutrino observatory has also derived upper limits on the DM annihilation cross section in different
channels. Those are especially stringent at high masses, however they are only in tension with the positron anomaly best fit with MED
propagation parameters when DM annihilates into $W^{+}W^{-}$\citepads{2013PhRvD..88l2001A}.
In summary, we have not found a single channel case that accounts for the positron data and still survives the severe tests presented
above.

\vskip 0.1cm
Combining the leptonic channels together lessens the tension with the data only for a DM mass of 500~GeV. The limits
set by MAGIC from Segue~1 and Fermi-LAT from dwarf galaxies are, however, a factor 7 to 10 below the required annihilation
cross section into $\tau^{+} \tau^{-}$ pairs and exclude a pure leptonic mixture.
Adding $b$ quarks significantly improves the fit and allows us to accomodate DM masses from 500~GeV up to 40~TeV.
However, the observations of Segue 1 by MAGIC set stringent constraints on that possibility. DM species up to 20~TeV are
excluded because of their large branching ratios into $\tau^{+} \tau^{-}$ pairs, whereas heavier particles exceed the upper
limit set on the annihilation cross section into $b$ quarks.
Inspired by universal extra dimension models, we also impose equal branching ratios for the three lepton families while allowing a $b$ quark admixture.
Excellent agreement with the data is found for a DM mass between 5 and 40~TeV. Once again, dwarf satellites severely constrain
that possibility, which is excluded by at least one order of magnitude by the recent analyses by MAGIC and\citetads{2014arXiv1410.2589H}.
%
The case where DM annihilates into four leptons is much less constrained by dwarf galaxies. Only H.E.S.S. and VERITAS analysed
that possibility for the electron and muon channels, but no limit has been derived so far on DM annihilating into $4\tau$. This
case is nevertheless constrained by the CMB analysis of\citetads{2013JCAP...03..044C}.
A combination between the four-tau (75\%) and four-electron (25\%) channels turns out to provide a good fit to the AMS-02 data
for a DM mass between 0.5 and 1~TeV, and evades the above mentioned bounds. This is the only viable case where a DM species
accounts for the positron anomaly while satisfying all known constraints from gamma ray or cosmological measurements. The
corresponding positron signal is featured in the right panel of Fig.~\ref{fig:4leptons_case} for a 600~GeV DM particle.

\vskip 0.1cm
In the same way, we have shown that the rise of the positron fraction can be alternatively explained by an additional contribution from a single pulsar. Indeed, five pulsars from the ATNF catalogue have been identified to satisfy the experimental measurements within their distance uncertainties. 
For all the selected pulsars we obtain an excellent fit result even though the adjustment of the last few high-energy data points is unsatisfactory. However, this can be improved by decreasing the pulsar distance within its uncertainty.
AMS-02 is expected to take data for more than ten years reducing considerably its statistical uncertainties especially for the highest energies. If the trend of the positron fraction remains the same, our analysis shows that ten years of data could completely exclude the single pulsar hypothesis.
Naturally, assuming a pulsar origin for the positron fraction rise leads to a cumulative contribution from all detected and yet undiscovered pulsars. Nevertheless, demonstrating that the positron fraction can be explained by a unique pulsar contribution provides us with a valid alternative to the DM explanation of this anomaly. 
As a matter of fact, if the single pulsar hypothesis is viable, the entirety of detected pulsars is hence capable of reproducing the experimental data.
However, since the normalisation of the pulsar source term and the annihilation cross-section of dark matter are treated as free parameters, both pulsars and dark matter could contribute to the positron anomaly.
\vskip 0.1cm
The above conclusions were drawn assuming a given set of cosmic ray transport parameters derived from the boron-to-carbon analysis of~\citetads{2001ApJ...555..585M} and dubbed MED in \citetads{2004PhRvD..69f3501D}. However, the transport mechanisms of charged cosmic rays are still poorly understood, necessitating the inclusion of their uncertainties in the studies of the rise of the positron fraction. In this work, we use 1623 different transport parameter sets, all in good agreement with nuclear measurements. We observe that the error arising from the propagation uncertainties is much larger than the statistical uncertainty on the fitted parameters. In conclusion, the ignorance of the exact transport parameter values is the main limitation of such analyses. Henceforth, the study of cosmic ray propagation should be the main focus of future experiments.

\vskip 0.1cm
We have used the recent positron fraction measurement performed by the AMS-02 collaboration \citep{Accardo:2014lma}.
However,  AMS-02  recently published a new measurement of the electron and positron fluxes\citepads{2014PhRvL.113l1102A} up to an energy
of 700 and 500~GeV, respectively\footnote{During the refereeing process of this paper, the AMS-02 collaboration published the all lepton flux \citep{Aguilar:2014fea}. Since the final flux is consistent with the preliminary flux used here~\citep{leptons_ICHEP_2014}, the results and conclusions of our analysis hold.}.
We intend to repeat this analysis using the positron flux to check that our conclusions hold as they should.
Modelling the positron energy losses differently, changing the primary proton and helium fluxes, as well as taking
the actual gas distribution into account in the Galactic disc may also have an impact that needs to be assessed.

%
\vskip 1.0cm
\begin{acknowledgements}
We would like to thank P.~Serpico for his kind help in the first stages of this work and for enlightening discussions.
We are also greatly indebted to G.~Giesen and M.~Cirelli who have helped us with the four-lepton spectra from
WIMP annihilation.
Part of this work was supported by the French \emph{Institut universitaire de France}, by the French
\emph{Agence Nationale de la Recherche} under contract  12-BS05-0006 DMAstroLHC,
and by the \emph{Investissements d'avenir}, Labex ENIGMASS.
\end{acknowledgements}



%
\bibliographystyle{aa}
\bibliography{CRAC_1_biblio}

\end{document}